\documentclass[twocolumn,twocolappendix]{aastex631}

\usepackage{newtxtext,newtxmath}
\usepackage[T1]{fontenc}

\newcommand{\refadd}[1]{#1}

\shorttitle{Symfind: The Durability of Subhalos}
\shortauthors{Mansfield et al.}

\usepackage{graphicx}
\usepackage{amsmath}
\usepackage{amssymb}	
\usepackage{wasysym}
\usepackage{xcolor}

\begin{document}
\title{Symfind: Addressing the Fragility of Subhalo Finders and Revealing the Durability of Subhalos}

\author[0000-0001-9863-5394]{Philip Mansfield}
\affiliation{Kavli Institute for Particle Astrophysics \& Cosmology, P. O. Box 2450, Stanford University, Stanford, CA 94305, USA}
\affiliation{SLAC National Accelerator Laboratory, Menlo Park, CA 94025, USA}

\author[0000-0002-8800-5652]{Elise Darragh-Ford}
\affiliation{Kavli Institute for Particle Astrophysics \& Cosmology, P. O. Box 2450, Stanford University, Stanford, CA 94305, USA}
\affiliation{SLAC National Accelerator Laboratory, Menlo Park, CA 94025, USA}
\affiliation{Department of Physics, Stanford University, 382 Via Pueblo Mall, Stanford, CA 94305, USA}

\author[0000-0001-8913-626X]{Yunchong Wang}
\affiliation{Kavli Institute for Particle Astrophysics \& Cosmology, P. O. Box 2450, Stanford University, Stanford, CA 94305, USA}
\affiliation{SLAC National Accelerator Laboratory, Menlo Park, CA 94025, USA}
\affiliation{Department of Physics, Stanford University, 382 Via Pueblo Mall, Stanford, CA 94305, USA}

\author[0000-0002-1182-3825]{Ethan O.~Nadler}
\affiliation{Carnegie Observatories, 813 Santa Barbara Street, Pasadena, CA 91101, USA}
\affiliation{Department of Physics $\&$ Astronomy, University of Southern California, Los Angeles, CA, 90007, USA}

\author[0000-0001-9568-7287]{Benedikt Diemer}
\affiliation{Department of Astronomy, University of Maryland, College Park, MD 20742, USA}

\author[0000-0003-2229-011X]{Risa H.~Wechsler}
\affiliation{Kavli Institute for Particle Astrophysics \& Cosmology, P. O. Box 2450, Stanford University, Stanford, CA 94305, USA}
\affiliation{SLAC National Accelerator Laboratory, Menlo Park, CA 94025, USA}
\affiliation{Department of Physics, Stanford University, 382 Via Pueblo Mall, Stanford, CA 94305, USA}

\correspondingauthor{Philip Mansfield}
\email{phil1@stanford.edu}


\begin{abstract}
A major question in $\Lambda$CDM is what this theory \emph{actually} predicts for the properties of subhalo populations. Subhalos are difficult to \refadd{accurately} simulate and to find within simulations, and this propagates into uncertainty in theoretical predictions for satellite galaxies. We present \textsc{Symfind}, a new particle-tracking-based subhalo finder, and demonstrate that it can track subhalos to orders-of-magnitude lower masses than commonly used halo-finding tools, with a focus on \textsc{Rockstar} and \textsc{consistent-trees}. These longer survival times mean that at a fixed peak subhalo mass, we find $\approx\,15\%{-}40\%$ more subhalos within the virial radius, $R_\textrm{vir}$, and $\approx\,35\%{-}120\%$ more subhalos within $R_\textrm{vir}/4$ in the Symphony dark-matter-only simulation suite. More subhalos are found as resolution is increased\refadd{, in contrast to the \textsc{Rockstar} halo finder, which appears to be converged at smaller subhalo counts}. We perform extensive numerical testing. In agreement with idealized simulations, we show that the $v_{\rm\,max}$\refadd{, the maximum circular velocity, is systematically biased low until high resolutions ($n_\textrm{peak}\gtrsim3\times\,10^4$) are achieved}, but that mass loss itself can be resolved at much more modest resolutions ($n_\textrm{peak}\gtrsim4\times\,10^3$). We show that \textsc{Rockstar} converges to false solutions for the mass function, radial distribution, and disruption masses of subhalos. We argue that our new method can trace resolved subhalos until the point of typical galaxy disruption without invoking \refadd{post-hoc} ``orphan'' modeling. We outline a concrete set of steps for determining whether other subhalo finders meet the same criteria. We publicly release \textsc{Symfind} catalogs and particle data for the Symphony simulation suite at \url{http://web.stanford.edu/group/gfc/symphony}.
\end{abstract}




\section{Introduction}
\label{sec:intro}

Many open questions in cosmology are about the state of the universe: questions like ``what is the nature of dark matter?'' or ``how has the clustering of matter evolved over time?'' However, many are also questions about the predictions of a specific model. This second class of questions is interesting because they hamper our ability to answer the first type of question. At present, $\Lambda$CDM --- i.e., a popular class of cosmological models that contain both ``cold'' dark matter and a cosmological constant --- suffers from an open question of this second kind: there is substantial uncertainty over how subhalos behave and disrupt in $\Lambda$CDM. This uncertainty is a major systematic in numerous cutting-edge cosmological probes.

In $\Lambda$CDM, all galaxies form within massive dark matter structures known as {\em halos} \citep{white_reese_1978_core_condensation,wechsler_tinker_2018_connection}. Large galaxies are surrounded by swarms of small {\em satellite} galaxies that inhabit their own dark matter {\em subhalos}. The lives of satellite galaxies are dramatic: originally isolated galaxies in their own right, they are accreted onto hosts and pulled into orbits that can vary between leisurely transits across the host halo's outskirts to rapid, disruptive encounters with the host's center. Throughout its orbit, mass in a subhalo's outskirts is pulled away by the gravitational field of the host, and this decrease in mass causes the region of the subhalo protected from the host's gravity --- the region inside the {\em tidal radius} \citep[see review in][]{vandenbosch_2018_disruption} --- to decrease. Simultaneously, tidal shocks heat the interior of subhalos at each pericentric passage, causing the interior mass to expand outwards towards the encroaching tidal radius (e.g., \citealp{hayahi_2003_structural}; also see historical review in \citealp{moore_2000_brief_history}).
This leads to runaway, exponential mass loss \citep{tormen_1998_survival,klypin_1999_overmerging}. Although the satellite galaxy is initially insulated from this mass loss, eventually, its tidal radius decreases so much that even the galaxy is torn apart \citep{penarrubia_2008_tidal_evolution,smith_2016_preferential}.

Modifying the nature of dark matter can change the abundance, properties, and durability of subhalos \citep[e.g.,][]{lovell_2014_properties,nadler_2021_effects}, which makes observations of satellite galaxies a rich cosmological probe. Tantalizingly, there is no shortage of apparent conflicts between observed satellite galaxies and the predictions of $\Lambda$CDM \citep[see reviews in][]{bullock_2017_small_scale,bechtol_2022_snowmass}. Relative to observations, simulated satellite groups {\em appear} to be too diffusely concentrated around their hosts \citep[e.g.][see Section \ref{sec:radius_cdf} for further review]{carlsten_2022_elves}, to be too isotropically distributed \citep[e.g.][]{pawlowski_2018_planes}, and to potentially have incorrect dark matter distributions \citep[e.g.,][]{oman_2015_unexpected,hayashi_2020_diversity}. Historically, a large body of literature has been written on the potential tension between the observed abundance of satellite galaxies and the simulated abundance of subhalos \citep[e.g.,][]{moore_dark_1999,klypin_1999_where,bolyan_kolchin_2011_tbtf}, but both the original formulation of this problem (the ``missing satellites problem'') and a more challenging reformulation (``too big to fail'') seem to be resolved by a combination of improved observational programs and better modeling of selection effects \citep[e.g.,][]{newton_2018_total_sat_pop,kim_2018_midding_sats,drlica_wagner_2020_census,nadler_2020_census}, more realistic treatments of galaxy formation physics \citep[e.g.,][]{benson_2002_missing_sats,somerville_2002_missing_sats,kravtsov_2004_tumultuous,wetzel_2016_reconciling,lovell_2017_tbtf}, and accounting for the impact of the central galaxy's potential on satellite disruption \citep[e.g.,][]{brooks_2013_tbtf,garrison_kimmel_2017_lumpy}.

\refadd{Accurate satellite modeling} also impacts our ability to infer cosmological parameters from large-scale clustering statistics (e.g., the probability of pairs of galaxies being separated by a given distance). Altering the properties of dark energy and other cosmological parameters changes the rate that structure forms and can lead to large changes in these statistics, particularly at ``small'' scales ($r\lesssim$ 10 Mpc; e.g., \citealp{wechsler_tinker_2018_connection}). However, the durability of satellites directly impacts these clustering statistics. This is true even at scales well beyond the radius of an individual halo due to satellites in neighboring halos adding weight to the correlation function \citep[the so-called ``two-halo term,'' e.g.,][]{cooray_sheth_2002PhR_halo_model}. As a result, the impact of cosmology on some clustering statistics can be canceled out by corresponding changes in the model used for satellites \citep[e.g.,][]{wechsler_tinker_2018_connection}. Some popular models that attempt to match these clustering statistics by populating simulated subhalos with galaxies cannot match observations self-consistently unless one assumes that galaxies and their subhalos far outsurvive their simulated counterparts (\citealp{campbell_2018_crisis}, Appendix B in \citealp{behroozi_2019_universemachine}). There is controversy over whether similar assumptions are needed to match the radial distribution of satellites (see overview in Section \ref{sec:radius_cdf}).

Does this mean that satellite galaxies \refadd{signal the end} for $\Lambda$CDM? Has such a far-reaching theory been undone by its smallest and most humble predictions? Not necessarily. The behavior of satellite galaxies and subhalos is a non-linear process that is primarily understood through numerical simulations. To make predictions for satellite populations from simulations, one must be able to configure those simulations in a way that allows for subhalo evolution to be properly resolved and must be able to reliably extract subhalo information from simulation outputs. Both steps are non-trivial.

The first step, running numerically reliable simulations, relies on {\em convergence testing}. Convergence testing is a form of correctness testing that compares simulation behavior at varying resolutions \citep[e.g.,][]{ludlow_2019_numerical}. The true behavior of a system cannot depend on any purely numerical parameter, so simulation results are not correct in any region of numerical parameter space where small changes in numerical parameters lead to meaningful changes in those results. {\em Convergence} --- agreement between resolution levels --- is a necessary but insufficient condition for correctness. Convergence without correctness is called {\em false convergence}, and false convergence has been observed even in some of the largest cosmological simulations ever run \citep{mansfield_avestruz_2021_biased}. Currently, one of the most pressing issues in this form of testing is whether subhalo populations converge at the resolution levels typically analyzed in cosmological simulations. Subhalos in cosmological simulations tend to disrupt quickly, even at high resolutions \citep{vdb_2017_dissecting,jiang_vdb_2017_statistics_iii,han_2016_unified,behroozi_2019_universemachine,diemer_2023_haunted}, but this does not seem to be the correct behavior of subhalos in $\Lambda$CDM.

Rapid disruption is certainly the correct behavior for very large subhalos ($m_{\rm peak}/M_{\rm vir} \gtrsim 0.1$; Darragh-Ford et al., in prep). Dynamical friction quickly saps orbital energy from these subhalos, causing them to sink to the centers of their hosts within a few orbits \citep[e.g.,][]{vasiliev_2022_radialization} and to meld into their hosts' smooth matter distributions. However, dynamical friction is far weaker for low-mass subhalos (e.g., \citealp{vdb_2016_segregation}; see also Section \ref{sec:radius_cdf} for extended discussion) and generally does not cause these subhalos to sink to the host's center on observationally relevant timescales. These low-mass subhalos can still experience true disruption under certain alternative cosmologies and baryonic physics formulations that cause low-density central cores in subhalos, as the lowered density makes them less resilient to  tidal fields \citep[e.g.,][]{penarrubia_2010_impact,errani_2023_core_survival}. Historically, there has been some debate over whether the same can occur in the ``cuspy'' high-density centers of pure-$\Lambda$CDM subhalos (see \citealp[e.g.,][]{errani_navarro_2021_asymptotic} for review), but modern high-resolution, idealized simulations strongly predict that this is not the case: low-mass subhalos can survive as shrinking bound remnants for essentially arbitrarily long periods of time \citep{penarrubia_2010_impact,vandenbosch_2018_disruption,errani_penarrubia_2020_can_tides,errani_navarro_2021_asymptotic}. Even tidal shocks from disc potentials are not able to fully disrupt these subhalos \citep{green_2022_disc}. Because changes in cosmology and galaxy formation physics can decrease the durability of subhalos, simulation techniques that erroneously lead to rapid disruption are particularly pernicious, allowing pure-$\Lambda$CDM simulations to falsely emulate these effects and thus hampering analysis that compares these types of models. 

To make matters worse, convergence testing between cosmological simulations suggests that modest and easily achieved resolution levels are enough to ensure convergence in subhalo abundances \citep[e.g.,][$\gtrsim10^2$ to $10^3$ particles depending on the statistic]{mansfield_avestruz_2021_biased,nadler_2022_symphony}. But idealized simulations suggest that orders of magnitude more particles are needed to prevent numerical effects from causing subhalos to lose mass too quickly, especially for old subhalos \citep[e.g.,][$\approx10^5$ particles needed, see also Sections \ref{sec:numerical_limits}, \ref{sec:structural}, and \ref{sec:mass_loss_rates} for extended discussion]{vandenbosch_2018_disruption}. This tension leads to an uncomfortable question: is the apparent reliability of subhalo analysis in cosmological simulations merely false convergence?

Assuming that subhalos can be simulated reliably, they must also be identified within simulation outputs. {\em Halo finders} are software packages that attempt to identify halos and their subhalos within simulations. Finding isolated halos is mostly a solved problem \citep{knebe_2011_MAD} --- except for ambiguities about halo boundaries \citep[e.g.,][]{more_2011_fof,more_2015_splashback,diemer_2021_flybys} and about the complexity of mergers between equal-mass halos \citep[e.g.,][]{behroozi_et_al_2014_mergers} --- so one of the most important properties of a halo finder is how effective they are at identifying and measuring the properties of subhalos. Most halo finders work primarily within a single snapshot, requiring a second tool, a {\em merger-tree} code that connects halos and subhalos across time. 
The split between the halo finder and merger tree is not always clear: some halo finders use information from previous timesteps or may even explicitly track a halo/subhalo's particles over time (see Section \ref{sec:method_comp} for more details). The wide variety of subhalo finders is \refadd{at} least partly caused by inherent difficulty in finding subhalos. Subhalos are enveloped in dense streams of their own \refadd{tidally stripped} matter, they must be identified against the complex background of the host's density field and can be confused with non-subhalo structure within the host halo, such as fluctuations and the sloshing of dark matter as the host settles into equilibrium after a major merger. Testing halo finders is also difficult: beyond convergence testing (see above), one option is to test whether finders can recover idealized subhalos manually placed into a host halo \citep[e.g.,][]{knebe_2011_MAD}, and a second is to compare the performance of different halo finders across realistic halos \citep[e.g.,][]{knebe_2011_MAD,onions_2012_notts,onions_2013_notts_spin,srisawat_2013_sussing,avila_2014_sussing, behroozi_et_al_2014_mergers,elahi_2019_velociraptor}. The former method suffers from the fact that much of the difficulty in subhalo finding comes from the complex interplay between host and subhalo, or subhalo and subhalo remnant, meaning that idealized tests will overestimate a tool's reliability. The latter method suffers from the fact that the researcher usually does not know the correct answer ahead of time. If two packages disagree, how does one know if one is under-predicting, the other is over-predicting, or both are wrong?

In this paper, we aim to make significant progress on these questions. After outlining the data, tools, and definitions used in this paper in Section \ref{sec:sims}, we present a new subhalo-finding method based on ``particle-tracking'' in Section \ref{sec:methods}, \textsc{Symfind}. In Section \ref{sec:reliability}, we perform extensive testing on the reliability of this method and on the convergence properties of subhalos in general. In Section \ref{sec:subhalo_pops}, we investigate the impact of our method on subhalo populations. In Section \ref{sec:good_enough}, we argue that our subhalo finder (and any subhalo finder with similar performance) will no longer be the limiting factor for analyzing the abundance of satellite galaxy populations. In Section \ref{sec:method_comp}, we compare with other methods, and in Section \ref{sec:conclusions}, we provide our conclusions.

Throughout this paper, we use lower-case letters to label the properties of subhalos (e.g., $m$, $n$, $v_{\rm max}$, $r$), and upper-case letters to label the properties of central/host halos (e.g., $M_{\rm vir}$, $R_{\rm vir}$). These central halos are sometimes referred to as ``main subhalos'' in the literature.

\section{Simulations, Codes, and Definitions}
\label{sec:sims}

The analysis in this paper makes extensive use of five of the Symphony simulation suites: SymphonyLMC, SymphonyMilkyWay, SymphonyMilkyWayHR, SymphonyGroup, and SymphonyL-Cluster \citep{mao_2015_dependence,bhattacharyya_2022_lcluster,nadler_2022_symphony}. The full details of these simulations can be found in \citet{nadler_2022_symphony}, and we list the most important parameters of these simulations in Table \ref{tab:symphony}. 

Our scientific results focus primarily on characterizing the  average subhalo populations in SymphonyMilkyWay. In some cases where numerical behavior does not depend on central halo mass or particle mass, we stack all the suites together to improve number statistics. Some analysis requires isolating the impact of resolution from subhalo mass, in which case we compare the high-resolution resimulations in SymphonyMilkyWayHR with a subset of SymphonyMilkyWay.

SymphonyMilkyWayHR consists of the four central halos in SymphonyMilkyWay with the smallest Lagrangian regions. These halos were resimulated with particle masses that were eight times smaller and force-softening scales that were two times smaller than the fiducial suite. This selection process allowed for lower cost resimulations, but also means that these four objects are not representative of the entire Milky Way-mass sample. Most notably, our testing found that this high-resolution subsample has fewer high-mass subhalos and a different distribution of subhalo mass loss rates than the full suite. This means that for some resolution tests, this suite cannot be directly compared against the full SymphonyMilkyWay suite and needs to be compared against only their fiducial-resolution re-simulations.

The original SymphonyMilkyWayHR suite contained five hosts, but we remove the fourth SymphonyMilkyWayHR host, Halo530, from both the fiducial and high-resolution simulation sets when performing matched analysis. Several high-mass subhalos and their sub-subhalos that were accreted in the high-resolution run were never accreted in the fiducial resolution run, as determined by manual inspection and position-based cross-matching. \refadd{This is likely a manifestation of the ``butterfly effect,'' in which two simulations with minor differences in initial conditions can lead to surprisingly different outcomes \citep{genel_2019_butterfly,borrow_2023_stochastic}}. This difference leads to very different subhalo populations. No similar mismatches were found in any other host pairs. 

\begin{table}
\centering
\begin{tabular}{l|c|c|c|c}
\hline
Simulation & $N_{\rm host}$ & $M_{\rm vir}$ & $m_p$ & $\epsilon$ \\ 
& & ($M_\odot$) & ($M_\odot$) & (kpc)\\ [0.5ex]
\hline\hline
SymphonyLMC & 39 & $10^{11.02}$ & $5.0\times 10^4$ & 0.08 \\
SymphonyMilkyWay & 45 & $10^{12.09}$ & $4.0\times 10^5$ & 0.17 \\
SymphonyMilkyWayHR & 4 & $10^{12.07}$ & $5.0\times 10^4$ & 0.08 \\
SymphonyGroup & 49 & $10^{13.12}$ & $3.3\times 10^6$ & 0.36 \\
SymphonyL-Cluster & 33 & $10^{14.62}$ & $2.2\times 10^8$ & 1.2 \\
\hline
\end{tabular}
\caption{The most important parameters of the simulation suites used in this paper. The first column gives the name of the simulation, the second gives the number of unique hosts in the suite, the third gives the median host mass, and the final two columns give the particle mass and comoving, Plummer-equivalent force softening scale, respectively.}
\label{tab:symphony}
\end{table}

We make substantial use of the \textsc{Rockstar} subhalo finder \citep{behroozi_2013_rockstar} and \textsc{consistent-trees} merger tree code \citep{behroozi_2013_consistent}. Both tools are widely used, and a common reading of the testing literature is that they perform at least as well as most other subhalo finders and merger tree codes, respectively (see discussion and caveats in Section \ref{sec:method_comp}). To simplify language, we refer to the combined \textsc{Rockstar}+\textsc{consistent-trees} pipeline as ``RCT'' and both steps simply as \textsc{Rockstar} in figures, as is commonly done in the literature. We also make use of the \textsc{Subfind} halo finder \citep{springel_2001_populating}.

\subsection{Halo Property Definitions}
\label{sec:definitions}

We define a halo as becoming a {\em subhalo} at its snapshot of first infall and as a {\em central halo} before this point. A {\em host halo} is a central halo that a subhalo has fallen into at some point in the past. {\em First infall} is defined as the first snapshot at which a subhalo is within the {\em virial radius} ($R_{\rm vir},$ see below) of a more massive halo. \refadd{In practice, this definition becomes complicated in the presence of halo finder errors which can cause subhalos to briefly be misidentified as the host halo \cite{behroozi_2015_major_mergers}.} We correct these issues using the methods described in Appendix \ref{sec:tree_post_processing}. A consequence of this definition is that it includes {\em splashback} subhalos (subhalos whose orbits have temporarily taken them outside the virial radius of their host halo, e.g., \citealp{diemer_2021_flybys} and references therein) and flyby subhalos (former subhalos who have truly been ejected from their host, often due to three-body interactions; e.g., \citealp{ludlow_2009_unorthodox}). True ``flyby'' subhalos are rare: only 1-2\% of all subhalos that have left their host's virial radius are outside the extended splashback surface \citep{mansfield_2020_three}, so classifying both objects as subhalos is a reasonable approximation.

We take two definitions of halo mass. Central halos are characterized by their virial mass, $M_{\rm vir}$, the total bound mass within the virial radius, $R_{\rm vir},$ defined relative to a characteristic density, $\rho_{\rm vir}$, such that
\begin{align}
    M_{\rm vir} = \frac{4\pi}{3}\rho_{\rm vir}R_{\rm vir}^3.
\end{align}
We adopt the \citet{bryen_norman_1998_statistical} definition of $\rho_{\rm vir}.$ For the cosmology used by SymphonyMilkyWay, $\rho_{\rm vir}=99.2\rho_c$ at $z=0.$ For subhalos, our definition of subhalo mass is dependent on the subhalo finder. RCT subhalo masses are virial masses computed using only the bound particles within that subhalo's local phase-space overdensity. \textsc{Symfind} subhalo masses are the sum of the masses of all bound particles within that subhalo's tracked particle set. We label both masses as $m$. There are meaningful differences between these two mass definitions (see Appendix \ref{sec:iterative}). RCT only uses a single ``unbinding pass'' to compute boundedness, increasing masses by $\approx 5\%$ relative to \textsc{Symfind}'s full unbinding, and RCT also tends to include some host particles within the subhalo, further increasing masses by $\approx5\%$. The difference between the total bound mass and the total bound mass within the virial radius is small for subhalos: we find an $\approx 2\%$ effect in \textsc{Symfind}.

In some places, we also characterize subhalo masses via $v_{\rm max}$, the maximum value of the rotational velocity $v_{\rm rot}(r) = \sqrt{G\,m(<r)}/r$ for $r>\epsilon$. $v_{\rm max}$ is closely related to $m$ in central halos, but decreases more slowly than $m$ in disrupting subhalos (see Section \ref{sec:structural}).

There are numerous ways to characterize a subhalo's mass prior to infall. In this paper, we use $m_{\rm peak}$ and $v_{\rm peak}$, the maximum values of $m$ and $v_{\rm max}$, respectively, {\em prior to the snapshot when the subhalo first became a subhalo}. This latter condition is non-standard and has been introduced to avoid certain halo finder errors (see Fig.~\ref{fig:example_halo}, Section \ref{sec:stitching_errors}, and Appendix \ref{sec:tree_post_processing}). In some places in this paper, we compare against studies that used alternative definitions of a halo's pre-infall mass such as $v_{\rm infall}$ (the value of $v_{\rm max}$ at the snapshot of first infall), $m_{\rm infall}$ (the value of $m$ at the snapshot of first infall), and $v_{\rm Mpeak}$ (the value of $v_{\rm max}$ at the snapshot when $m$ reaches its maximum pre-infall value). The distinction between these different definitions is at the few-to-ten percent level and matters for certain classes of empirical models \citep[e.g.,][]{reddick_2013_connection}, so we switch to the appropriate definition when necessary.

\subsection{Merger Tree Terminology}
\label{sec:merger_tree_term}

\begin{figure}
\hspace{-3.5mm}
\includegraphics[width=0.49\textwidth]{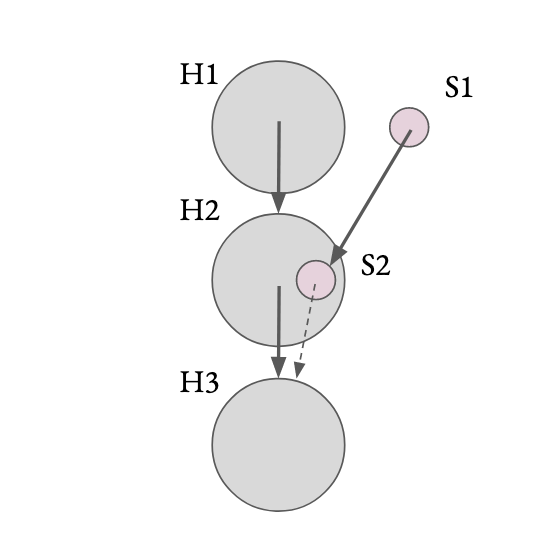}
\caption{\refadd{A cartoon illustrating the terminology used to discuss merger trees in this paper (see Section \ref{sec:merger_tree_term} for definitions). Shown is a three snapshot simulation containing a grey host (H1, H2, H3) and pink subhalo (S1, S2) which disrupts between snapshots 2 and 3. Each circle is a {\em halo}, and each arrow points from a {\em progenitor} to {\em descendant}. H1 and S1 are {\em leaf} halos and H3 is a {\em root} halo. The connection between S2 and H3 due to disruption is a {\em tree-merger}, while the infall of the intact S2 into H2 is a {\em merger}. Only the thick, solid arrows are {\em branches} of the merger tree. We do {\em not} refer to the arrow connecting S2 to H3 as a branch, although it remains part of the tree connectivity.}}
\label{fig:merger_tree_def}
\end{figure}

The evolution of halos over time is represented by a structure called a {\em merger tree}. The structure is tree-shaped with respect to time because halos can merge together over time but generally do not split apart unless a serious error has occurred in the halo finder. We briefly define the most important terminology here. A {\em halo} is a structure that is found within a single snapshot. Every halo is matched with at most one halo in a subsequent snapshot: the former halo is called a {\em progenitor}, and the latter is called a {\em descendant}. A halo with no progenitors is called a {\em leaf} halo, and one with no descendants is called a {\em root} halo. Some unbroken paths which start at leaves and progressively pass from descendant to descendant are called {\em branches} (see below). 

A halo can have multiple progenitors, an event called a {\em tree-merger}. A common source of confusion is that there are three similar but distinct events commonly referred to as ``mergers'' in the literature. The first is when a subhalo first falls into a host halo. The second is after this subhalo has lost so much mass that the halo finder cannot track it anymore. This second event can occur many Gyr later and is highly dependent on the halo finder and merger tree code used. The third is when the galaxy hosted by a subhalo merges with its host galaxy's stellar halo. We refer to the first type of events as ``mergers'' and the second type as ``tree-mergers.'' We do not analyze galaxy mergers directly in this paper. Still, their existence is quite important to evaluating the quality of merger tree codes, as we discuss in Section \ref{sec:good_enough}.

All merger trees have a method for choosing which of a halo's progenitors are disrupting subhalos and which progenitor is the same halo at a previous time. This latter progenitor is called its {\em main progenitor} and is usually the more massive of the two halos. A branch consisting of only main progenitors is called a {\em main branch}. Qualitatively, a main branch represents the evolution of a single halo over time.

Different authors use different terminology when defining which linkages can be considered part of the same branch and how large branches are (for example, if A is the main progenitor of B, but a separate halo, C, merges with A to form B, is C part of the same branch as B? What about B's descendants?). In this paper, we take the convention that a halo can only be a member of one branch and that linkages coming from tree mergers are not part of any branch, even though they are part of the connectivity of the tree. A consequence of this definition is that the term branch can {\em only} refer to paths along main branches, allowing us to use the terms ``branch'' and ``main branch'' interchangeably.

\section{Methods}
\label{sec:methods}

\begin{figure*}
\hspace{-3.5mm}
\includegraphics[width=0.95\textwidth]{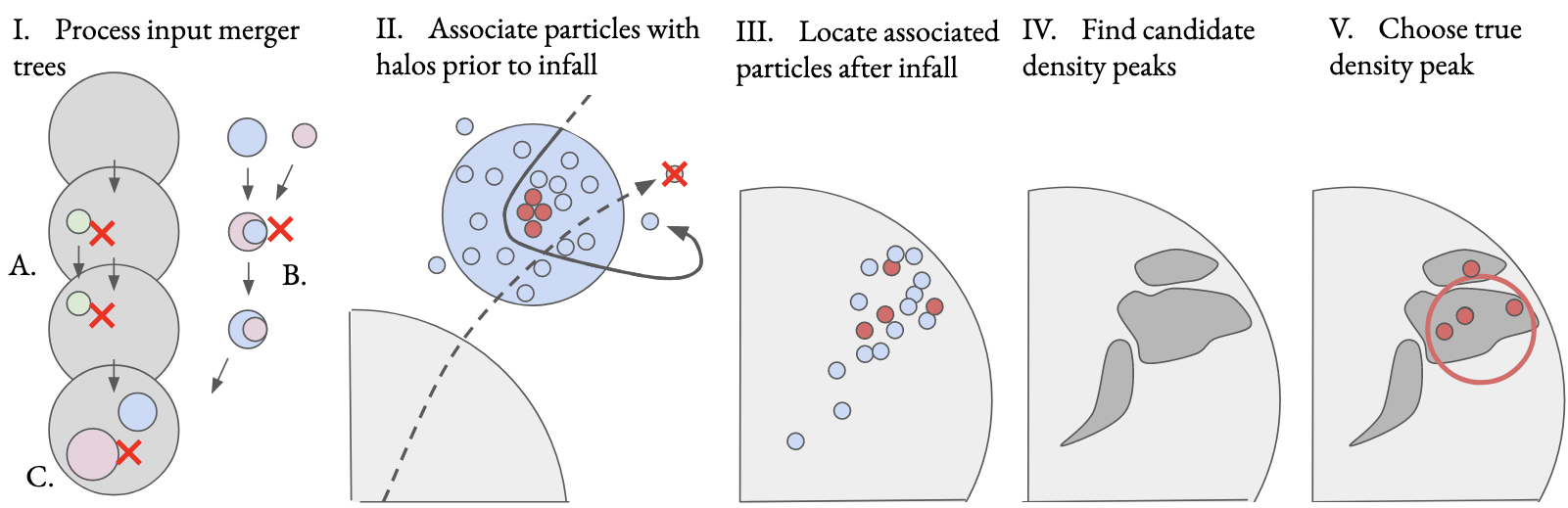}
\caption{Cartoon illustrating the major steps in our subhalo-finding method, \textsc{Symfind}. {\em Panel I}: First, we annotate an input merger tree (for this paper, input catalogs are generated by \textsc{Rockstar}), identifying and correcting various errors (Appendix \ref{sec:tree_post_processing}). Here, the red X's indicate portions of the input merger tree that would be removed or corrected. \refadd{Three events are labeled: A, a subhalo ``formed'' within its host, B, numerical mass switching during pre-infall mergers, and C, numerical fluctuations in mass post-infall.} {\em Panel II:} Next, using this annotated tree, we find the first halo branch that every particle ever ``smoothly'' accreted onto. We also track the ``non-smooth'' accretion of particles from lower mass halos to higher mass halos, but not the inverse (Appendix \ref{sec:particle_associations}). Here, solid lines show particles that are tracked for this subhalo and dashed show particles that are untracked. {\em Panel III:} Once a halo becomes a subhalo, we stop using the input merger tree and compute properties from the tracked particles. At the snapshot of the first infall, we flag a subset of highly bound particles as ``core'' particles (Appendix \ref{sec:identifying_substructure}). Here, core particles are in red. {\em Panel IV:} Using an existing halo finder (currently \textsc{Subfind}), we identify density peaks within the tracked, smoothly accreted particles (Appendix \ref{sec:identifying_substructure}). {\em Panel V:} The density peak with the largest number of core particles is taken to be the true density peak (Appendix \ref{sec:subhalo_center}). Properties of the subhalo are then calculated relative to this true density peak using all gravitationally bound, tracked particles (Appendix \ref{sec:subhalo_properties}), and several tests are run to assess whether the subhalo has disrupted/merged with its host (Appendix \ref{sec:subhalo_disruption}).}
\label{fig:method_cartoon}
\end{figure*}

\subsection{Subhalo Finding Overview}
\label{sec:method_overview}

At a high level of abstraction, our subhalo-finding method, \textsc{Symfind}, has three steps. In the first step, we use an existing halo catalog to identify and track all the particles associated with a subhalo prior to infall. In the second step, we use an existing subhalo finder \refadd{to} re-identify the subhalo {\em using only its tracked particles} after infall, rather than trying to find it within the background of host particles. Finally, the position and velocity identified by that halo finder are used to calculate subhalo properties with our own methods using only the tracked particles.

This approach falls within a larger family of similar techniques which are generally called ``particle-tracking'' subhalo finders. These use a subhalo's pre-infall particles to find that subhalo after infall \citep[e.g.,][see also Section \ref{sec:other_particle_tracking}]{tormen_1998_survival,kravtsov_2004_tumultuous,han_2012_hbt,han_2018_hbt_plus,springel_2021_gadget_4,diemer_2023_haunted}. They stand in contrast to ``single-epoch'' subhalo finders, which identify subhalos in a single snapshot and then attempt to connect objects over time afterward (see Section \ref{sec:single_epoch}). Particle-tracking is generally expected to be an effective subhalo-finding method because focusing only on previously accreted particles removes all host particles from consideration, vastly simplifying subhalo finding and reducing the chance of errors.

Although our general framework is simple, there are many questions that need to be addressed. Could numerical artifacts in the input halo catalogs hamper our ability to associate particles with halos? What exactly does it mean for a particle to be ``associated'' with a halo? If there are multiple density peaks in the tracked particles, how do we decide which one belongs to the subhalo? Is it possible for us to find a density peak that isn't actually a subhalo? There are a large number of design decisions in \textsc{Symfind} which are devoted to addressing and resolving these concerns, and many of these decisions are quite different from existing particle-tracking methods. In the list below, we outline the general structure of these decisions and point the reader to the relevant portions of Appendix \ref{sec:subhalo_algo_appendix}, where each point is discussed in detail. Fig.~\ref{fig:method_cartoon} illustrates the major steps of our algorithm.
\begin{enumerate}
\item ({\em Appendix \ref{sec:tree_post_processing}; Fig.~\ref{fig:method_cartoon}, Panel I}) We re-analyze input RCT merger trees to identify and correct for various errors: spurious phase space overdensities that are misidentified as subhalos (see Appendix \ref{sec:false_mergers in Rockstar}), errors during subhalo disruption that can cause subhalos to spike in mass, and errors during major mergers that can cause central halos to appear to become subhalos too quickly due to aphysical switching of mass between the primary and secondary halo during the merger.
\item ({\em Appendix \ref{sec:particle_associations}; Fig.~\ref{fig:method_cartoon}, Panel II}) \refadd{For each subhalo, we find all the particles which ever passed within that halo's $R_{\rm vir}$ at any point in time. We remove all particles which entered a larger halo before entering the subhalo in question so that subhalos don't accrete their hosts' particles. We track all the remaining particles. At the moment of accretion, we break tracked particles into ``{\em smoothly}'' and ``{\em non-smoothly}'' accreted sets. Smoothly accreted particles are ones that have never been accreted by another halo and non-smoothly accreted particles are ones that have.}
\item ({\em Appendix \ref{sec:identifying_substructure}; Fig.~\ref{fig:method_cartoon}, Panel III}) Once a central halo becomes a subhalo, we stop using RCT catalogs and calculate subhalo properties from the particles associated with that subhalo branch. Non-smoothly accreted particles can be associated with multiple branches, meaning our method can analyze nested substructures. At the snapshot of infall, we identify the $N_{\rm core}$ most gravitationally bound smoothly accreted particles (``{\em core}'' particles). These particles will be used in later snapshots to confirm the location of the subhalo. 
\item ({\em Appendix \ref{sec:identifying_substructure}; Fig.~\ref{fig:method_cartoon}, Panel IV}) We use an existing subhalo finder to identify density peaks within the tracked particles for each subhalo. Currently, we use \textsc{Subfind}, \citealp{springel_2001_populating}, using a density kernel over the $k$ nearest neighbors to estimate densities. A re-implementation of \textsc{Subfind} is used chiefly due to its simplicity, and we expect to replace this with \textsc{Rockstar} in the future. \refadd{This internal halo finder is only used for identifying density peaks and its own unbinding/mass association routines are not used.}
\item ({\em Appendix \ref{sec:subhalo_center}; Fig.~\ref{fig:method_cartoon}, Panel V}) We find which density peak each of the original $N_{\rm core}$ core particles is contained within. We take the subhalo's true position and velocity as the position and velocity of the peak containing the most core particles. The core particles are only used to select the peak and could, hypothetically, be located in the peak's outskirts.
\item ({\em Appendix \ref{sec:subhalo_properties}}) Using this peak's position and velocity, we calculate subhalo properties for all tracked particles that remain gravitationally bound after iterative unbinding. Both smoothly and non-smoothly accreted particles are used to calculate subhalo properties and binding energies.
\item ({\em Appendix \ref{sec:subhalo_disruption}}) We count a subhalo as disrupted/merged if it contains no bound core particles within its half-mass radius or if its half-mass radius intersects with its host's center. This first condition generally means either (i) that the core particles have completely dispersed and the ``peak'' is some random fluctuation in the extended tidal tail, (ii) that the subhalo has lost so much mass that its the core particles have become unbound, or (iii) that the peak's velocity is poorly determined. The latter of these two conditions generally means that the subhalo has sunk to its host's center through dynamical friction. There are some other very rare conditions that can lead to subhalo disruption, as well. After disruption, we continue trying to re-find the subhalo for the rest of the simulation and interpolate its properties during snapshots when it was erroneously marked as disrupted.
\end{enumerate}

\subsection{Fiducial Values, Application to Symphony, and Data Release}
\label{sec:data_release}

We have applied this subhalo-finding method to the SymphonyLMC, SymphonyMilkyWay, SymphonyGroup, and SymphonyL-Cluster zoom-in suites \citep{nadler_2022_symphony} with $k=16$ and $N_{\rm core} = 32$. These values were chosen by searching a wide range of parameter values, as described in Appendix \ref{sec:parameter_selection}. Only subhalos with $n_{\rm peak} > 300$ are included. We have made the resulting halo catalogs publicly available at \url{http://web.stanford.edu/group/gfc/symphony/}. This website also contains \textsc{Rockstar}+\textsc{consistent-trees} catalogs processed with the steps described in Appendix \ref{sec:tree_post_processing} and partial particle snapshots containing the tracking and halo association information described in Appendix \ref{sec:particle_associations}. This website also contains extensive documentation and tutorials on using these data.

The pipeline for generating these catalogs will be made available upon request. We are not making a general public code release at this time, but researchers interested in assigning a name to this specific subhalo-finding method can refer to it as \textsc{Symfind}, the Symphony halo finder. We are not making a public code release because \textsc{Symfind} currently does not have runtime performance that would allow it to be run on moderate-size cosmological simulations. This is not a fundamental limitation in the algorithm and will be addressed in future work. Furthermore, there are some algorithmic changes to \textsc{Symfind} that may make it well-suited to being run efficiently on very large cosmological simulations, as we discuss briefly in Appendix \ref{sec:iterative}.

Finally, we remind readers that these catalogs make heavy use of output from \textsc{Rockstar} \citep{behroozi_2013_rockstar}, \textsc{consistent-trees} \citep{behroozi_2013_consistent}, and some algorithms from \textsc{Subfind} \citep{springel_2001_populating}.

\section{The reliability of tracked subhalos}
\label{sec:reliability}

As we discussed in Section \ref{sec:intro} and as we will further discuss in Section \ref{sec:single_epoch}, there are substantial holes in our current ability to quantify the reliability of subhalo finders. Existing tests typically look at the performance of subhalo finders on statistics where one does not already know the expected result, such as a subhalo mass function. Because the values of these statistics are unknown {\em a priori}, these tests generally take the form of either checking for internal consistency/convergence in subhalo finder results as resolution is increased \citep[e.g.,][]{nadler_2022_symphony} or as a comparison between the results of different subhalo finders \citep[e.g.,][]{onions_2012_notts}. But convergence is not the same as correctness, and noting that two subhalo finders arrive at the same (or different) results cannot prove that either is correct.

In this Section, we lay out a series of systematic tests which do not fall victim to these issues and apply those tests to \textsc{Symfind} and RCT. These tests are a combination of qualitative inspection (Section \ref{sec:case_study} and Appendix \ref{sec:additional_halos}), characterization of the conditions that cause a subhalo finder to lose track of subhalos (Section \ref{sec:survial}), and quantification of when subhalos deviate from the predictions of high-resolution idealized simulations (Sections \ref{sec:numerical_limits}, \ref{sec:structural}, and \ref{sec:mass_loss_rates}). By combining these tests, we are able to identify subhalo populations which we can guarantee that a subhalo finder will be able to locate and can also guarantee that certain properties of these subhalos will be correctly recovered. We argue that the same level of guarantees cannot be made with traditional convergence testing.

With these tests, we demonstrate  that \textsc{Symfind} does not falsely converge and is capable of following subhalos to orders-of-magnitude smaller masses than RCT is capable of. RCT, regrettably, does falsely converge.

\subsection{Evolution of subhalo properties: a case study}
\label{sec:case_study}

\begin{figure}
\hspace{-3.5mm}
\includegraphics[width=0.495\textwidth]{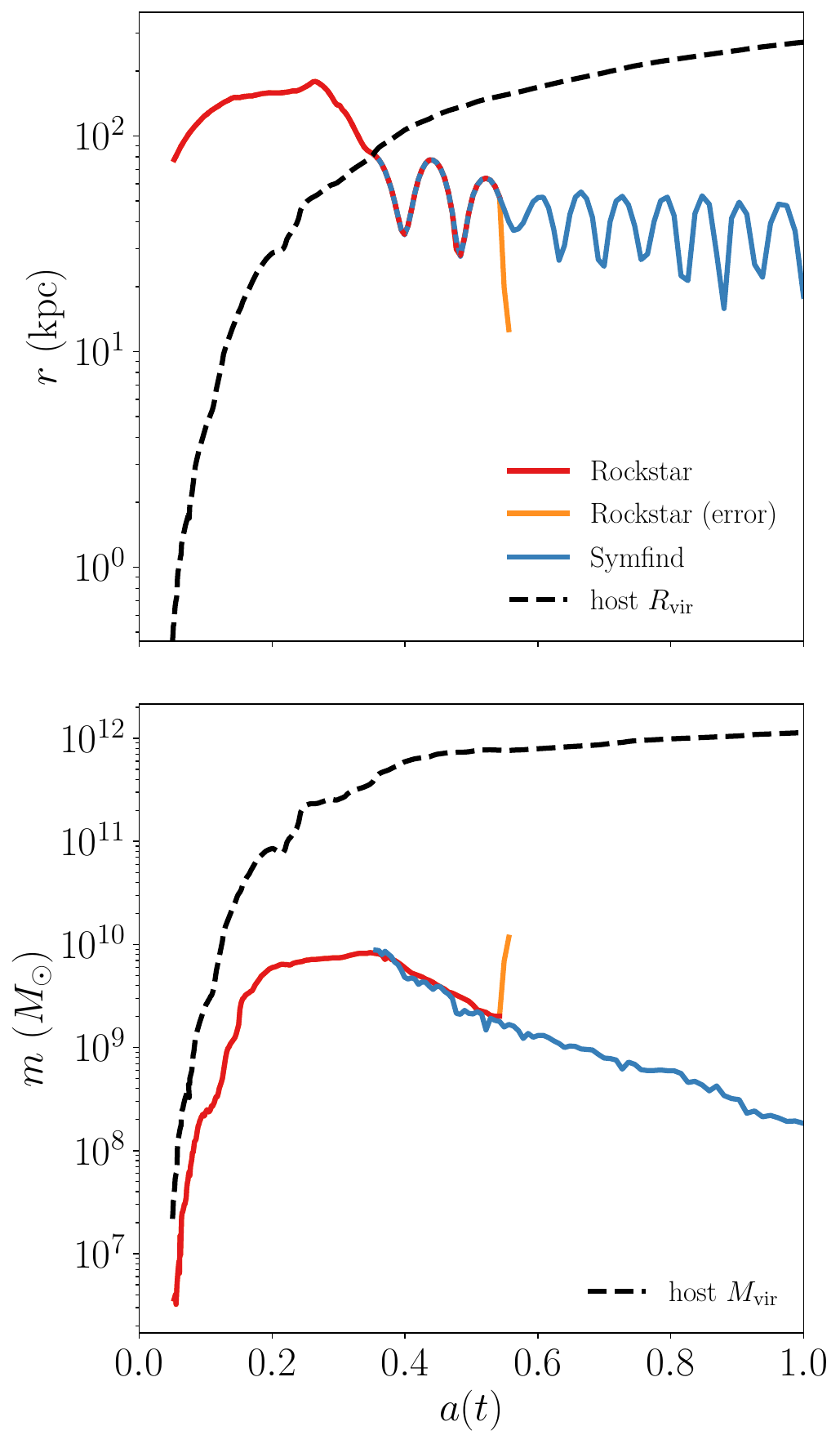}
\caption{
The evolution of a representative subhalo over time, as measured by both \textsc{Rockstar} (red) and \textsc{Symfind} (blue). In the top panel, the \textsc{Rockstar} curve is dashed during the period when it overlaps with the \textsc{Symfnd} curve. Snapshots during which \textsc{Rockstar} has identified an incorrect subhalo center are shown in orange (see Appendix \ref{sec:parameter_selection}), and the virial mass/virial radius of the host halo is shown in black. When followed with \textsc{Rockstar}, the halo survives two orbits before disrupting. \textsc{Rockstar} incorrectly associates this subhalo's branch with an unrelated density peak for its last few snapshots, leading to an \refadd{unphysical} change in mass and position. \textsc{Symfind} agrees with \textsc{Rockstar} while \textsc{Rockstar} reliably tracks the subhalo and continues to follow the halo for many more orbits and a further factor of ten in mass loss. Note that without post-processing checks, the \textsc{Rockstar} branch reaches $m_{\rm peak}$ during its final, erroneous snapshot. More example halo trajectories can be found in Appendix \ref{sec:additional_halos}.
}
\label{fig:example_halo}
\end{figure}

\begin{figure}
\hspace{-3.5mm}
\includegraphics[width=0.475\textwidth]{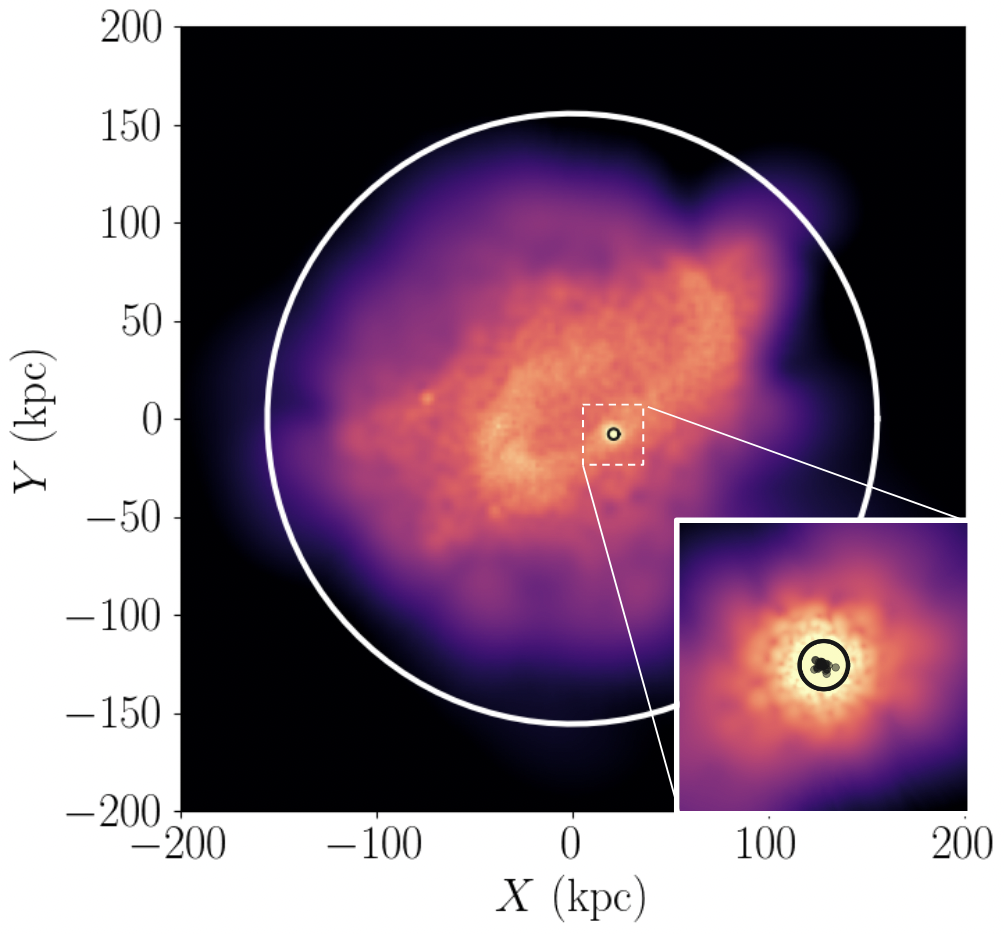}
\caption{Projected dark matter density of the subhalo shown in Fig.~\ref{fig:example_halo} immediately after it disrupts within the \textsc{Rockstar} catalog. In the main image, the white circle is the virial radius of the host, the black circle is the half-mass radius of the subhalo according to \textsc{Symfind}, and the color map shows the logarithm of the projected density of all particles that fell in with the subhalo, estimated by an SPH kernel. The majority of this subhalo's particles have been lost into tidal tails. The inset panel shows the region around the subhalo in more detail. The black dots show the 32 most-bound particles identified during the snapshot when the subhalo was first accreted. These ``core'' particles are still well-clustered and sit in the center of a fully bound, five-thousand-particle structure; thus the object identified by \textsc{Symfind} is a real subhalo.}
\label{fig:example_halo_image}
\end{figure}

Before performing statistically rigorous tests on large populations of subhalos, we first consider the qualitative behavior of a representative subhalo drawn from a pool of thousands of subhalos that we have visually inspected. Visual inspection is not a novel test, but it is an important one because it can show that a given subhalo finder is not misidentifying random phase-space detritus as true subhalos. It also qualitatively demonstrates many of the issues that we will quantify in later testing.

Fig.~\ref{fig:example_halo} shows the evolution of a typical subhalo resolved with $\approx 2\times 10^4$ particles at its peak mass. The top panel shows the distance between the subhalo and its host over time. The dashed black line shows the virial radius of the host, and the colored lines show the positions of the subhalo tracked by RCT and \textsc{Symfind}. There are several snapshots where the RCT catalog continues to have entries for this halo. Still, manual inspection and core-particle-based tests described in Appendix \ref{sec:parameter_selection} show that these ``halos'' are unassociated with the subhalo's particles. 

During the period where RCT has reliable estimates of the subhalo's position, it agrees exactly with the position of the \textsc{Symfind} subhalo. As the subhalo approaches its third pericenter, RCT experiences an error that causes a sudden jump in the subhalo's apparent position. This is soon followed by the apparent disruption of the subhalo. However, \textsc{Symfind} continues to follow the halo for many more orbits.

The bottom panel of Fig.~\ref{fig:example_halo} shows the same subhalo with the same color scheme, except that it shows the subhalo's mass. During the period where RCT reliably tracks the subhalo, both it and \textsc{Symfind} find that the subhalo mass is decreasing approximately exponentially. \textsc{Symfind} masses are slightly noisier and lower than RCT masses (see Appendix \ref{sec:iterative}). RCT's third-pericenter error corresponds to a sharp increase in mass. \textsc{Symfind} continues to follow the mass loss unabated past the point where this error occurs and the subhalo continues to lose mass at the same exponential rate through the following orbits.

Fig.~\ref{fig:example_halo_image} shows an image of this subhalo several snapshots after the RCT branch disrupts. The projected density field generated by all particles that fell into the host halo with this subhalo is shown in pink. This density field is estimated through a standard 2D SPH density kernel applied to the 128 nearest particles \citep[e.g.,][]{springel_2010_sph}. The radius of the host halo is shown in white, and the half-mass radius of the subhalo of bound particles is shown in black. The inset shows a zoomed-in view of the subhalo, except that only bound particles are shown, and the SPH kernel is now applied to the 32 nearest particles. The 32 most-bound particles during the infall snapshot are shown as black dots. The inset shows a self-bound, roughly spherical structure that contains the same particles that have been at the center of the subhalo since infall. In short, particle-tracking is following a real subhalo, and it is the target subhalo.

To summarize, \textsc{Symfind} is capable of following this subhalo far longer than RCT and the long-term evolution of its inferred properties is reasonable. Manual inspection of the particle distribution shows that the object being followed really is the original subhalo. Taken together, this means that the difference between RCT and \textsc{Symfind} is not simply an unresolvable difference in definitions: this subhalo {\em does} out-survive its RCT branch, and particle-tracking correctly follows this subhalo.

An extensive manual review of individual subhalos shows that the qualitative behavior of this subhalo is typical. In Appendix \ref{sec:additional_halos}, we show eight randomly selected subhalo trajectories that are all qualitatively similar to Fig.~\ref{fig:example_halo}. Beyond this, we have manually inspected several thousand subhalo trajectories, two hundred images of heavily disrupted subhalos, and several dozen movies. The longer survival time of this subhalo and the fact that it tracks a well-defined, bound remnant containing the most-bound particles identified at infall are representative of other subhalos with similar peak resolution levels. The same is generally true at other resolution levels, although the relative advantage of \textsc{Symfind} decreases as particle counts decrease.

Qualitative assessment has limitations, so we also perform quantitative analysis on the minimum masses reached prior to disruption by subhalos followed by RCT and \textsc{Symfind} in Section \ref{sec:survial}. A typical subhalo at this resolution level survives to masses $\approx$ 30 to 100 smaller than subhalos tracked by RCT, meaning that the difference in final masses seen in Fig.~\ref{fig:example_halo} is close to the typical difference in final masses that one would see in a subhalo dataset that was not right-censored by the end of the simulation.  As we discuss in Section \ref{sec:stitching_errors}, at the resolution level of the subhalo shown in Fig.~\ref{fig:example_halo}, roughly a third of all RCT subhalos experience a similar error during their final snapshot. So while such an error is common, it does not always occur during RCT disruption.

\subsection{Subhalo survival thresholds}
\label{sec:survial}

\begin{figure*}
\hspace{-3.5mm}
\includegraphics[width=0.475\textwidth]{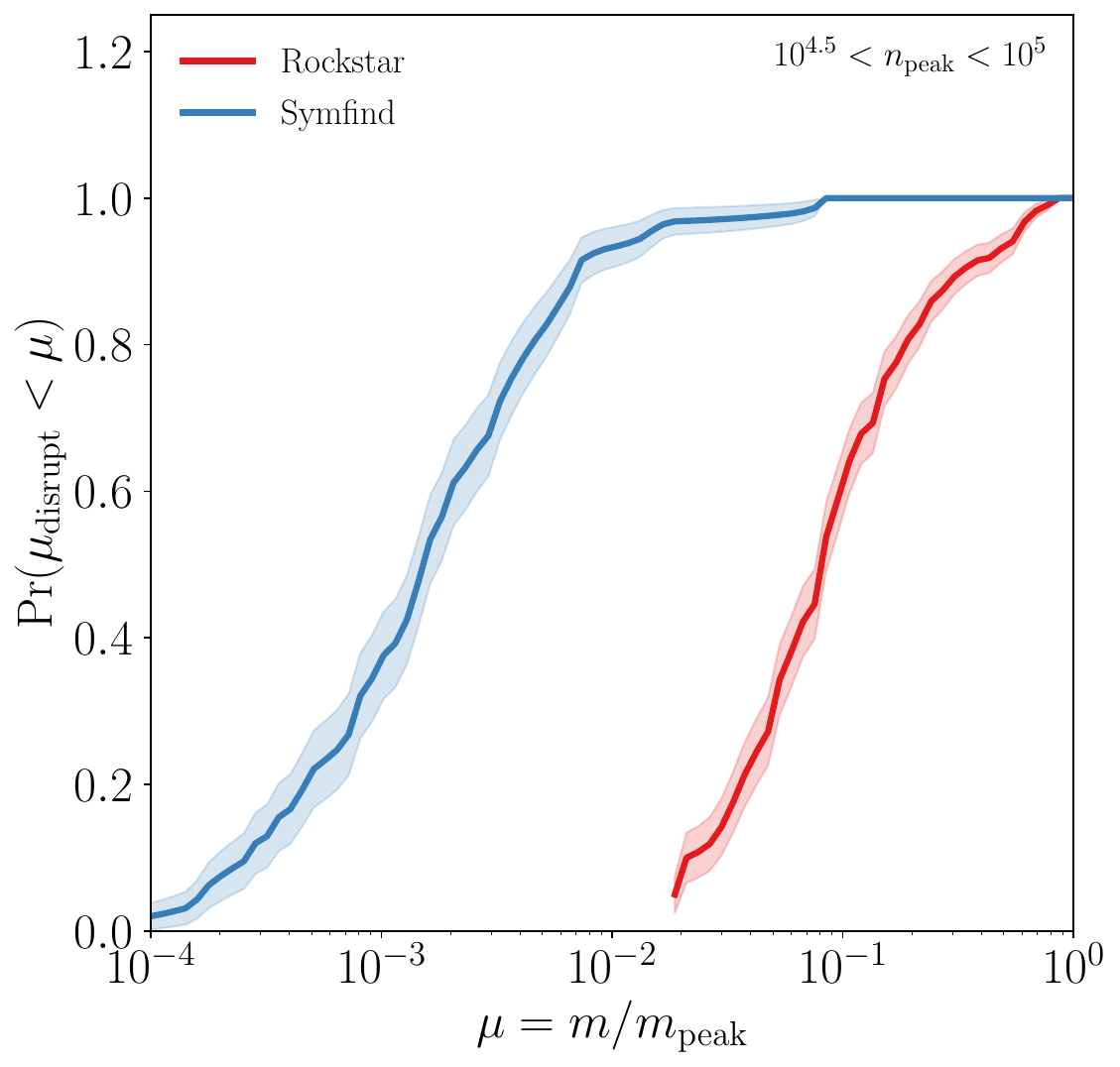}
\includegraphics[width=0.475\textwidth]{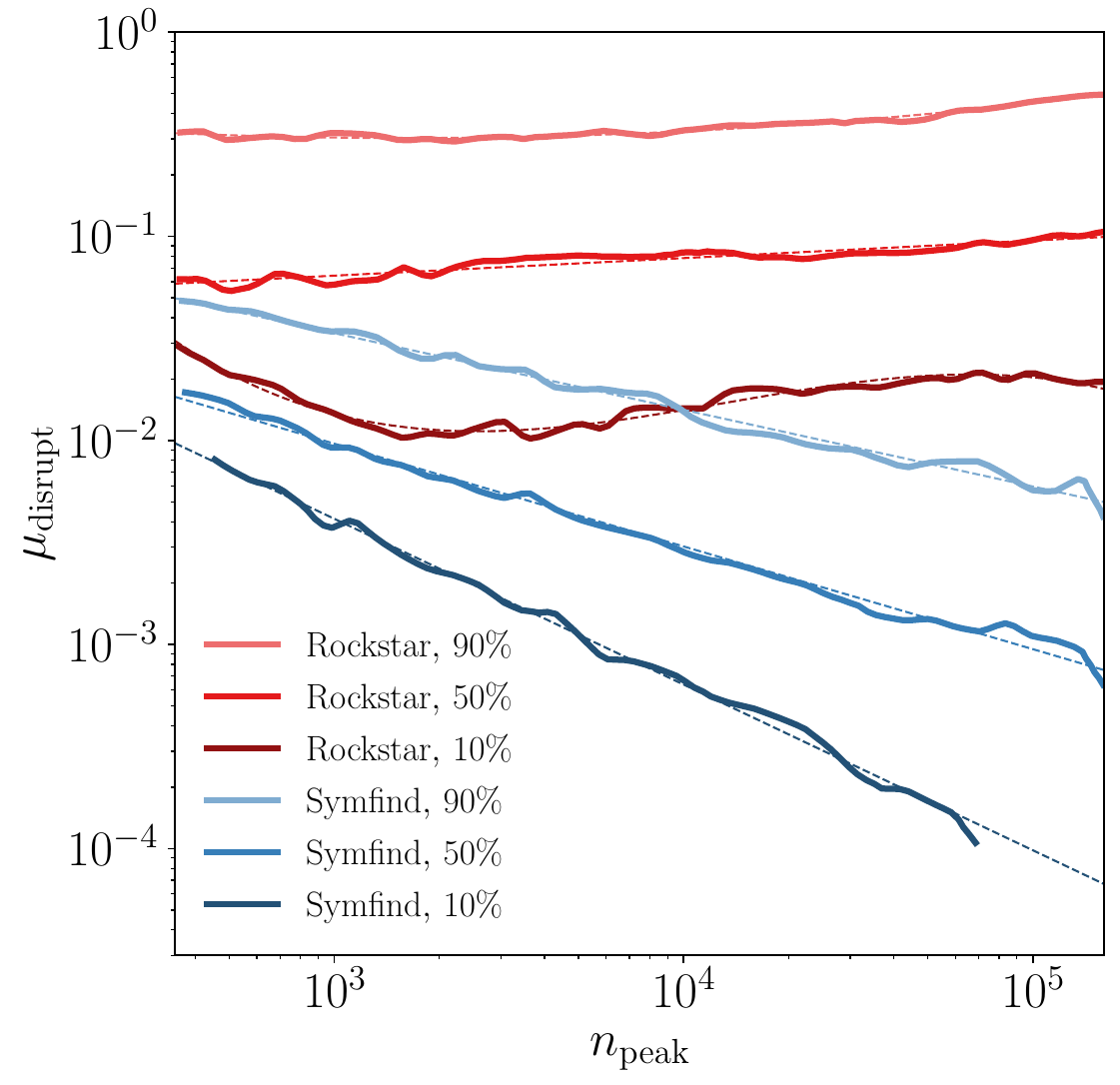}
\caption{The amount of mass subhalos lose before subhalo finders lose track of them, measured across the entire Symphony suite. {\em Left:} The probability that subhalos with $10^{4.5} < n_{\rm peak} < 10^5$ disrupt before reaching a given mass loss ratio, $\mu \equiv m/m_{\rm peak},$ when subhalos are followed by \textsc{Rockstar} (red) and by \textsc{Symfind} (blue). Survivor bias is accounted for using the Kaplan--Meier estimator, and 1-$\sigma$ confidence intervals (shaded bands) are calculated through Greenwood's formula. At this resolution level, \textsc{Symfind} can follow subhalos to masses about 30{-}100 times smaller than \textsc{Rockstar}, depending on the quantile of the distribution. {\em Right:} The distribution of $\mu_{\rm disrupt}$ as a function of $n_{\rm peak}$ for \textsc{Rockstar} and \textsc{Symfind}. Different quantiles in the $\mu_{\rm disrupt}$ distribution are shown as shades of red and blue. Fits according to Eq.~\ref{eq:disrupt_limit_fit} and Table \ref{tab:disrupt_limit_table} are shown as dashed lines. At a fixed $n_{\rm peak}$ and quantile in the $\mu_{\rm disrupt}$ distribution, \textsc{Symfind} can follow subhalos to factors of three-to-one-hundred times lower in mass than \textsc{Rockstar}. \textsc{Rockstar} disruption masses do not decrease as resolution increases, which can lead to the false impression of convergence in many types of numerical tests. Only minor mergers with $m_{\rm peak}/M_{\rm vir} < 0.1$ are shown; major mergers experience substantial dynamical friction and physical disruption on short time scales and require separate, dedicated analysis.}
\label{fig:survival}
\end{figure*}

Idealized, high-resolution simulations show that, in the absence of strong dynamical friction, some portion of low-mass subhalos should survive as bound remnants for essentially arbitrarily long periods of time \citep[][and references therein]{errani_2023_core_survival}. Dynamical friction is quite weak for low-mass subhalos  \citep[e.g.,][]{vdb_2016_segregation}, giving us our first opportunity to perform a quantitative correctness test where we already know what the correct behavior of the finder should be. In this Section, we determine how long RCT and \textsc{Symfind} can follow subhalos before losing track of them. When this analysis is restricted to a low-mass subhalo population, all such losses will be artificial in origin, caused by some combination of subhalo finder algorithm and simulation numerics. This means that this analysis can put limits on which subhalo populations are safe to treat as complete.

We estimate ``survival curves'' for subhalos followed by both RCT and particle-tracking. The survival curve is the probability that a subhalo will disrupt below some mass ratio
\begin{equation}
\mu\equiv m/m_{\rm peak}
\end{equation}
and can be computed via the Kaplan--Meier estimator \citep{kaplan_meier_1958_nonparametric}. As discussed in Section \ref{sec:definitions}, $m_{\rm peak}$ is defined using only masses prior to first infall, meaning that it does not suffer from the sort of RCT mass fluctuation error shown in Fig.~\ref{fig:example_halo}. This gives the expected distribution of {\em disruption mass ratios}, $\mu_{\rm disrupt}$, the smallest mass fractions subhalos achieve before dropping out of the catalog. Survival curves are a standard analysis tool in the medical sciences and in engineering analysis, where they are used, for example, to estimate the distribution of lifespans of a set of patients or the distribution of times-until-failure for a set of machines.

The biggest problem that one encounters when constructing a survival curve is statistical censoring. In $\Lambda$CDM simulations, many subhalos survive past the last snapshot, meaning that simply building a histogram of the minimum mass reached by every subhalo will underestimate the typical disruption mass because it mixes disruption masses with the surviving $\mu$ distribution in the final snapshot. Restricting the analysis to subhalos that disrupt selects for a sample with shorter survival times than average. To address this problem, the Kaplan--Meier estimator breaks up the minimum mass that any subhalo was observed to achieve into intervals, estimates the instantaneous probability of failure within each interval, and multiplies those instantaneous probabilities together to get a cumulative probability. The estimator is more accurate with smaller intervals, so one typically sorts the data and inserts one interval between each consecutive pair of measurements.

This estimator can be written as,
\begin{equation}
    \label{eq:kaplan_meier}
    \widehat{{\rm Pr}}(\mu_{\rm disrupt}<\mu_{f,i}) = \prod^j_{0 \leq j \leq i} \left(1 - \frac{d_j}{N(\leq\mu_{f,j})}\right).
\end{equation}
Here, $\widehat{{\rm Pr}}(\mu_{\rm disrupt} < \mu_{f,i})$, is the probability that a subhalo will have a disruption mass ratio, $\mu_{\rm disrupt}$, less than some value $\mu_{f,i}$, the final mass ratio of one of the subhalos in the dataset. To estimate this, one iterates over all the final mass ratios of subhalos with $\mu_{f,j}\geq\mu_{f,i}$, indexed by $j$ in order of decreasing $\mu_{f,j},$ $d_j$ is an indicator variable that is 1 if the final mass of the subhalo, $j$, is caused by disruption and 0 if it is caused by the end of the simulation, and $N(\leq\mu_{f,j})$ is the number of subhalos where $\mu_{f} \leq \mu_{f,j}$. We then interpolate $\widehat{{\rm Pr}}(<\mu_{f,i})$ to get a function that is continuous in $\mu$, giving us ${\rm Pr}(\mu_{\rm disrupt} < \mu)$.

We estimate the standard error on survival probabilities with Greenwood's formula \citep{greenwoods_formula_1928}
\begin{align}
\label{eq:greenwood}
\widehat{{\rm Var}}[\widehat{\rm Pr}(< \mu_{f,i})] =& \widehat{\rm Pr}(< \mu_{f,i})^2 \times\\ 
\nonumber
&\sum^j_{0\leq j \leq i} \frac{d_j}{N(\leq\mu_{f,j})(N(\leq\mu_{f,j}) - d_j)}.
\end{align}
$\widehat{{\rm Var}}[\widehat{\rm Pr}(< \mu_{f,i})]$ is similarly interpolated to get a function that is continuous in $\mu.$

The left panel of Fig.~\ref{fig:survival} shows survival curves for RCT subhalos and \textsc{Symfind} subhalos at high resolutions ($10^{4.5}<n_{\rm peak}<10^5$). Here, we have combined all the Symphony simulation suites for improved number statistics because testing shows that the shape of survival curves depends only on $n_{\rm peak}$ and not on $m_{\rm peak}$ (see Appendix \ref{sec:survival_mass_dependence}). Both methods have wide distributions of $\mu_{\rm disrupt}$, spanning about two decades in $\mu$. However, RCT subhalos disrupt at much higher masses than \textsc{Symfind} subhalos, and even with this width, almost all \textsc{Symfind} subhalos outsurvive even the longest-lasting RCT subhalos. The distribution of disruption masses is about 30 to 100 times lower when using \textsc{Symfind} than when using RCT.

To characterize these survival curves with a single number, we compute the 10\%, 50\%, and 90\% quantiles in the $\mu_{\rm disrupt}$ distribution as a function of $n_{\rm peak}$ for both RCT and \textsc{Symfind}. We once again combine the four Symphony simulation suites. For each value of $n_{\rm peak}$, we select the 2,000 subhalos with $n_{\rm peak}$ closest to each target value. For each method and quantile, we fit a low-order exponential polynomial to $\mu_{\rm disrupt}$ as a function of $n_{\rm peak}:$
\begin{align}
\label{eq:disrupt_limit_fit}
\mu_{\rm disrupt}(n_{\rm peak};\,q) = 10^{a_3x^3 + a_2x^2 + a_1x + a_0}.
\end{align}
Here, $q$ is the quantile in the distribution, $x\equiv \log_{10}(n_{\rm peak})$ and $a_0$ through $a_3$ are fit parameters. The fits were performed through least squares minimization in $\log_{10}(\mu)$-space via the Levenberg–Marquardt algorithm. Higher-order terms in the fit were manually set to zero in cases where reasonable qualitative agreement could be achieved with lower-order fits. We show the best-fitting values for each method and target quantile in Table \ref{tab:disrupt_limit_table}.

\begin{table}
\centering
\begin{tabular}{c|c|c|c|c|c}
\hline
Method & $q$ & $a_3$ & $a_2$ & $a_1$ & $a_0$ \\ [0.5ex]
\hline\hline
\textsc{Rockstar} & 0.9 & --- & 0.0532 & -0.3415 & 0.0301 \\
& 0.5 & --- & --- & 0.0860 & -1.4505 \\
& 0.1 & -0.1969 & 2.4327 & -9.7238 & 10.7231 \\
\hline
\textsc{Symfind} & 0.9 & --- & --- & -0.3756 & -0.3473 \\
& 0.5 & --- & --- & -0.5034 & -0.5054\\
& 0.1 & --- & --- & -0.8121 & 0.0526 \\
\hline
\end{tabular}
\caption{Best fitting values for different quantiles, $q$, of the $n_{\rm peak}$-dependent $\mu_{\rm disrupt}$ distribution for both Rockstar and \textsc{Symfind} according to Eq.~\ref{eq:disrupt_limit_fit}. Fits are only calibrated to the range $300 \leq n_{\rm peak} \leq 2\times 10^5$ and it is unlikely that the fits extrapolate straightforwardly outside this range.}
\label{tab:disrupt_limit_table}
\end{table}

These $n_{\rm peak}$-dependent quantiles of the $\mu_{\rm disrupt}$ distribution and their fits are shown in the right panel of Fig.~\ref{fig:survival}. RCT disruption thresholds are essentially independent of $n_{\rm peak}$, with the median subhalo being lost from the catalog at roughly a tenth of its peak mass and a sizable sub-population of subhalos being lost after losing only a third of its peak mass. This independence from $n_{\rm peak}$ will result in false convergence: resimulating a subhalo with more particles will, on average, not change the mass at which it drops out of the RCT catalog. Because of this, some subhalo statistics (e.g., Section \ref{sec:radius_cdf}, Appendix \ref{sec:shmf_converge}) may appear not to change with increasing resolution, giving the impression of numerical reliability when one is actually only seeing resolution-independent limitations in the subhalo finder.

\textsc{Symfind} can follow subhalos substantially longer, following subhalos to masses that are $\approx3$-to-6 times smaller than RCT for $n_{\rm peak}=300$ subhalos and roughly a hundred times smaller for $n_{\rm peak} = 10^5$ subhalos. \textsc{Symfind} tracks subhalos to smaller masses as resolution increases and thus does not suffer from this form of false convergence.

In Section \ref{sec:good_enough} we discuss what mass scales one would want a subhalo finder to be able to follow subhalos to if one wishes to study the population statistics of satellite galaxies and avoid the use of ``orphan'' modeling. We note that in this case, one is not interested in the median $\mu_{\rm disrupt}$, but in a higher quantile of the $\mu_{\rm disrupt}$ distribution, such as 90\%. If, for example, the median subhalo is being lost from the catalog at the same time one would expect its satellite galaxy to disrupt, that still means that half of all subhalos are being lost too quickly and would require orphan modeling to account for. This does not mean that the lower quantiles in the $\mu_{\rm disrupt}$ distribution are useless: for studies that do not need to track individual subhalos, one could weight the contribution of a given subhalo to the statistic of choice by the inverse of Pr$(\mu_{\rm disrupt} < \mu).$ But doing so would require confirming that the $\mu_{\rm disrupt}$ distribution does not depend on any quantities of interest for this statistic (e.g., Appendix \ref{sec:survival_mass_dependence}). One would also need to be careful of numerical (as opposed to purely halo-finding-based) effects in the deep mass-loss regime when doing so.

Although the analysis in this Section has shown the minimum masses that can be resolved with our method and with RCT, merely being able to resolve a subhalo with a subhalo finder does not mean that the subhalo is a numerically reliable analysis target. In Sections \ref{sec:numerical_limits}, \ref{sec:mass_loss_rates}, and \ref{sec:structural}, we establish the regimes where subhalo masses, abundances, and $v_{\rm max}$ values are well-resolved.

\subsection{Idealized Numerical Reliability Limits}
\label{sec:numerical_limits}

\begin{figure*}
\hspace{-3.5mm}
\includegraphics[width=0.475\textwidth]{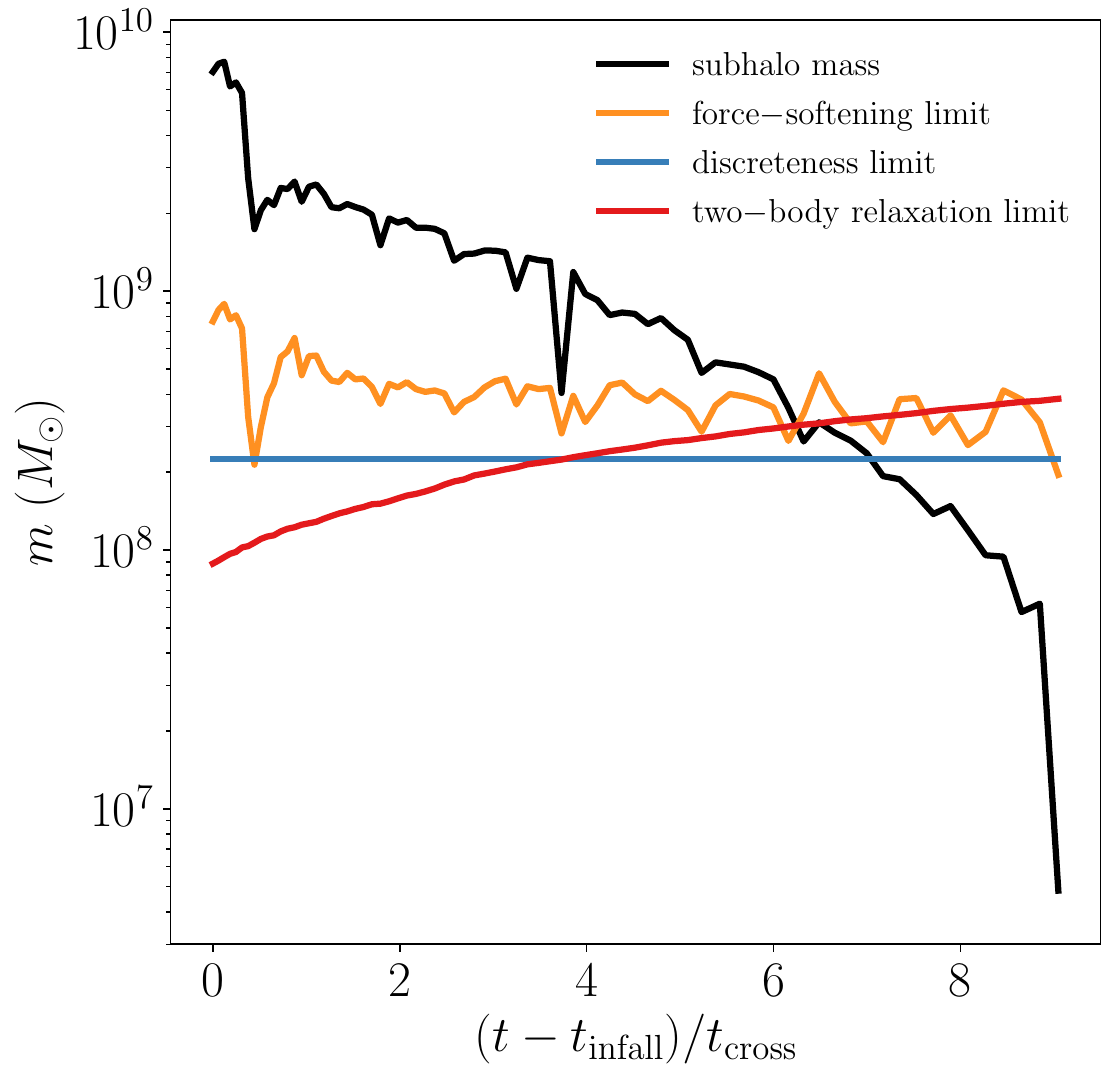}
\includegraphics[width=0.467\textwidth]{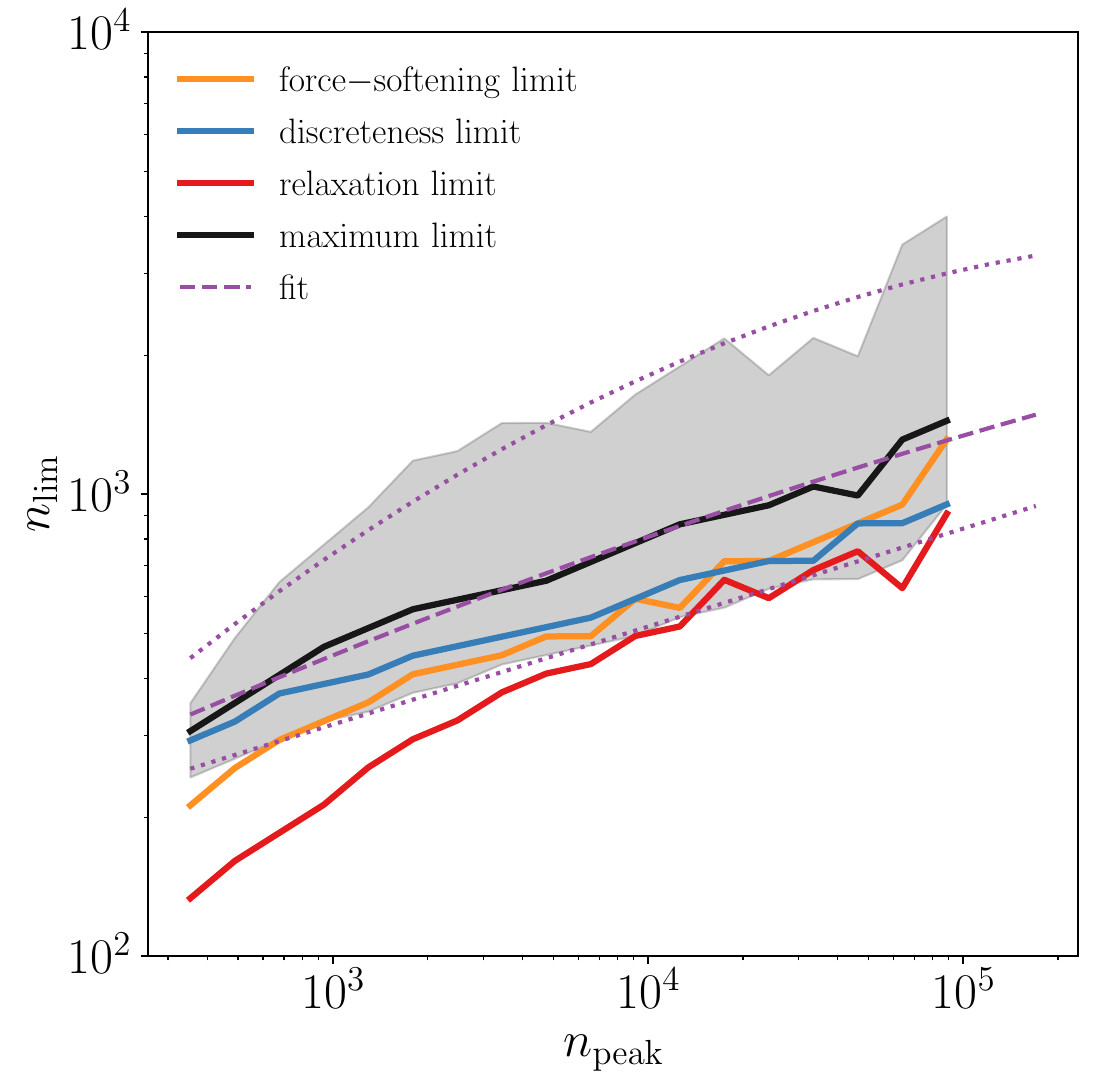}
\caption{The convergence limits of subhalos in \textsc{Symfind}. {\em Left:} Comparison between several idealized convergence limits and the mass-loss-history of a fairly well-resolved $n_{\rm peak} \approx 10^4$ subhalo. Lines compare the force-softening limit (Eq.~\ref{eq:limit_1}; orange), the discreteness limit (Eq.~\ref{eq:limit_2}; blue), and the two-body relaxation limit (Eq.~\ref{eq:limit_3}; red). Each limit predicts that the subhalo is only converged when the black curve is above the respective colored curve. {\em Right:} The median limiting particle count, $n_{\rm lim,ideal}$ for subhalos as a function of $n_{\rm peak}$. All Symphony suites are combined in this Figure. The colored curves show the limiting particle counts for each method shown in the left panel and the black curve shows the maximum limit across the three on a halo-by-halo basis. The 68\% spread around the black curve is shown as a gray-shaded region. The purple curves show fits of Eq.~\ref{eq:max_conv_limit_fit} to the median (dashed purple line) and 68\% spread (dotted purple lines).}
\label{fig:conv_limits}
\end{figure*}

Having established how long \textsc{Symfind} and RCT can follow a subhalo, we move on to testing when the properties of these subhalos can no longer be properly measured, either due to failures in the simulation or failures in the subhalo finder. Before performing any empirical testing, we first review the main causes of non-convergence of subhalos and combine several estimators of these effects based on idealized simulations and first-principles arguments and summarize these limits with Eq.~\ref{eq:max_conv_limit_fit}. These combined limits will be used as a component of the empirical testing in Sections \ref{sec:structural} and \ref{sec:mass_loss_rates}, although we establish that these limits are too conservative for some subhalo properties.

For a simulation with well-calibrated time stepping \citep[see Section 6.1 in][for discussion]{mansfield_avestruz_2021_biased}, three major issues impact the disruption of subhalos. The first is excessive force softening. Force softening suppresses rotation curves at scales many times larger than $\epsilon$ \citep[e.g., Appendix B in][]{mansfield_avestruz_2021_biased}, and this suppression means that subhalos have smaller enclosed densities at fixed radii, leading to smaller tidal radii and more rapid mass loss. Using idealized simulations, \citet{vandenbosch_2018_disruption} find that subhalos above the limit
\begin{equation}
    \label{eq:limit_1}
    \frac{m}{m_{\rm infall}} > \frac{1.79}{1.284}\left(\frac{\epsilon\,r_{1/2}}{f(c_{\rm infall})\,r_{s,{\rm infall}}^2}\right)\\
\end{equation}
are largely unaffected by this process. Here, $\epsilon$ and $r_{\rm 1/2}$ are the instantaneous Plummer-equivalent force-softening scale and the half-mass radius of the subhalo, respectively, while $c_{\rm infall}$ and $r_{s,{\rm infall}}$ are the subhalo's NFW concentration, and NFW scale radius, respectively. $f(x)=\ln{(1+x) + 1/(1+x)}.$ A corrective factor of 1.284 has been applied to account for the conversion between the Plummer force kernels used in \citet{vandenbosch_2018_disruption} and the Gadget spline-based force kernels used by Symphony. Gadget force kernels are already expressed in ``Plummer-equivalent'' units, but the traditional conversion factor is based on matching the depth of particles' potentials at small radii \citep{springel_2001_gadget1} and does not do a good job describing the large-radius impact of force softening \citep{mansfield_avestruz_2021_biased}.

The second source of numerical biases is discreteness noise. Once a subhalo has sufficiently few particles, Poisson fluctuations cause the mass loss rate to experience excessive noise. Fluctuations that temporarily increase the mass loss rate cause the subhalo to expand in response to the excess mass loss, which leads to smaller tidal radii and larger future Poisson fluctuations. Meanwhile, fluctuations that decrease the mass loss rate leave the subhalo relatively unchanged and thus have little impact on its future evolution. The asymmetric impact of fluctuations on the future evolution of the subhalo leads to an instability that can cause runaway mass loss at low resolutions. Using idealized simulations, \citet{vandenbosch_2018_disruption} find that subhalos above the limit
\begin{equation}
    \label{eq:limit_2}
    \frac{m}{m_{\rm infall}} > 0.32 \left(\frac{n_{\rm infall}}{10^3}\right)^{-0.8}
\end{equation}
are largely unaffected by this process. Here, $n_{\rm infall}$ is the number of particles the subhalo had at infall.

The third source of numerical biases is numerical relaxation. Dark-matter-only simulations are designed to approximate a perfectly collisionless fluid, but their discretization into particles means that simulated particles can scatter off one another, allowing flows of energy and mass across a halo over the relaxation timescale $t_{\rm relax}(r)$. Generally, regions of the halo where $t_{\rm relax}(r)$ is less than the age of the system are considered unresolved \citep[e.g.,][]{power_2003_inner}. Excessively small force-softening scales can lead to catastrophic ``large-angle'' scattering (e.g., Fig. 6 in \citealp{knebe_2000_resolution}) and can lead simulations to fail to conserve energy. However, for simulations like the Symphony suite where the force softening scale has been set large enough to suppress this effect, $t_{\rm relax}$ is set by the superposition of many large-distance ``small-angle'' scatterings \citep[e.g.,][]{ludlow_2019_numerical} and the primary effect of force softening becomes a weak, logarithmic suppression of the Coulomb logarithm as $\epsilon$ becomes larger. Following \citet{ludlow_2019_numerical}, this suppression leads to relaxation times that scale with the local orbital time as
\begin{equation}
\frac{t_{\rm relax}(r)}{t_{\rm orbit}(r)} = \frac{N(<r)}{4}\left(\ln\left(\frac{r^2}{\epsilon^2} + 1\right) + \frac{\epsilon^2 -2r^2}{3(\epsilon^2 + r^2)} - \ln\left(\frac{3}{2}\right) \right)^{-1}.
\end{equation}
Here, $t_{\rm orbit}$ is the circular orbit time, $t_{\rm orbit} \equiv 2\pi r^{3/2}/\sqrt{GM(<r)},$ and $N(<r)$ is the number of particles with radii smaller than $r.$

For each particle in the halo, we calculate $t_{\rm relax}(r)$ using that particle's position at the snapshot of the subhalo's infall. From this, we can calculate a relaxation limit
\begin{equation}
\label{eq:limit_3}
m(t_0) > \sum_{i\in {\rm bound}} m_p\cdot H(t_0 - t_{{\rm infall},i} - t_{{\rm relax},i}).
\end{equation}
Here, the sum goes over all particles that were bound to the subhalo at the subhalo's snapshot of first infall, $H(x)$ is a Heaviside step function that is 0 for $x\leq0$ and 1 otherwise, $t_0$ is the current age of the universe, and $t_{\rm infall}$ is the particle accreted onto the subhalo (not the snapshot that the subhalo accreted onto the host). In other words, when the total relaxed mass within the halo is equal to its current subhalo mass, we consider its mass loss to be unconverged.

The use of particle-tracking allows for more accurate estimates of how long a particle has been orbiting a halo, but a similar methodology that approximates the orbiting time of particles has been widely used to model the unconverged inner regions of high-resolution, isolated halos \citep[e.g.,][]{power_2003_inner,springel_2008_aquarius,ludlow_2019_numerical}. Because these studies do not track individual particles, they generally assume that particles have been orbiting their host halo for a timescale comparable to the age of the universe and correct for this assumption with an empirical multiplicative factor. (This empirical factor also effectively corrects for any order-unity inaccuracies in the Coulomb logarithm, so in principle, one would want to re-calibrate this corrective factor using the true orbital times of particles, although we do not attempt to do so here.)

It is particularly important to account for the individual orbiting times of particles for subhalos because an old subhalo with a small $m/m_{\rm peak}$ will have more relaxed mass than a recently accreted subhalo with large $m/m_{\rm peak}$, even if the two have identical instantaneous properties. Therefore, failing to account for individual orbiting times would lead to incorrect convergence limit trends with $n_{\rm peak}.$

Each of these limits is different on a halo-by-halo basis. To account for noise in $m(t)$, we count a subhalo as having passed a particular convergence limit if there are no subsequent snapshots where $m(t)$ is above that limit. We show a typical high-resolution ($n_{\rm peak} > 10^4$) subhalo in Fig.~\ref{fig:conv_limits}. Time is normalized by the crossing time at infall, $t_{\rm cross} \equiv 2\,R_{\rm vir}(t_{\rm infall})/V_{\rm vir}(t_{\rm infall})$.

For each subhalo in all the Symphony hosts considered in this paper, we calculate all three limits, find the most restrictive of the three and show the results in the right panel of Fig.~\ref{fig:conv_limits}. We then use the Kaplan--Meier method (see Section \ref{sec:survial}) to estimate the median values and 68\% scatter of the particle counts at which subhalos pass these limits as a function of $n_{\rm peak}.$ We show the median number of particles in a subhalo at the time that it crosses under each limit are shown as well as the median and distribution of the most restrictive of the three limits. We have constructed these curves for each individual Symphony suite and found them to be in good agreement with one another, justifying the choice to combine all four suites together.

Comparing the three individual limits, we see that Symphony's $\epsilon$ values were reasonably well-calibrated for subhalo studies. Increasing $\epsilon$ increases the amplitude of Eq.~\ref{eq:limit_1} while decreasing Eq.~\ref{eq:limit_3}. Both are less restrictive than Eq.~\ref{eq:limit_2} across essentially our entire resolution range. This means that increasing or decreasing $\epsilon$ would likely only worsen the convergence properties of our subhalos. As we discuss in Sections \ref{sec:structural} and \ref{sec:mass_loss_rates}, there are some subhalo properties whose reliability is well described by these limits, but there are other properties that appear to be reliable below these limits.

We fit the moments of the combined idealized limit distribution against the form
\begin{equation}
   \label{eq:max_conv_limit_fit}
   n_{\rm lim,ideal}(n_\star;\,q) = 10^{b_2x^2 + b_1x + b_0}
\end{equation}
Here $q$ is the target quantile, $x\equiv\log_{10}(n_\star)$, $n_\star$ is either $n_{\rm peak}$ or $n_{\rm infall}$, and $b_2,$ $b_1,$ and $b_0$ are fit parameters. The best-fitting parameters for $q$ = 0.16, 0.50, and 0.84 are shown in Table \ref{tab:conv_fits}. We caution readers before using these fits: the relative importance of different numerical limits depends strongly on force softening, so this fit may not be applicable to all simulations. That said, it is likely that this fit is correct for other simulations that are also primarily limited by discreteness noise.

Whether or not these predicted limits are consistent with the picture painted by RCT convergence tests depends on how the convergence tests are performed. Convergence tests performed on the {\em instantaneous} SHMF show that convergence is achieved at  $\approx 300{-}1000$ particles \citep[e.g.][]{nadler_2022_symphony}, which approximately matches the range spanned by $n_{\rm lim,ideal}.$ The picture painted by $m_{\rm peak}$ SHMFs is more complicated because RCT $m_{\rm peak}$ SHMFs appear to very roughly converge at modest particle counts, while $n_{\rm lim,ideal}$ should impact convergence behavior at all $n_{\rm peak}$ values. We discuss this in detail in Appendix \ref{sec:shmf_converge}, but put briefly: the fact that RCT rapidly loses track of subhalos prevents non-convergence from being relevant at high $n_{\rm peak}$ values. This Appendix should be read in concert with Section \ref{sec:shmf}.

In the following Sections, we test how well $n_{\rm lim,ideal}$ describes the convergence behavior of subhalos in practice.

\begin{table}
\centering
\begin{tabular}{c|c|c|c|c}
\hline
$n_\star$ & $q$ & $b_2$ & $b_1$ & $b_0$ \\ [0.5ex]
\hline\hline

$n_{\rm peak}$ & 0.16 & 0.00013 & 0.2109 & 1.8675 \\
& 0.50 & -0.01853 & 0.3861 & 1.6597 \\
& 0.86  & -0.07829 & 0.9338 & 0.7743 \\
\hline
$n_{\rm infall}$ & 0.16 & -0.01294 & 0.2938 & 1.7376 \\
& 0.50 & -0.01853 & 0.3861 & 1.6597 \\
& 0.84 & -0.07829 & 0.9338 & 0.7743 \\
\hline
\end{tabular}
\caption{Best fitting-values for several moments of the resolution-dependent $n_{\rm lim,ideal}$ distribution (Eq.~\ref{eq:max_conv_limit_fit}). $n_\star$ is the definition of subhalo resolution used for the fit, $q$ is the quantile of the $n_{\rm lim,ideal}$ distribution, and $b_2,$ $b_1,$ and $b_0$ are parameters in a log-quadratic fit (Eq.~\ref{eq:max_conv_limit_fit}). We caution readers that these parameters may change for simulations where $\epsilon$ is set too large or too small or simulations where time-stepping is too coarse.}
\label{tab:conv_fits}
\end{table}

\subsection{The numerical reliability of $v_{max}$ in disrupting subhalos}
\label{sec:structural}

\begin{figure}
\hspace{-3.5mm}
\includegraphics[width=0.475\textwidth]{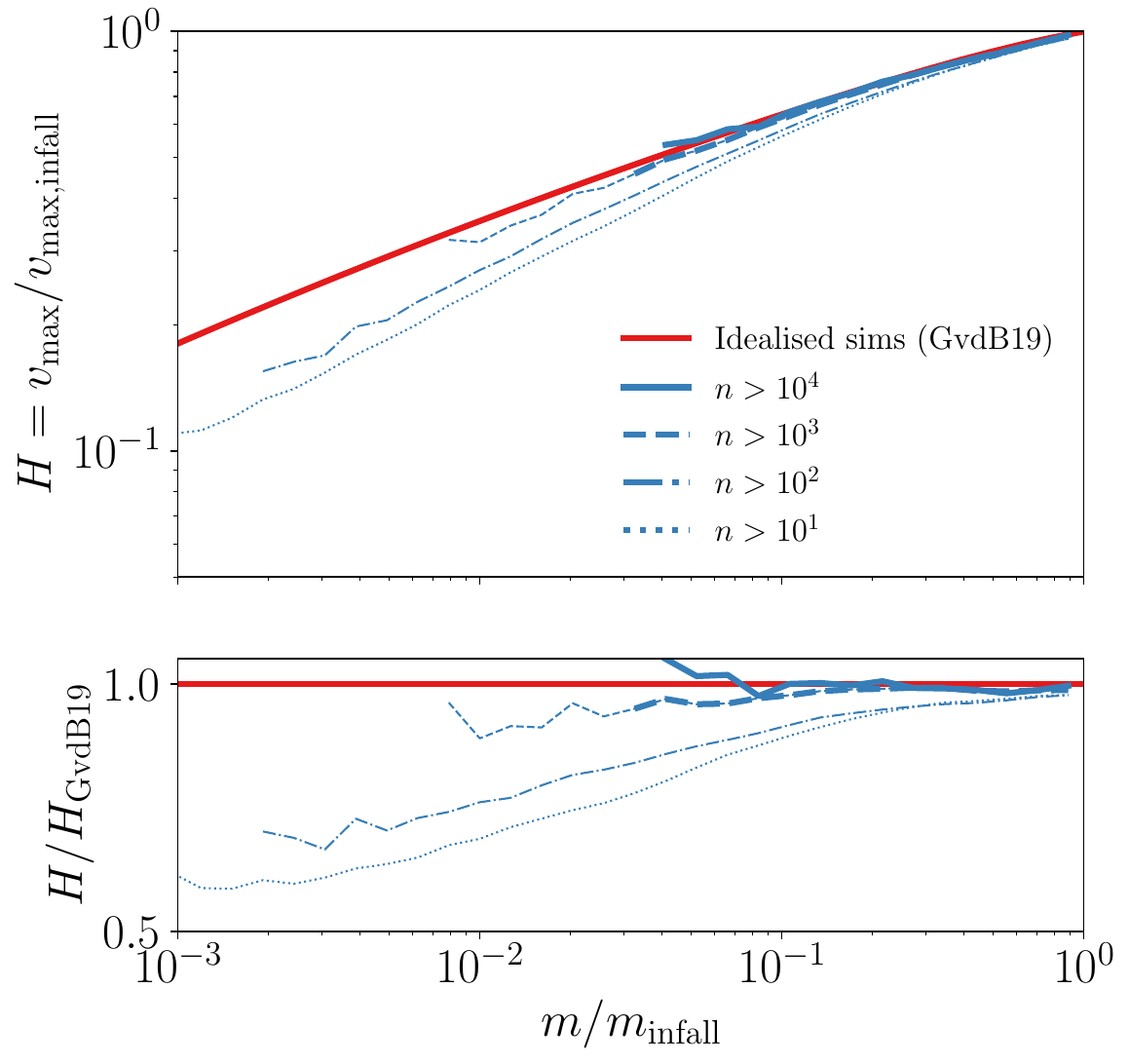}
\caption{The median relationship between $v_{\rm max}/v_{\rm max,infall}$ and $m/m_{\rm infall}$ for subhalos, stacked across the entire Symphony suite, as measured by \textsc{Symfind}. The red curve shows the predictions of high-resolution idealized simulations \citep{green_vdb_2019_tidal}, performed over a range of orbital parameters and matched to the infall concentration distribution of our subhalo sample. The blue curves show the relation when calculated above different particle-count cutoffs. The ratios between these two curves are shown in the bottom panel. For simplicity, we only plot the \citet{green_vdb_2019_tidal} model for the $n > 10^4$ bin in the top panel, but in the bottom panel, each bin is compared against models with a matching concentration distribution. Curves transition from thick to thin at the $m/m_{\rm infall}$ values where Eq.~\ref{eq:max_conv_limit_fit} predicts that the median halo is no longer converged. Our simulations are in good agreement with idealized simulations for converged subhalos; $v_{\rm max}$ is biased low when numerical models predict resolution effects should be important.}
\label{fig:vacc_macc}
\end{figure}

In this Section, we show that Eq.~\ref{eq:max_conv_limit_fit} does a good job at describing when the $v_{\rm max}$ values of subhalos are properly recovered by \textsc{Symfind} and show that above this limit, our halo catalogs are in good agreement with the predictions of idealized simulations.

As subhalos orbit their hosts, they lose high-energy particles close to their tidal radii before losing small-radius, low-energy particles \citep[e.g.,][]{penarrubia_2008_tidal_evolution,green_vdb_2019_tidal,errani_navarro_2021_asymptotic}. To characterize the relative mass loss at small and large radii, authors often consider the functional form
\begin{equation}
\label{eq:vacc_macc}
\frac{v_{\rm max}}{v_{\rm max,infall}} = 2^\xi v_{\rm max,infall} \frac{(m_/m_{\rm infall})^\nu}{(1 + m/m_{\rm infall})^\xi},
\end{equation}
following \citep{penarrubia_2008_tidal_evolution,penarrubia_2010_impact}. Initially, studies of this relation relied on relatively small samples of idealized simulations of orbiting subhalos that were run at very high resolutions. These studies convincingly showed that the relationship between $v_{\rm max}$ and $m$ remains unchanged at a fixed subhalo mass loss fraction under changes in host properties and subhalo mass loss rate \citep{penarrubia_2010_impact}, in agreement with full dark matter zoom-in simulations \citep{kravtsov_2004_tumultuous}. However, the relation depends on subhalo concentration and remained poorly understood due to the limited sizes of these simulation suites and the limited ranges of infall concentrations employed by them. This was rectified with the running of DASH \citep{ogiya_2019_dash}, a collection of thousands of idealized simulations that span a wide range of initial subhalo parameters. Using DASH, \citet{green_vdb_2019_tidal} developed an accurate model for $\xi(c_{\rm infall},\,m/m_{\rm infall})$ and $\nu(c_{\rm infall},\,m/m_{\rm infall})$, where $c_{\rm infall}$ is the concentration of the subhalo at infall. Note that in \citet{green_vdb_2019_tidal}, the variable we have labeled $\xi$ was labeled as $\mu$. Given the massive size of the simulation suite used to calibrate this model, the range of subhalo parameters explored, and the detailed tests of this model, we consider it to be the most accurate representation of the ``true'' predictions of idealized simulations. 

In Fig.~\ref{fig:vacc_macc}, we compare the evolution  of $v_{\rm max}$ predicted by these models and what we find in our \textsc{Symfind} catalogs. We break our subhalos into groups according to the instantaneous number of particles, $n=m/m_p.$ For each bin, we evaluate at Eq.~\ref{eq:vacc_macc} using the model presented in \citet{green_vdb_2019_tidal} for the distribution of infall concentrations in each resolution bin. For simplicity, we only show this model for the highest resolution bin in the top panel, but each curve is compared against its own concentration distribution in the bottom bin. We have constructed this plot separately for each Symphony simulation, and the ratio between $v_{\rm max}/v_{\rm max,infall}$ in our simulated subhalos and in the idealized models does not depend on subhalo mass, only resolution, so we stack all the suites together to improve number statistics.

Above the resolution limits predicted by Eq.~\ref{eq:max_conv_limit_fit}, $v_{\rm max}$ evolution is consistent with the predictions of idealized simulations. Below these limits, $v_{\rm max}$ skews low relative to these predictions, and the bias increases as the resolution is decreased. In other words, our simulations converge towards the behavior of high-resolution idealized simulations and reach agreement at the resolution limits where these idealized simulations predict they should.

The idealized simulations described above also make strong predictions for the evolution of $r_{\rm max}$, the radius at which the circular velocity profile is highest. We find that this quantity is very noisy in our catalogs, even for isolated halos and subhalos experiencing slow mass loss, regardless of whether we use \textsc{Symfind} or RCT. Thus we do not perform tests on it here. 

In summary, the tests performed in this Section support restricting analysis that relies on the $v_{\rm max}$ values of subhalos to the regime suggested by numerical limits discussed in Section \ref{sec:numerical_limits}:
\begin{align}
\label{eq:n_lim_vmax}
    n_{\rm lim,vmax}(n_{\rm peak}) = n_{\rm lim,ideal}(n_{\rm peak})
\end{align}
We discuss how this translates to constraints on a subhalo population selected by pre-infall masses in Section \ref{sec:good_enough}. We argue that using current models of galaxy disruption, only subhalos with $n_{\rm peak} > 3\times 10^4$ will have $n>n_{\rm lim,vmax}$ (and thus resolved $v_{\rm max}$) until the point of likely galaxy disruption.

\subsection{Mass loss rates}
\label{sec:mass_loss_rates}

\begin{figure}
\hspace{-3.5mm}
\includegraphics[width=0.475\textwidth]{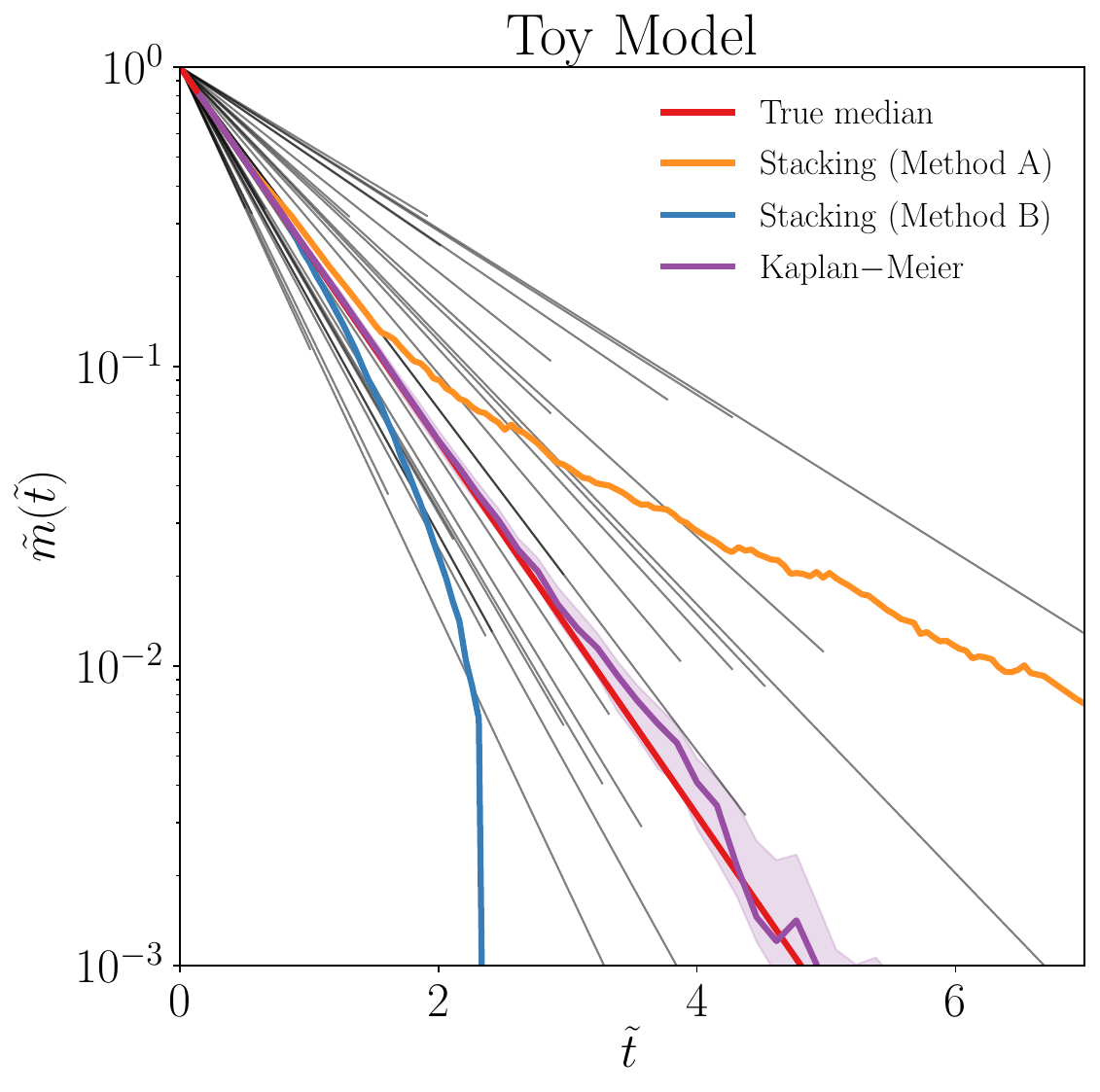}
\caption{A toy model for the evolution of a toy mass, $\tilde{m}$, as a function of toy time, $\tilde{t}$, which illustrates the impact of survivor bias. A population of $10^3$ mock subhalos was generated with random exponential timescales and random mass scales at which they are lost by the subhalo finder. 30 random subhalos are shown in black. The true median of the population is shown in red. In orange and blue, we show two flawed methods for estimating the population median. Orange shows the median mass of all surviving subhalos, and blue shows the median assuming that $\tilde{m}(\tilde{t})=0$ after censoring (disruption). Both methods are heavily biased, and this bias begins well before the median disruption mass and disruption time for the sample ($\tilde{m}=10^{-1.5}$ and $\tilde{t}=2.5$, respectively). The purple curve shows the Kaplan--Meier-based estimator described in Section \ref{sec:mass_loss_rates} and the purple shaded contour shows the 68\% confidence interval on this median. Kaplan--Meier remains an unbiased estimator to far lower masses, and we thus use it in Fig.~\ref{fig:mass_loss_conv}.}
\label{fig:survivor_bias}
\end{figure}

\begin{figure*}
\hspace{-3.5mm}
\includegraphics[width=0.32\textwidth]{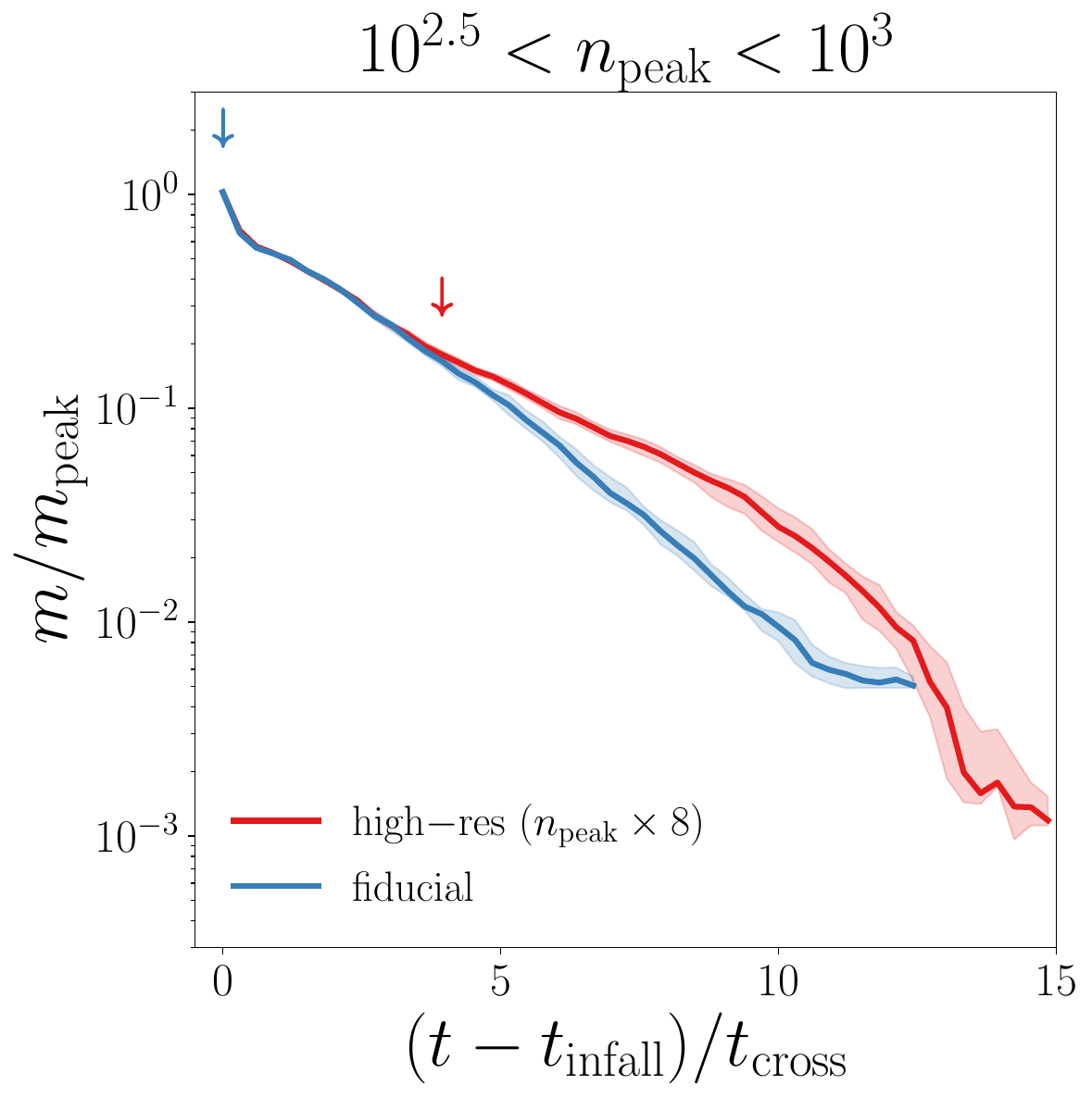}
\includegraphics[width=0.32\textwidth]{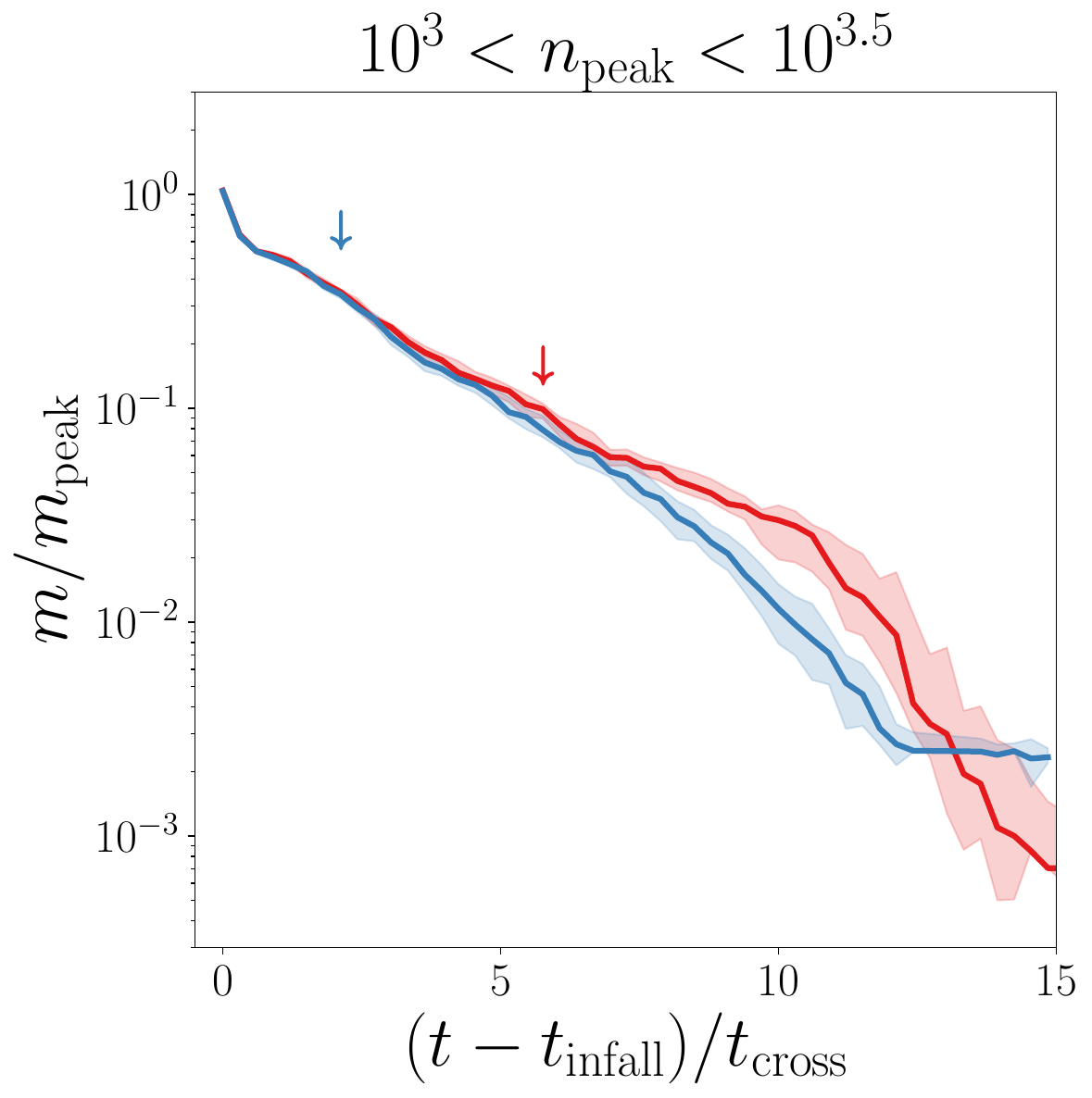}
\includegraphics[width=0.32\textwidth]{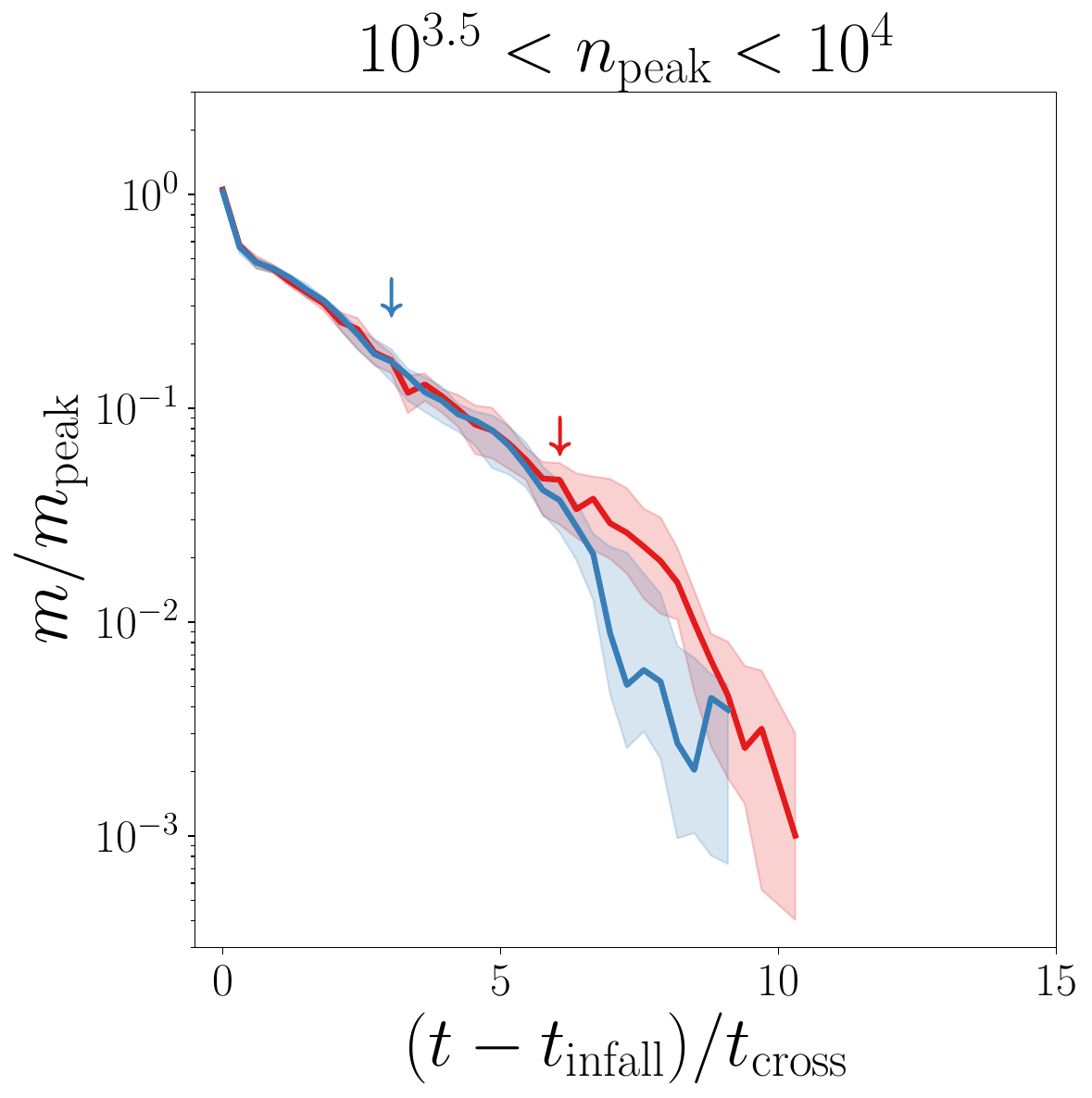}
\caption{Subhalo mass loss rates compared against idealized numerical limits. Each panel shows a set of \textsc{Symfind} subhalos from the SymphonyMilkyWayHR suite (red) and their mass-matched subhalos from the fiducial-resolution SymphonyMilkyWay (blue), grouped by $n_{\rm peak}$ of the fiducial-resolution subhalos. The Kaplan-Meier estimator has been used to correct for survivor bias and to estimate $1\,\sigma$ uncertainties (shaded bands). The red and blue arrows show the times at which mass loss rates in the corresponding simulations are predicted to become unconverged according to Eq.~\ref{eq:max_conv_limit_fit}  a fit against idealized numerical limits. These limits underestimate how long mass loss rates are converged. Instead, our fiducial simulations stay converged until the predicted limit for the {\em high-resolution} simulations, motivating the less conservative empirical limit given in Eq.~\ref{eq:n_lim_mass}.
}
\label{fig:mass_loss_conv}
\end{figure*}

The previous Section established limits where the $v_{\rm max}$ of subhaloes can be properly resolved, and in this Section, we construct similar limits for subhalo mass loss rates. Unfortunately, subhalo mass loss rates are more complicated than the connection between mass loss and the decrease in $v_{\rm max},$ so there is no idealized equivalent to a tidal track that we can simply compare against, as we did in Section \ref{sec:structural}. Instead, we compare against a matched sample of higher resolution subhaloes \refadd{from SymphonyMilkyWay} {\em which are predicted to be converged according to Eq.~\ref{eq:max_conv_limit_fit}}. As we demonstrated in Section \ref{sec:structural}, the evolution of $v_{\rm max}$ --- a more numerically challenging problem than mass loss rates --- is converged and in agreement with idealized simulations above this threshold, so it is highly unlikely that this comparison could suffer from false convergence.

Subhalo mass loss rates are particularly susceptible to survivor bias. Survivor bias is an effect one finds in samples with statistical censoring (see Section \ref{sec:survial}) where censoring is connected with the statistic being measured in some significant way. To review, in the case of Section \ref{sec:survial}, the statistic we wanted was the distribution of disruption masses, but some long-lived subhalos survived past the end of the simulation, meaning that long-lived subhalos were more likely to be censored, biasing the distribution of fully disrupted subhalos to higher disruption masses than the full sample. The solution to this was to use the Kaplan--Meier estimator (Eq.~\ref{eq:kaplan_meier}).

Subhalo mass loss rates face a similar, but more subtle version of survivor bias, as  noted in \citet{han_2016_unified}.\footnote{Survivor bias can take many forms and is often very subtle. For example, during World War II, the statistician Abraham Wald wrote a series of memos designed to assess what portions of Allied bomber planes needed to be reinforced with more armored plating \citep{mangel_1984_wald}. He developed a sophisticated statistical system for this, with the conclusion that it was most important to reinforce parts that were rarely shot on returning bombers. The planes that {\em did} get shot in those places never returned.} This is because long-lived subhalos will tend to have slower mass loss rates, thus correlating survival times and $m(t)$. To illustrate this, we construct a toy model for subhalo mass loss, where all subhalos have an exponential mass loss rate $\tilde{m}(\tilde{t}) = \exp{(-\alpha \tilde{t})}$, where $\tilde{m}(\tilde{t})$ is the unitless toy mass relative to the infall mass, $\tilde{t}$ is the unitless toy time since accretion, and $\alpha$ is a free parameter controlling the mass loss rate. We allow for random scatter in $\alpha$ and cause subhalos to be censored at random $\tilde{m}(\tilde{t})$ values. The specifics of the distributions used do not change any conclusions that we draw from this toy model, so we pick them for visual clarity. In this case, we generate $\alpha$ uniformly at random between 0.25 and 1, and generate the censoring masses from a log-normal distribution with median $10^{-1.5}$ and $\sigma = \ln{(10)}$ and do not allow subhalos to have disruption masses greater than 1. This leads to a median disruption mass of approximately $10^{-1.5}$ and a median disruption time of approximately 2.5. We draw $10^3$ random realizations from this model.

Fig.~\ref{fig:survivor_bias} shows a subset of 30 realizations, along with what the median of the $\tilde{m}(\tilde{t})$ distribution would be without censoring. We also show two simple methods for stacking $\tilde{m}(\tilde{t})$ to estimate the median: (1) taking the median of all uncensored subhalos, and (2) taking the median of all subhalos regardless of censoring, but setting $\tilde{m}(\tilde{t})=0$ after censoring. The former method biases high, and the latter method biases low. The same effect can be seen in the radial distribution of $m/m_{\rm infall}$ for subhalos in Fig.~9 of \citet{han_2016_unified}. These biases relative to the true median of $\tilde{m}(\tilde{t})$ start early, well before the median disruption mass. This is problematic if one wants to characterize, e.g., the sharp downturn seen in Fig.~\ref{fig:conv_limits} for a statistical sample of subhalos. when present, this sharp downturn is always terminated by the subhalo finder losing track of the object, meaning that the location of the feature is very close to the median disruption mass. This is also problematic for resolution tests: increasing resolution will change the median disruption mass and even far above the typical disruption mass this could lead to a slope change that could be interpreted as non-convergence.

To account for this bias, we once again use the Kaplan--Meier estimator (Eq.~\ref{eq:kaplan_meier}) and Greenwood's formula (Eq.~\ref{eq:greenwood}). We aim to measure the distribution of 
\begin{equation}
m((t-t_{\rm infall})/t_{\rm cross})/m_{\rm peak}\equiv\mu((t - t_{\rm infall})/t_{\rm cross})
\end{equation}
for a range of fixed values of $t/t_{\rm cross},$ where $t_{\rm cross} \equiv 2\,R_{\rm vir}(t_{\rm infall})/V_{\rm vir}(t_{\rm infall})$ is the crossing time at infall. In this case, the measurement is considered uncensored if the subhalo survives until $(t - t_{\rm infall})/{\rm t_{\rm cross}}$, and is otherwise censored at the final $\mu$ value achieved prior to the simulation ending or the halo finder losing the subhalo. We estimate a 68\% confidence interval by constructing two bounding probability distributions $\widehat{\rm Pr}(<\mu) \pm \widehat{{\rm Var}}[\widehat{\rm Pr}(<\mu)]$, both bounded between 0 and 1. We then estimate the confidence interval of ${\rm \widehat{Pr}(<\mu)}$ as the medians of these two distributions. We show this estimator in purple in Fig.~\ref{fig:survivor_bias}. It is able to recover the median for our toy model to well below the median disruption mass.

Using this Kaplan--Meier-based method, we find the median $m/m_{\rm peak}((t-t_{\rm infall})/t_{\rm cross})$ evolution in three resolution bins across the range $10^{2.5} < n_{\rm peak} < 10^4$. We measure these distributions for both our high-resolution SymphonyMilkyWayHR halos and their paired fiducial-resolution resimulations and show the results in Fig.~\ref{fig:mass_loss_conv}.

For all three mass bins, we find that mass loss rates are converged below the mass scales predicted by Eq.~\ref{eq:max_conv_limit_fit} and are consistent with being reliable to much lower resolutions: until $\approx n_{\rm lim,ideal}(8\, n_{\rm npeak})$.

We do not attempt to develop a model that explains this fortunate, but unexpected, convergence in this paper. These limits are a combination of three independent effects that all select similar mass scales, and these limits were effective at predicting when internal subhalo densities are biased low. Thus, it is surprising to not see a similar bias in mass loss rates given the dependence of tidal forces on enclosed density. However, we note that:
\begin{enumerate}
\item This is unlikely to be an example of false convergence (i.e. a situation where both the high- and low-resolution simulations are both incorrect but happen to agree). This is because the high-resolution curves are above their own formal resolution requirements in the region of agreement, and those formal limits were shown to be good predictors of $v_{\rm max}$ convergence behavior in Section \ref{sec:structural}.
\item Formal convergence and practical convergence are not the same as one another. Slight differences between the fiducial and high-resolution curves that are smaller than our $1\,\sigma$ error contours can be seen as $t$ increases, particularly in the $10^3 < n_{\rm peak} < 10^{3.5}$. It is possible that non-convergence in internal structure translates into a weak non-convergence in mass-loss rates that is consistent with our error bars.
\item The majority of the simulations used to calibrate Eq.~\ref{eq:limit_1} and Eq.~\ref{eq:limit_2} in \citet{vandenbosch_2018_disruption} had $n_{\rm peak} > 10^4$, so it is possible these limits scale more gently with resolution in the low particle-count regime.
\end{enumerate}

Because of this third point, we strongly urge readers not to extrapolate these results to higher resolutions without further testing.

In summary, the tests performed in this Section support restricting analysis that relies on the instantaneous masses of moderate-resolution subhalos to objects where $n > n_{\rm lim,mass}$, defined as
\begin{align}
\label{eq:n_lim_mass}
    n_{\rm lim,mass}(n_{\rm peak}) = n_{\rm lim,ideal}(8\, n_{\rm peak}).
\end{align}
We discuss how this translates to constraints on a subhalo population selected at by pre-infall masses in Section \ref{sec:good_enough}. We argue that using current models of galaxy disruption, only subhalos with $n_{\rm peak} > 4\times 10^3$ will have $n>n_{\rm lim,mass}$ (and thus resolved $m(t)$ and abundances) until the point of likely galaxy disruption. This is roughly an order of magnitude less demanding in particle count than the requirement for converged $v_{\rm max}$ across the same mass loss range.

\section{Subhalo population statistics}
\label{sec:subhalo_pops}

As discussed in the Introduction, there are good reasons to wonder whether some current subhalo finders may have falsely converged and are missing an appreciable number of real subhalos. In Section \ref{sec:reliability}, we showed that \textsc{Symfind} is able to track subhalos to much smaller masses than one current cutting-edge subhalo finder, RCT, and that RCT exhibits signs of false convergence. In this Section, we show how these two findings propagate into well-worn subhalo statistics: the subhalo mass function and the subhalo radial distribution. We show that, as one would expect from the previous Section, at a fixed $m_{\rm peak},$ \textsc{Symfind} recovers substantially more subhalos than RCT, particularly at small radii and high resolutions, even though RCT appears to be converged.

\subsection{The subhalo mass function}
\label{sec:shmf}

\begin{figure*}
\hspace{-3.5mm}
\includegraphics[width=0.475\textwidth]{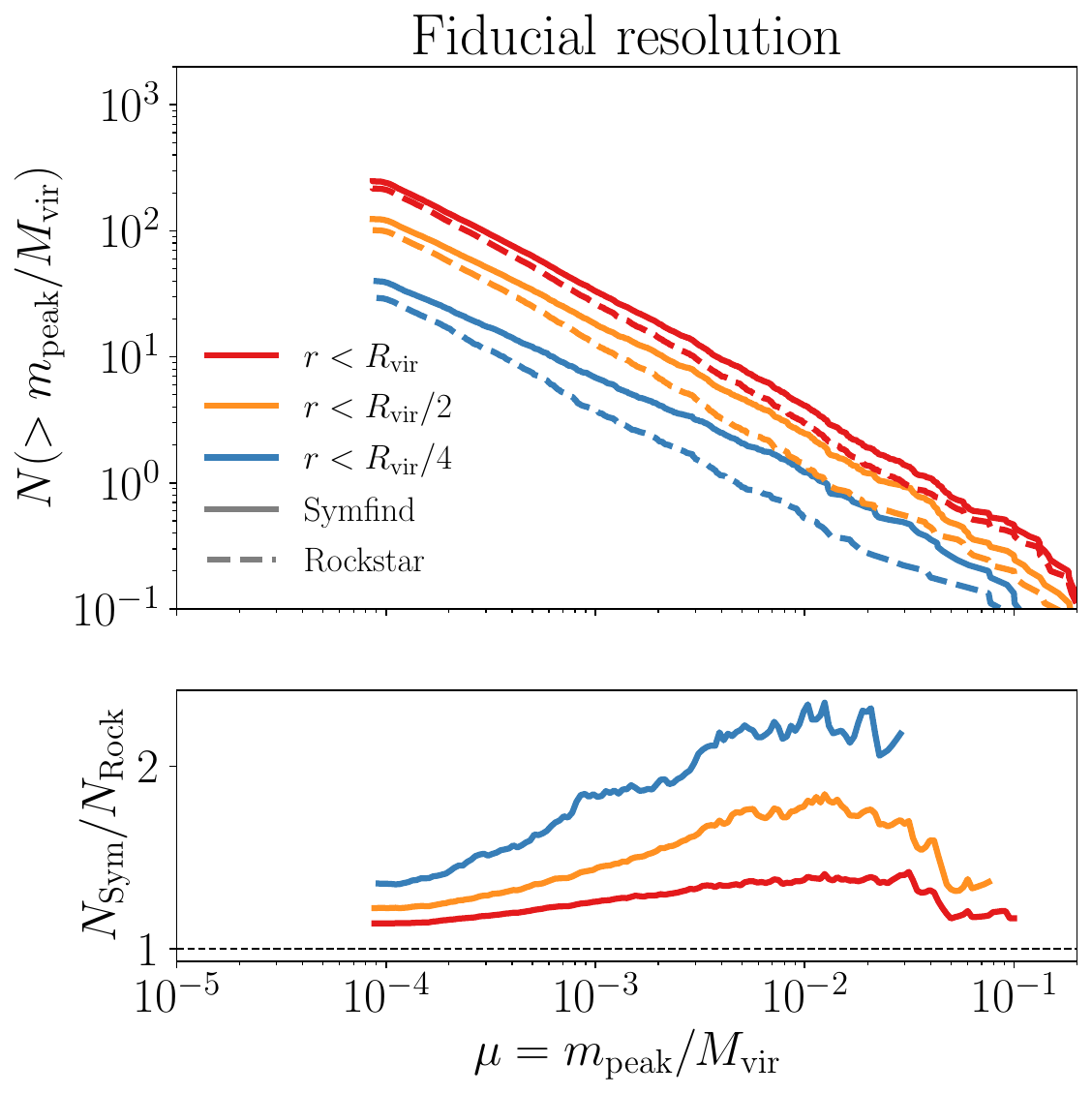}
\includegraphics[width=0.475\textwidth]{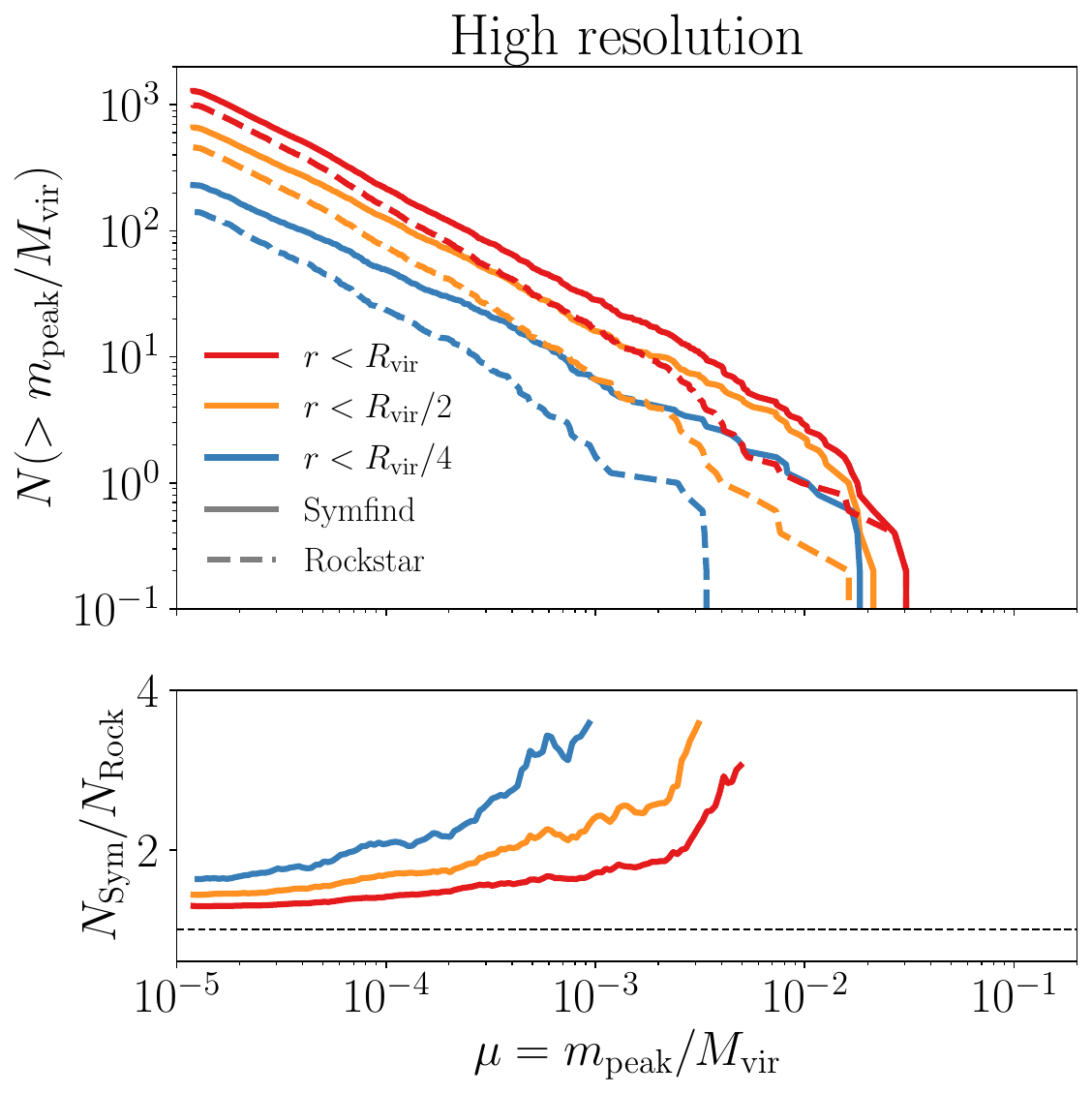}
\caption{
The $z=0$ subhalo $m_{\rm peak}$ functions of Milky Way-mass halos, normalized by hosts masses, $M_{\rm vir}$. {\em Left:} Comparison between the subhalo mass functions measured by \textsc{Rockstar} (dashed) and \textsc{Symfind} (solid) for the fiducial-resolution SymphonyMilkyWay suite. Mass functions within different radii are shown as different colors, and the bottom panel shows the ratio between \textsc{Symfind} mass functions and \textsc{Rockstar} mass functions. \textsc{Symfind} recovers more subhalos than \textsc{Rockstar}, with the difference increasing at smaller radii and higher $m_{\rm peak}$. {\em Right:} The same as the left panel, except that only mass functions for the four high-resolution SymphonyMilkyWayHR halos are shown. Note that the range of the $y$-axis of the lower panel has been expanded due to the larger differences between the two subhalo-finding methods. The same qualitative trends seen at fiducial resolutions are seen at higher resolutions, but roughly significantly more subhalos are recovered by \textsc{Symfind} compared to the fiducial case. See Section \ref{sec:shmf} for discussion.
}
\label{fig:shmf}
\end{figure*}

First, we consider the subhalo mass function (SHMF), a measure of the abundance of subhalos as a function of subhalo mass. The RCT SHMF is generally thought to be converged (e.g. \citealp{nadler_2022_symphony}) and we explicitly demonstrate that there appears to be qualitative convergence in the RCT SHMF in Appendix \ref{sec:shmf_converge} (although, as we discuss in that Appendix, the issue is subtle).

The left panel of Fig.~\ref{fig:shmf} compares SHMFs for the SymphonyMilkyWay suite, measured by RCT and by \textsc{Symfind} as a function of $m_{\rm peak}$. The SHMF is measured within three different radii: $R_{\rm vir}$, $R_{\rm vir}/2$, and $R_{\rm vir}/4$. We also plot the \refadd{ratio} between \textsc{Symfind} and RCT SHMFs, terminating the curve when the RCT sample has fewer than 10 subhalos in it to prevent excess shot noise.

The difference between RCT and \textsc{Symfind} significantly changes three of the most important aspects of the SHMF: its amplitude, slope, and radial dependence. \textsc{Symfind} recovers more subhalos than RCT at all masses. The effect is stronger at smaller radii and for higher mass subhalos. Within $R_{\rm vir}$, 15\%-to-40\% more subhalos are recovered, depending on subhalo mass. The logarithmic slope of the mass function at low masses, $\alpha$, is shallower, with $\alpha_{\rm sym}-\alpha_{\rm RCT} = -0.04$. Within $R_{\rm vir}/4$, 35\% to 120\% more subhalos are recovered, and $\alpha_{\rm sym}-\alpha_{\rm RCT} = -0.11$

These trends occur because of \textsc{Symfind}'s ability to follow subhalos to smaller $m/m_{\rm peak}$ ratios than RCT. Because the radii of low-mass subhalo orbits usually do not evolve much with time, subhalos at small orbits tend to be older, and because they are closer to the center of the host halo, tidal fields are more intense. This means that small-radius subhalos are more likely to have lost a large amount of mass than large-radius subhalos \citep[e.g.,][]{vdb_2016_segregation,han_2016_unified}. In contrast, high-mass subhalos are typically more recent accretions than low-mass subhalos but are also much more highly resolved. As we discussed in Section \ref{sec:survial}, improving subhalo resolution does not reduce the minimum $m/m_{\rm peak}$ values that RCT can reach before disruption, while increasing resolution {\em does} reduce this threshold for particle-tracking (Fig.~\ref{fig:survival}).

Increasing resolution increases the number of recovered subhalos and makes the slope of the SHMF shallower, as we show in the right panel of Fig.~\ref{fig:shmf}. This panel shows subhalo mass functions for the SymphonyMilkyWayHR suite, which has particle masses eight times smaller than SymphonyMilkyWay. As discussed in Section \ref{sec:sims}, these hosts were originally selected with criteria that biased against the presence of high-mass subhalos, as can be seen in Fig.~\ref{fig:shmf}. Compared to RCT, \textsc{Symfind} finds between 25\% and 130\% more subhalos within $R_{\rm vir}$ and between 60\% and upwards of 250\% more subhalos within $R_{\rm vir}/4$. The corresponding $\alpha_{\rm sym}-\alpha_{\rm RCT}$ values for these two enclosing radii are -0.1 and -0.21, respectively. At these resolutions, the impact of the choice of subhalo finder on subhalo abundances is comparable to the impact of including an embedded high-mass disk potential in the center of the halo \citep[e.g.][]{garrison_kimmel_2017_lumpy}. \refadd{Given that the impact of the baryonic disk is understood to be a major contributing factor in alleged small-scale cosmological tensions like ``Too Big to Fail,'' it stands to reason that halo finder choice may have a comparably significant impact on theoretical predictions for subhalo populations. This prediction appears to be upheld in embedded disk simulations run with Symfind (Wang et al., in prep).}

Where does this strong resolution dependence come from? Decreasing particle mass decreases the minimum $m/m_{\rm peak}$ that low-mass subhalos can be tracked to with our particle-tracking method (Section \ref{sec:survial}). As this limit decreases, the low-mass end of the SHMF will converge towards the ``unevolved'' mass function, i.e., the $m_{\rm peak}$ mass function of all objects that have ever been accreted onto the host and its subhalos, regardless of disruption status. In contrast, because RCT's threshold for subhalo disruption slowly increases with increasing resolution, its $m_{\rm peak}$ functions decrease slightly as resolution improves. See Appendix \ref{sec:shmf_converge} for extended discussion.

One might be concerned by this resolution dependence: does this mean that $m_{\rm peak}$ functions are unconverged at all masses, even at high-resolution levels? Yes. Unless a simulation has such a heroic level of resolution that it and its subhalo finder are able to resolve the {\em entire} unevolved mass function, raw $m_{\rm peak}$ mass functions will always be formally unconverged, as there will always be some subhalos which have passed below the resolution floor that could be recovered if the simulation had more particles. But this is not as dire a circumstance as it might seem. The $m_{\rm peak}$ function is not an observable quantity: one of the primary reasons why simulated predictions for the $m_{\rm peak}$ function are interesting is because they are often used as a crucial component when modeling the stellar mass or luminosity of dark matter subhalos. Using a model where stellar masses are assigned to subhalos based purely on their $m_{\rm peak}$ values implicitly assumes that galaxies never disrupt and never lose mass as long as some portion of their original subhalo survives. But galaxies {\em do} eventually disrupt and lose mass themselves (see Section \ref{sec:galaxy_disruption}). Thus, the process of making reliable predictions for satellite galaxy abundances based on their $m_{\rm peak}$ values depends on simultaneously accounting for subhalo finder limitations, numerical limits, and one's explicit choice of a galaxy disruption model. We discuss the intersection between these three factors in Section \ref{sec:good_enough}. Given the unresolved modeling caveats discussed in this later Section, we defer quantitative predictions for the converged stellar mass function to future work.

\subsection{The subhalo radial distribution}
\label{sec:radius_cdf}

\begin{figure}
\hspace{-3.5mm}
\includegraphics[width=0.475\textwidth]{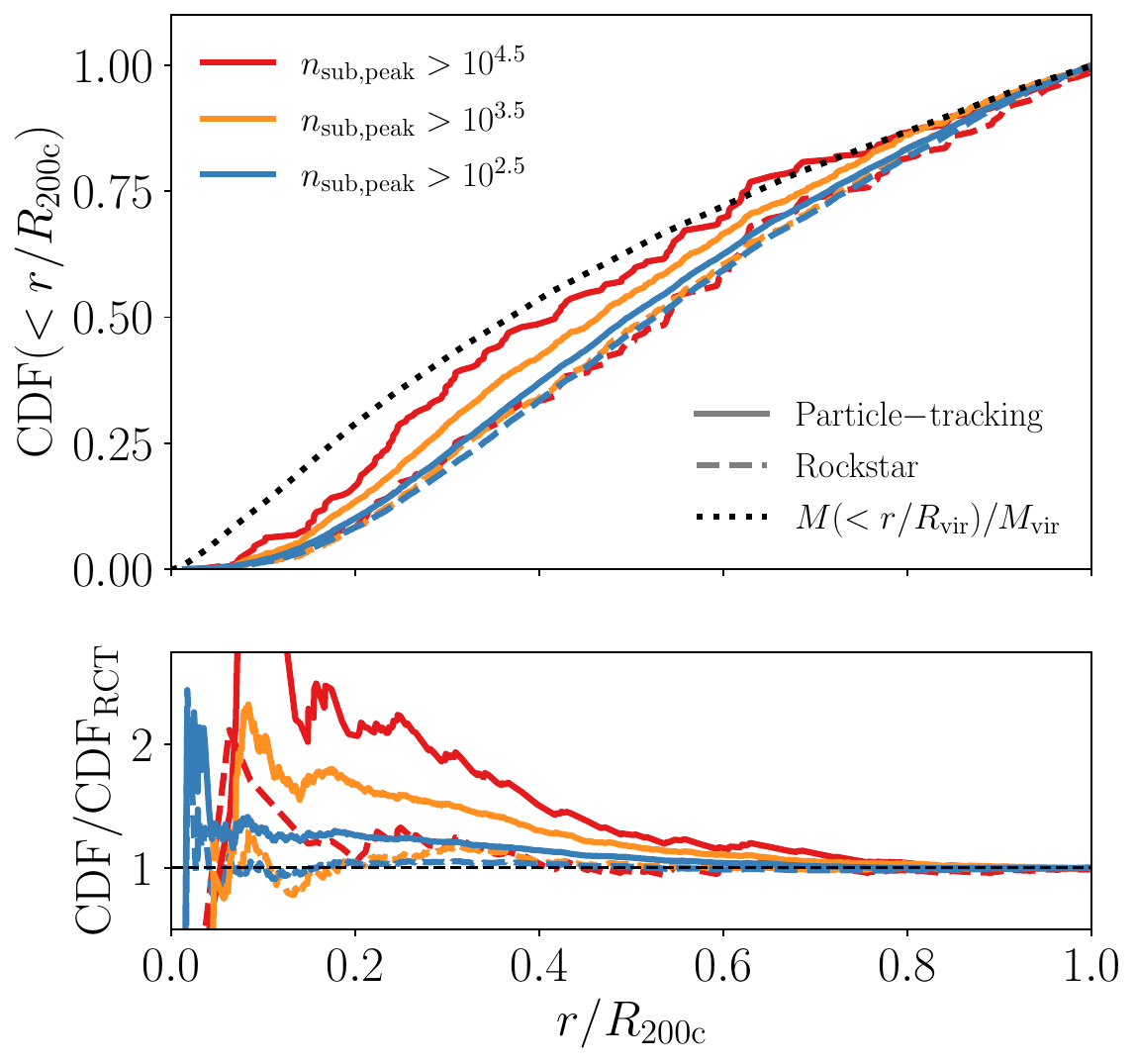}
\caption{Radial distribution of subhalos in the SymphonyMilkyWay suite. We compare the radial distribution of dark matter particles (dotted) against  \textsc{Symfind} (solid) and  \textsc{Rockstar} (dashed) . The bottom panel shows the ratios of these curves to a fit to the dashed blue curve, ${\rm CDF}_{\rm RCT}$ (Eq.~\ref{eq:cdf_rct}). \textsc{Rockstar} subhalos show no appreciable dependence on resolution. This is caused by the effect shown in the right panel of Fig.~\ref{fig:survival}: \textsc{Rockstar} survival mass ratios do not depend on resolution, meaning that high-resolution subhalos are no more adept at surviving in the inner regions of the host than low-resolution subhalos are, giving the false impression of convergence. \textsc{Symfind} finds more centrally concentrated subhalo distributions, and increasing resolution leads to radial profiles that increasingly approach the distribution of dark matter particles.}
\label{fig:radius_cdf}
\end{figure}

A great deal of literature has been written on whether the radial distribution of subhalos matches the observed distribution of satellite galaxies. At a fixed {\em present-day} subhalo mass, subhalo number densities are much less centrally concentrated than both satellite number densities and the total dark matter density profile of the host \citep[e.g.,][]{nagai_kravtsov_2005_radial,springel_2008_aquarius}. However, selecting subhalos at a fixed mass has little relevance to most observations of satellite radial distributions: satellites stay intact until their subhalos have lost large amounts of mass, meaning that any cut on a fixed present-day mass will necessarily underestimate the number densities of satellites at a fixed stellar mass at small radii where the ratio between stellar mass and subhalo mass will be abnormally high. It is well-known that number density profiles selected by infall masses --- which are more likely to trace galaxy stellar masses --- are more centrally concentrated than those selected by a present-day mass due to the larger amount of mass lost by small-radius subhalos \citep[e.g.,][]{nagai_kravtsov_2005_radial,han_2016_unified} and some authors claim that simply selecting subhalos by infall mass/velocity resolves the problem \citep[e.g.,][]{nagai_kravtsov_2005_radial,kuhlen_2007_shapes}. But others have found that agreement is only achieved for very high-mass subhalos with high resolutions that experience large amounts of dynamical friction \citep{ludlow_2009_unorthodox,bose_2019_revealing}. Some authors have argued that simulated and observed profiles cannot be brought to match without introducing ``orphan'' satellite galaxies in post-processing (i.e., model galaxies that outsurvive their subhalos; e.g., \citealp{gao_2004_galaxies_lcdm_clusters,newton_2018_total_sat_pop,carlsten_2022_elves}, see Section \ref{sec:method_comp}).

It is possible that much of the confusion comes from a combination of subhalo finding and resolution: \citet{green_202_tidal_evolution_ii} showed that idealized simulations predict that artificial disruption and subhalo finder limitations should have a large impact on subhalo number density profiles, and \citet{manwadkar_kravtsov_2022_grumpy,pham_2023_spatial} showed that with a modified version of RCT (see Section \ref{sec:caterpillar}), subhalo number density profiles converge towards the more highly concentrated dark matter density profiles of their host halos as resolution increases. In this Section, we come to a similar conclusion, following a procedure similar to \citet{manwadkar_kravtsov_2022_grumpy,pham_2023_spatial}.

In Fig.~\ref{fig:radius_cdf} we show the cumulative distribution of satellite radii, selected at different peak resolution levels, $n_{\rm peak},$ both with RCT and with \textsc{Symfind}. To aid in comparison against RCT, the bottom panels in Fig.~\ref{fig:radius_cdf} show the ratio of a given subhalo population's radial CDF against ${\rm CDF}_{\rm RCT}(r)$. ${\rm CDF}_{\rm RCT}(r)$ is a fit against the CDF of RCT subhalos with $n_{\rm sub,peak} > 10^{2.5}$ using the following expression:
\begin{equation}
\label{eq:cdf_rct}
    {\rm CDF}_{\rm RCT}(r) = 10^{c_4x^4 + c_3x^3 + c_2x^2 + c_1x}.
\end{equation}
Here, $x = \log_{10}(r/R_{\rm vir})$, $c_4 = -0.1434$, $c_3 = -0.6104$, $c_2 = - 1.6723$, and $c_1 = 0.6528$. This fit is accurate to the few-per-cent level within the range $10^{-1.5} < r/R_{\rm vir} < 1.$  We also show the average enclosed mass profile of the hosts in our sample, $M(<r/R_{\rm vir})/M_{\rm vir}.$

RCT profiles show no dependence on resolution, falling along a profile much less centrally concentrated than the underlying dark matter distribution. In contrast, \textsc{Symfind} leads to subhalo profiles that become more concentrated as resolution increases. Our highest resolution bin, $n_{\rm peak} > 10^{4.5}$ approaches the same shape as the underlying mass distribution, but is still less concentrated. The lack of convergence with increasing resolution makes it possible that even higher resolutions could lead to subhalo number density profiles that trace the underlying mass profile, as predicted by, e.g., \citet{han_2016_unified,green_202_tidal_evolution_ii}.

The agreement between RCT profiles at varying resolution levels is {\em false} convergence. As we showed in Section \ref{sec:reliability}, increasing resolution does not allow RCT to track subhalos to lower $m/m_{\rm peak}$ ratios, so multi-resolution tests do not result in different radial profiles, giving the incorrect impression that profiles are numerically reliable. 

There are two effects that could lead to increasingly concentrated subhalo number density profiles with increasing resolution. The first is physical: dynamical friction \citep[e.g.,][]{chandrasekhar_dyamical_1943} causes subhalos --- especially large subhalos --- to lose energy over time, leading to contracted orbits. This means that higher-mass subhalos will have smaller orbits than lower-mass subhalos accreted at the same time. The second is purely numerical: in the absence of dynamical friction, subhalos accreted at early times will have smaller orbits than those accreted at later times, which means that as resolution is increased and subhalos can be tracked for longer, number density profiles should become more concentrated. The fact that \textsc{Rockstar}'s number density profiles are fixed across resolution is false convergence in either case, but if the first effect dominates, the \textsc{Symfind} trend shown in Fig.~\ref{fig:radius_cdf} is a physical change that is missed by the false convergence; if the latter effect dominates, the particle-tracking trend is numerical and the curves should be interpreted as lower limits on the true CDF.

We do not believe Fig.~\ref{fig:radius_cdf} is showing the effects of dynamical friction. First, the effects of dynamical friction on subhalos in the mass range considered here are quite weak (e.g., Fig.~3 and Fig.~7 in \citealt{vdb_2016_segregation}). Second, we have re-created Fig.~\ref{fig:radius_cdf} using the smaller particle masses of SymphonyMilkyWayHR and did not see a significant difference in radial CDFs at a fixed resolution as the subhalo's $m_{\rm peak}/M_{\rm vir}$ changed. Third, \citet{manwadkar_kravtsov_2022_grumpy} showed that a similar convergence towards the mass profile can be found at a fixed subhalo mass by increasing resolution in a modified version of RCT (see discussion in Section \ref{sec:caterpillar}). 

In Appendix \ref{sec:subfind_radius_cdf}, we compare this analysis to \textsc{Subfind} radial number density profiles from \cite{han_2016_unified} and show that it is possible that \textsc{Subfind} does not suffer from this type of false convergence, but that it misses a large number of small-radius subhalos at moderate resolutions.

Lastly, as discussed previously in Section \ref{sec:shmf}, we note that simply making a cut at a fixed $m_{\rm peak}$ is insufficient to model the spatial distribution of galaxies. If a subhalo finder is capable of tracking a subhalo past the point at which its galaxy would have disrupted, some subset of very low $m/m_{\rm peak}$ subhalos would need to be removed from the sample (see Section \ref{sec:galaxy_disruption}). Given the connection between subhalo mass loss and subhalo position, this means that a flat $m_{\rm peak}$ or $v_{\rm peak}$ cutoff could overestimate the radial concentration of the satellite number density profile. We defer a complete treatment of this issue to future work.

\section{When is a subhalo finder ``good enough?''}
\label{sec:good_enough}

\begin{figure}
\hspace{-3.5mm}
\includegraphics[width=0.49\textwidth]{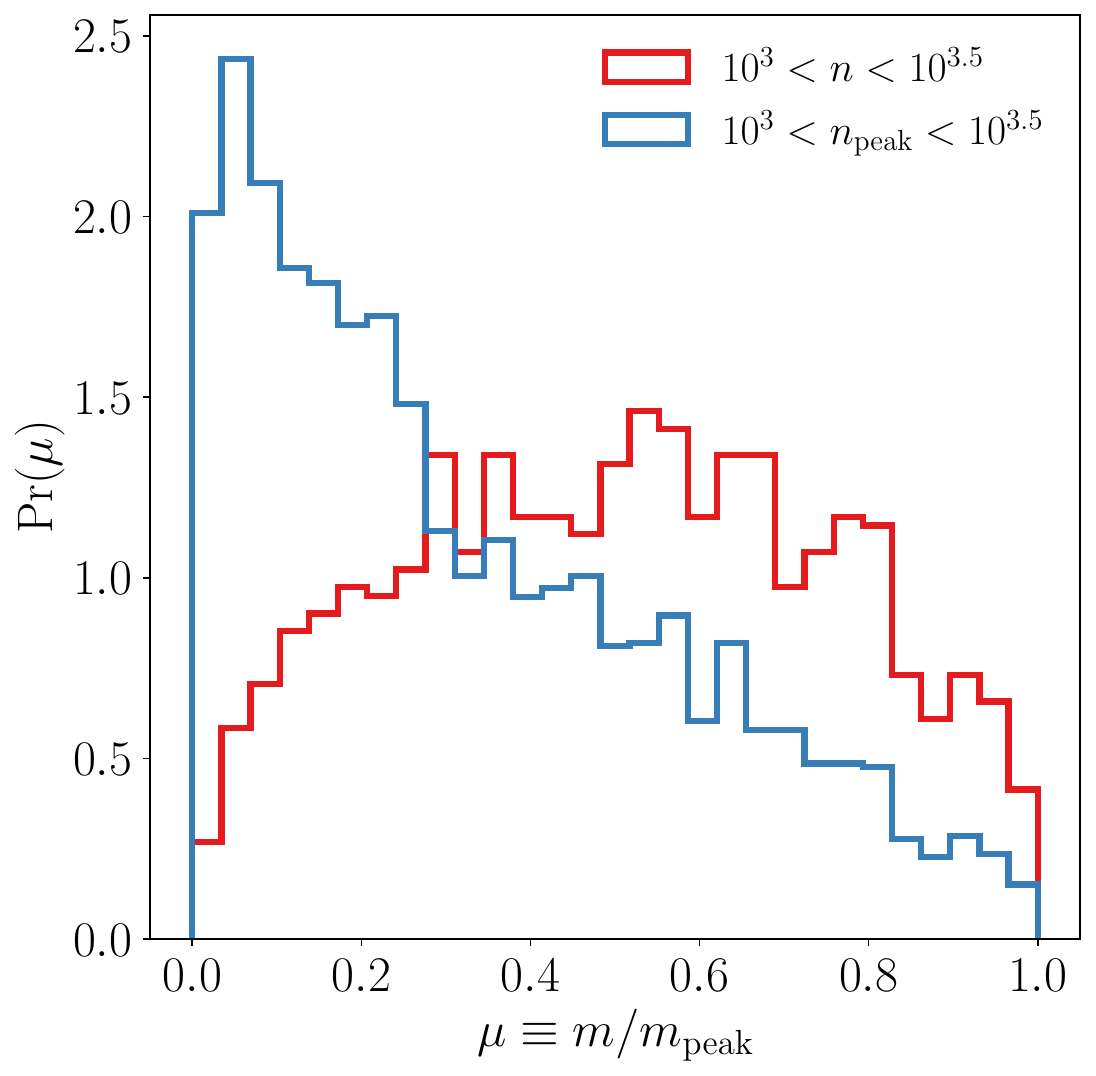}
\caption{\refadd{The distribution of $m/m_{\rm peak}$ for subhalos selected by current day particle count (red) and peak particle count (blue) in the SymphonyMilkyWay suite. Subhalos selected by peak quantities have experienced substantially more mass loss on average. Given the shallow slopes of disruption mass ratios as a function of $n_{\rm peak}$ shown in Fig.~\ref{fig:survival}, this means that at a fixed mass scale, peak-selected samples are more difficult for subhalo finders to resolve.}}
\label{fig:mloss_distr}
\end{figure}

\begin{figure}
\hspace{-3.5mm}
\includegraphics[width=0.475\textwidth]{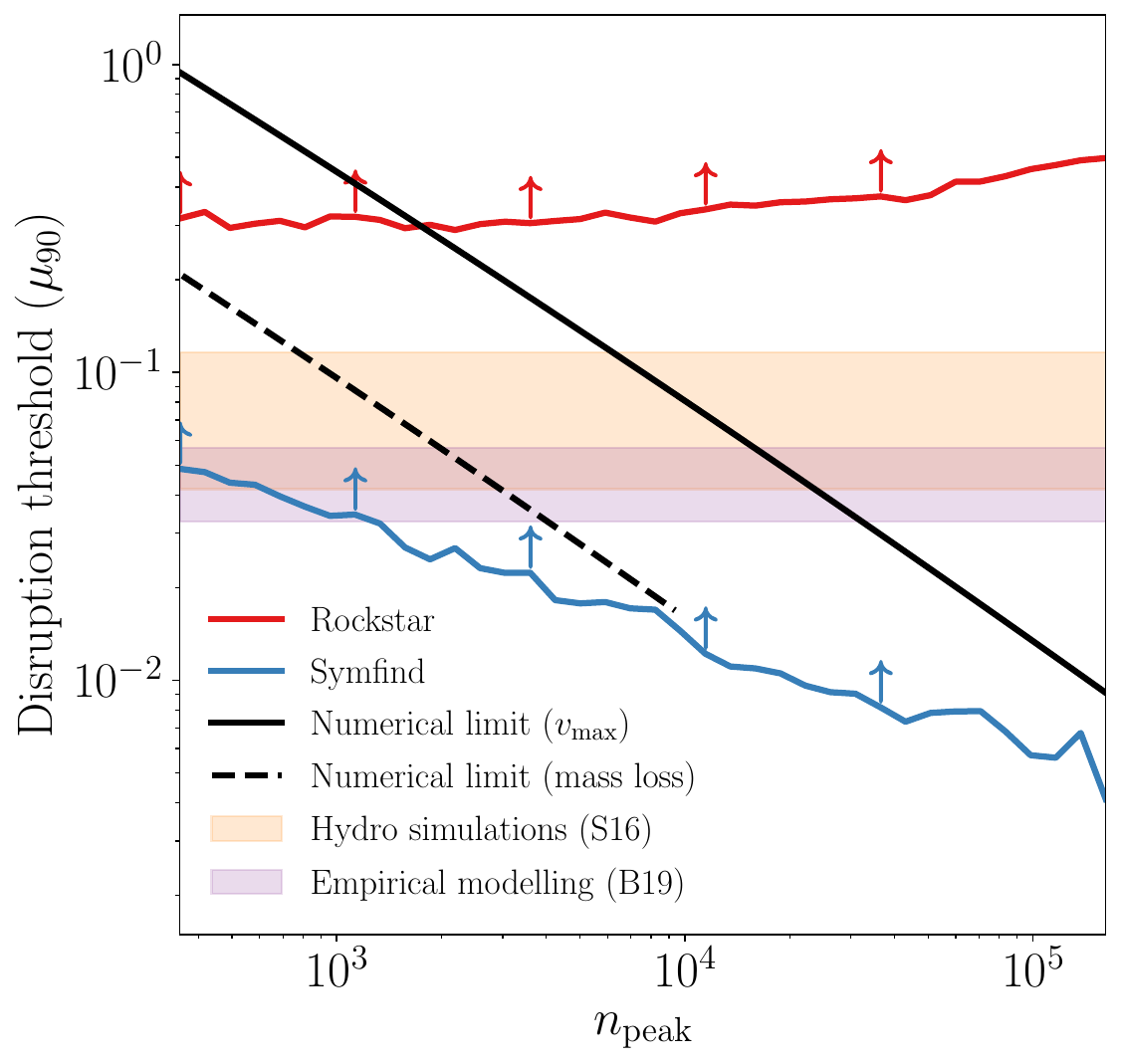}
\caption{
Comparison between subhalo finder disruption limits and other limiting factors for subhalo/satellite galaxy analysis.
The two colored curves show $\mu_{90}$, the $m/m_{\rm peak}$ ratio at which \textsc{Symfind} can still resolve 90\% of the subhalo population (reproduced from Fig.~\ref{fig:survival}). The orange and purple horizontal bands show the range of mass ratios at which galaxies are expected to disrupt from hydrodynamic simulations \citep[][orange]{smith_2013_impact} and from empirical modeling \citep[][purple]{behroozi_2019_universemachine}. To track all subhalos expected to host galaxies without ``orphan'' modeling, a subhalo finder must have disruption thresholds below these bands.
The solid black curve shows the minimum $\mu$ at which $v_{\rm max}$ is still resolved (Eq.~\ref{eq:n_lim_vmax}; \ref{sec:structural}) and the dashed black curve shows the minimum $\mu$ at which mass loss rates are still resolved (Eq.~\ref{eq:n_lim_mass}; Section \ref{sec:mass_loss_rates}). Satellite galaxy studies that only require resolved abundances and masses require $n_{\rm peak} \gtrsim 4\times 10^3$ and studies that require resolved rotation curves require $n_{\rm peak} \gtrsim 3\times 10^4$. 
\textsc{Symfind} is able to identify all resolved subhalos that are likely to host satellite galaxies.
}
\label{fig:survival_vmax}
\end{figure}

New subhalo-finding techniques have been developed nearly continuously for the past forty years. Many popular methods are described in \citet{knebe_2011_MAD}, and numerous sophisticated techniques have been developed in the subsequent years \citep[e.g.,][]{han_2018_hbt_plus,elahi_2019_velociraptor}. The development of these techniques even predates the time period when dark matter simulations were able to resolve substructure at all \citep{white_1987_galaxy_distribution}. Where does this end? Will we continue to make new halo finders for the next forty years? Does such a wide proliferation of methods mean that coming to some sort of consensus on the properties of the subhalos that host satellite galaxies is hopeless? Perhaps not. 

In this Section, we argue that \textsc{Symfind} {\em is able to track all numerically-resolved \refadd{dark matter-only} subhalos until the point of likely galaxy disruption}, according to current galaxy-disruption models. This means that improvements in halo finder techniques beyond our method would be unlikely to change predicted galaxy abundances. \refadd{We do not make claims about the performance of \textsc{Symfind} in hydrodynamic simulations or about the convergence properties of subhalos in hydrodynamic simulations.}

\refadd{Throughout this section we focus chiefly on $m_{\rm peak}$- and $n_{\rm peak}$-selected samples, rather than samples selected by their instantaneous mass. Both types of selections are important and relevant to different classes of observations. Pre-infall measures of mass like $m_{\rm peak}$ are generally more predictive of satellite stellar masses than current-day dark matter masses \citep[e.g.][]{reddick_2013_connection} and are thus important for modeling and predicting the properties of visible satellite galaxies. Current-day masses are instead important for gravitational probes of substructure (e.g., gaps in stellar streams, substructure lensing statistics; see, e.g., Section 5 of \citealp{bechtol_2022_snowmass} for review). Our focus on peak quantities is because this is a more stringent test for subhalo finders. Unless a subhalo finder specifically experiences unbinding errors that cause it to incorrectly lose mass in its outskirts, {\em a subhalo finder which can resolve a sample selected by peak mass will almost always be able to resolve an analogous sample selected by present day mass at a fixed mass scale}.

This statement has two components: one trivial and one non-trivial. The trivial component is that in the absence of mass errors, $m/m_{\rm peak} \leq 1$ for subhalos, meaning that an $m_{\rm peak}$-selected subhalo sample will always have fewer particles on average than an $m$-selected sample. The second, non-trivial component is illustrated in Fig.~\ref{fig:mloss_distr}. $m_{\rm peak}$-selected samples have lower average $m/m_{\rm peak}$ values than $m$-selected samples. This is because of the steep slope of the subhalo mass function. As high-mass subhalos lose mass, they move into lower mass bins where they find themselves increasingly outnumbered by subhalos which were accreted at lower infall resolutions but have only lost modest amounts of mass. As shown in Fig.~\ref{fig:survival}, the $m/m_{\rm peak}$ ratios at which either \textsc{Symfind} or \textsc{Rockstar} stop being able to track a fixed fraction of subhalos decreases slowly with $n_{\rm peak}$ with a slope shallower than -1.\footnote{As also seen in Fig.~\ref{fig:survival}, this slope is strongly dependent on halo finder. However, we find it {\em a priori} unlikely that any halo finder could have a slope steeper than -1. Highly disrupted subhalos are more difficult to distinguish from their tidal tails than lightly disrupted subhalos, are generally found in denser regions of their hosts, and have experienced more collisional relaxation in their centers.} Qualitatively, this means that heavily disrupted subhalos are harder to track at a fixed particle count than, and combined with Fig.~\ref{fig:mloss_distr}, means that even if the average {\em current day} mass of a $m_{\rm peak}$-selected sample matches a $m$-selected sample, the $m_{\rm peak}$-selected sample will be more difficult to track. }


\subsection{Disruption masses}

If idealized simulations are correct, low-mass \refadd{CDM} subhalos \refadd{without baryons} should survive in some form down to essentially arbitrarily small masses \citep{vandenbosch_2018_disruption,vdb_2018_tale_of_discreteness,errani_navarro_2021_asymptotic,green_2022_disc}. In this case, improvements in halo finders could likely always increase the abundance of subhalos one could find at a fixed $m_{\rm peak}$. However, galaxies lose mass along with their subhalos and at a certain point these subhalos no longer contain an observationally relevant galaxy for a given observation. When judging the utility of a halo finder, an important question is where this transition between relevant and irrelevant subhalos occurs for the target observable.

We argue that for the purposes of making predictions for visible satellite galaxy populations \refadd{with dark-matter-only simulations}, once a subhalo finder meets two criteria, improving subhalo finder performance no longer improves our ability to estimate the abundances of satellite galaxies. The first criterion is that the subhalo finder should be able to track subhalos until they are no longer numerically reliable. The second is that the subhalo finder should be able to track subhalos until the galaxies within them are likely to have disrupted.

\subsubsection{Numerical mass limits}

To estimate the impact of numerical resolution, we compare the 90\% $m/m_{\rm peak}$ disruption thresholds, $\mu_{90},$ from Fig.~\ref{fig:survival} for RCT and \textsc{Symfind} against two numerical limits in Fig.~\ref{fig:survival_vmax}. These are (1), the minimum $m/m_{\rm peak}$ at which $v_{\rm max}$ becomes unconverged ($n_{\rm lim,vmax}$; Eq.~\ref{eq:n_lim_vmax}) and (2), the limit at which subhalo mass loss rates become unconverged ($n_{\rm lim,mass}$; Eq.~\ref{eq:n_lim_mass}). Subhalo finders whose $\mu_{90}$ curve is below a particular limit are able to find all subhalos that can correctly resolve a particular subhalo property. We only plot $n_{\rm lim,mass}$ out to $n_{\rm peak} < 10^4$ because we do not know whether the relation is reliable beyond this resolution level. As discussed in Sections \ref{sec:structural} and \ref{sec:mass_loss_rates}, the limit at which $v_{\rm max}$ can be reliably measured, $n_{\rm lim,vmax}$, is higher at a fixed $n_{\rm peak}$ than $n_{\rm lim,mass}$. Both limits are shown in Fig.~\ref{fig:survival_vmax}.

RCT's $\mu_{90}$ disruption threshold is above both limits for most of the resolution range tested here, meaning that there are a large number of resolved subhalos that it cannot find.

\textsc{Symfind} has a $\mu_{90}$ disruption threshold below the mass-loss numerical limit while $n_{\rm peak} \lesssim 10^4$ and below the $v_{\rm max}$ numerical limit across the entire resolution range we can reliably probe (although extrapolation suggests that the disruption threshold will only stay below this numerical limit while $n_{\rm peak} \lesssim 3\times 10^5$). This means that above these two $n_{\rm peak}$ cutoffs, \textsc{Symfind} may start to miss some resolved subhalos with very low $m/m_{\rm peak}$.

However, these $m/m_{\rm peak}$ values are low enough that the galaxies that would be hosted by these subhalos may have either disrupted or experienced substantial mass loss. Because of this, the key question about these numerical limits is whether our method fails to find resolved subhalos {\em that still host galaxies}. We address this question in Section \ref{sec:galaxy_disruption}.

\subsubsection{Galaxy disruption mass loss limits}
\label{sec:galaxy_disruption}

To estimate the impact of galaxy disruption, we consider two estimates of when satellite galaxies disrupt. The first is calibrated off of hydrodynamic simulations, and the second is calibrated off of empirical models. There are several caveats associated with both estimates, which are discussed at the end of this subsection.

\citet{smith_2016_preferential} performed a study of galaxy mass loss as a function of subhalo mass loss in hydrodynamic simulations and found that galaxy masses exponentially decrease once $m/m_{\rm peak}$ crosses an exponential threshold, $f_0$. A reasonable approximation of this model is that galaxies disrupt when $m/m_{\rm peak} < f_0$. \citet{smith_2016_preferential} finds that small satellite galaxies ($0.025<r_{\rm eff}/r_{\rm vir}$ where $r_{\rm eff}$ is the satellite's effective radius at infall and $r_{\rm vir}$ is the subhalo's virial radius at infall) disrupt at relatively small masses, $f_0 = 0.0418,$ and that galaxies with large radii ($0.04>r_{\rm eff}/r_{\rm vir}$) disrupt quickly, $f_0=0.116$. This range is shown as an orange-shaded band in Fig.~\ref{fig:survival_vmax}.

\citet{behroozi_2019_universemachine} fit a galaxy disruption model against observational data that causes satellite galaxies to disrupt once their $v_{\rm max}$ values pass below some cutoff ratio $T_{\rm merge} \equiv v_{\rm max}/v_{\rm mpeak},$ where $v_{\rm mpeak}$ is the value of $v_{\rm max}$ during the snapshot that the subhalo achieved its maximum mass. $T_{\rm merge}$ was fit against:

\begin{align}
    \label{eq:um_disruption}
    T_{\rm merge} &= T_{\rm merge,300} + (T_{\rm merge,1000}-T_{\rm merger,300})\xi \\ 
    \xi &= 0.5 + 0.5\,{\rm erf}\left(\frac{\log_{10}{(V_{\rm Mpeak,host}/(1 {\rm km\,s^{-1})}) - 2.75}}{\sqrt{2}/4}\right).
\end{align}

Here, $V_{\rm Mpeak,host}$ is the $V_{\rm max}$ of the subhalo's host at the snapshot where the host achieved $M_{\rm peak}.$ \refadd{$T_{\rm merge,300}$ and $T_{\rm merge,1000}$ are \textsc{UniverseMachine} fit parameters intended to allow hosts with different masses to disrupt subhalos with different levels of efficiency.}

\citet{behroozi_2019_universemachine} found best-fitting values of 0.544 and 0.466 for $T_{\rm merge,300}$ and $T_{\rm merge,1000}$, respectively. Note that these limits are calibrated with respect to $V_{\rm peak}$ of the halo hosting the orphans, not the subhalos that sourced them. To convert these limits into the convention used by Fig.~\ref{fig:survival_vmax}, we invert the $v_{\rm max}/v_{\rm infall}$ to $m/m_{\rm infall}$ relation used by \textsc{UniverseMachine}  (Eq.~B5 in \citealp{behroozi_2019_universemachine}). The $v_{\rm max}$ values used in the fit to Eq.~\ref{eq:um_disruption} largely came from modeled ``orphan'' galaxies that were spawned after RCT lost track of their original dark matter subhalo and followed forward according to an analytic prescription that was used after the simulation finished running. From this point, subhalo masses were inferred from the orbit-averaged model in \citet{jiang_2016_statistics} and converted into $v_{\rm max}$ values by differentiating equation \ref{eq:vacc_macc}, adopting values for $\mu$ and $\nu$ from \citet{penarrubia_2010_impact}. As such, there is no loss of information by recasting these disruption limits in terms of subhalo mass. The resulting range of $m/m_{\rm peak}$ values at which this model predicts galaxies should disrupt is shown as a purple band in Fig.~\ref{fig:survival_vmax}.

RCT's disruption threshold is higher than both colored bands at all resolution levels, meaning that a substantial fraction of subhalos are lost before their satellite galaxies would have disrupted. This is unsurprising: galaxy models built on top of RCT catalogs generally need to generate post-disruption ``orphan'' galaxies to match large-scale clustering statistics \citep{pujol_2017_nifty,campbell_2018_crisis,behroozi_2019_universemachine}. In contrast, the disruption threshold for our particle-tracking method is lower than both colored bands for all subhalos with $n_{\rm peak} > 10^3$, meaning that --- if one ignores numerical effects --- no post-hoc orphan modeling would be needed to follow galaxies to the point of disruption for subhalos with $n_{\rm peak} > 10^3$.

However, numerical effects increase this lower limit on $n_{\rm peak}$ for all subhalo finders, including ours. Of the galaxy disruption models tested here, the one that requires the most subhalo mass loss prior to galaxy disruption is \textsc{UniverseMachine}'s high-galaxy-mass limit. $n_{\rm lim,mass}$ (Eq.~\ref{eq:n_lim_mass}) passes below this limit at $n_{\rm peak} > 4\times 10^3$, meaning that for generic galaxy populations, one would want at least this many particles for resolved masses and abundances. $n_{\rm lim,vmax}$ (Eq.~\ref{eq:n_lim_vmax}) passes below this limit at $n_{\rm peak} > 3\times 10^4$, meaning that for generic galaxy populations, one would want at least this many particles for resolved rotation curves. No improvements in subhalo finder techniques would reduce either limit because they come from a combination of the galaxy disruption model and numerics of the simulation. Note that, because this is the most conservative of the limits considered here, it may be possible that for {\em specific} galaxy populations less resolution is required. This would need to be calibrated explicitly.

Although the bands shown in Fig.~\ref{fig:survival_vmax} are likely qualitatively correct, more work needs to be done to understand their exact ranges. The best-fitting values in \citet{behroozi_2019_universemachine} are derived relative to the $v_{\rm max}$ values predicted by \textsc{UniverseMachine}'s post-disruption orphan model and are not directly simulated. Any biases in the predictions of this orphan model could lead to biases in the best-fitting disruption thresholds. \citet{smith_2016_preferential} calibrated their fit on a set of Milky Way-mass subhalos, meaning that the corresponding band does not account for any mass-dependence of disruption rates. Some level of mass dependence is expected because Milky Way-mass subhalos are more heavily star dominated than more/less massive subhalos \citep[e.g.,][]{wechsler_tinker_2018_connection}. The \citet{smith_2016_preferential} fits were only performed on a single hydrodynamical model \citep{dubois_2014_horizon_agn} using the \textsc{AdaptaHOP} subhalo finder \citep{aubert_2004_adaptahop} and did not account for survivor bias. Any systematic uncertainties associated with the hydrodynamic scheme or halo finder could propagate into biases in the fitted model.

Additionally, both studies were calibrated primarily on high-mass satellite galaxies, and it is unclear how strongly these disruption thresholds depend on galaxy mass. \citet{smith_2016_preferential} focused on Milky Way-mass subhalos, and \textsc{UniverseMachine} only included orphan galaxies for which $v_{\rm max}$ > 80 km s$^{-1}$ during the snapshot prior to disruption \citep{behroozi_2019_universemachine}. {\em A priori}, one would expect that galaxies with more self-gravity from stars 
would require more subhalo mass loss prior to disruption. If true, this would mean that subhalo finders capable of tracking the subhalos near the peak of the $m_{\star}/m$ relation would be even more capable of 
tracking the subhalos that host low-mass galaxies, all other galaxy properties being held equal. but other galaxy properties --- such as stellar radius --- also impact the ease of disruption, so this requires additional study.

\subsection{Error rates}
\label{sec:stitching_errors}
\begin{figure*}
\hspace{-3.5mm}
\includegraphics[width=0.475\textwidth]{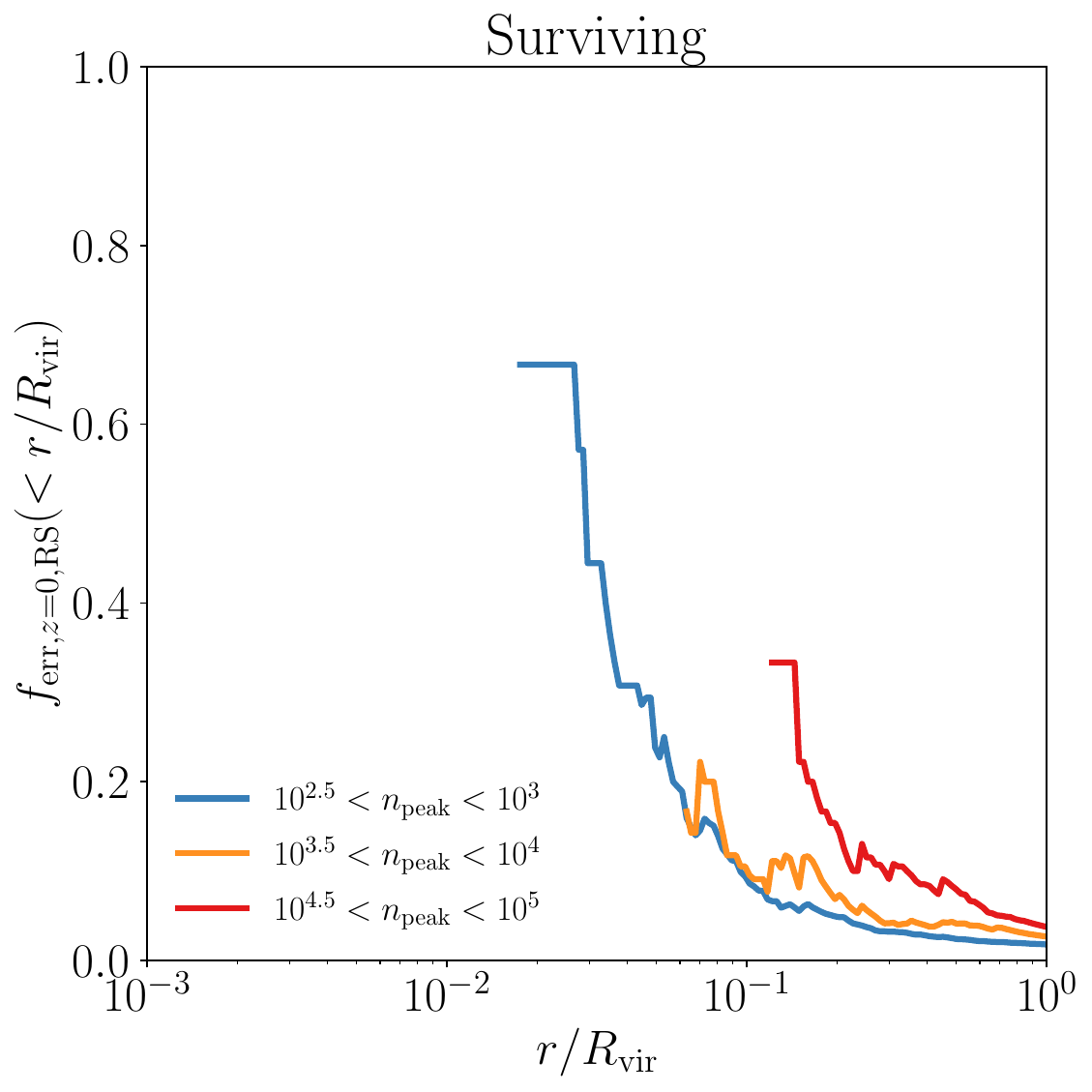}
\includegraphics[width=0.475\textwidth]{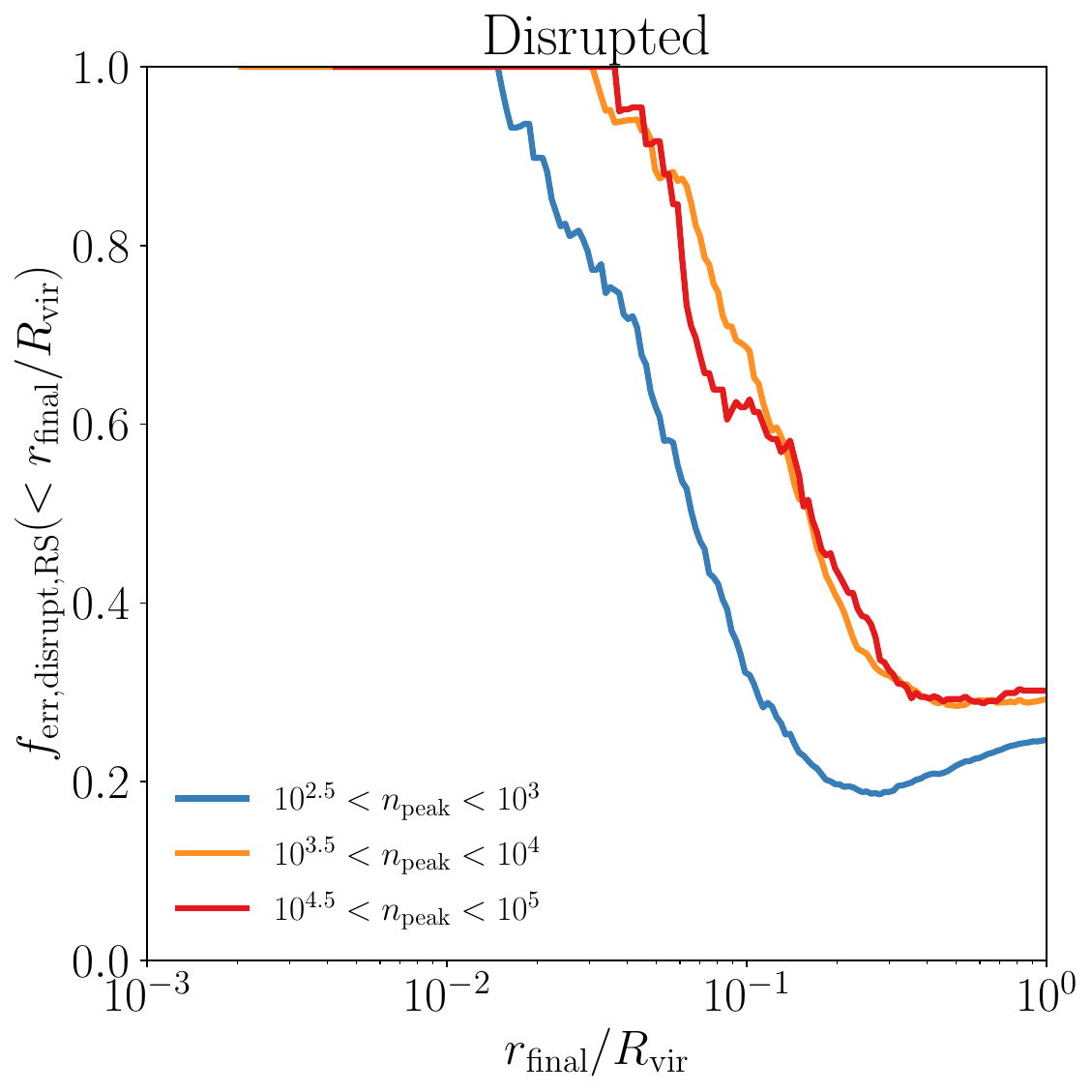}
\caption{The fraction of disrupted \textsc{Rockstar} subhalos that experienced an error where they were associated with an incorrect density peak. All the Symphony suites have been combined to improve number statistics. The left panel shows the error rate at $z=0$ as a function of radius among {\em surviving} subhalos, and the right panel shows the error rate among {\em disrupted} subhalos as a function of the final radius of the subhalo. This type of error is shown in orange in Fig.~\ref{fig:example_halo}. These errors can impact the $m_{\rm peak}$ value of a subhalo and can frustrate attempts to create ``orphan'' subhalos using the subhalo's last intact snapshot. A sizable fraction of all disrupted subhalos have experienced such an error. Errors are more common in disrupted subhalos than in surviving subhalos and become more likely at smaller radii. The majority of subhalos that disrupt at very small radii are errors.
}
\label{fig:rockstar_error_rate}
\end{figure*}

Beyond surviving to the masses needed to resolve galaxy disruption, a subhalo finder/merger tree must not attach a subhalo branch to incorrect density peaks and must continue to track the same subhalo through time.

As shown in Fig.~\ref{fig:example_halo}, some RCT subhalos experience an error during their final snapshot where incorrect positions and masses are assigned to the subhalo. To estimate the frequency of these errors, we follow Appendix \ref{sec:parameter_selection} and compare the locations of the subhalos found with RCT and with \textsc{Symfind} against one another and with the locations of the 32 most-bound ``core'' particles from the snapshot of first infall. If the spheres formed by the half-mass radii of the two subhalos do not intersect {\em and} the \textsc{Symfind} subhalo has more core particles within its half-mass radius than RCT does, we consider RCT to have encountered an error. All \textsc{Symfind} subhalos that fail the converse check are considered to be disrupted anyway, so a similar test cannot be performed in reverse. Regardless of this, RCT subhalos rarely outsurvive \textsc{Symfind} subhalos, making it difficult to obtain a statistical sample (see Section \ref{sec:survial}), and it is rare that \textsc{Symfind} subhalos are errors relative to RCT subhalos (see Appendix \ref{sec:parameter_selection}).

The fraction of surviving, $z=0$ RCT subhalos that encounter errors as a function of distance from their hosts' centers is shown in the left panel of Fig.~\ref{fig:rockstar_error_rate}. The same is shown in the right panel for subhalos that have disrupted before the end of the simulation. In the right panel, the final distance between the subhalo and its host is shown on the $x$-axis instead of the current-day distance. Averaged across the entire subhalo sample, errors are rare at any specific snapshot, staying at-or-below the 5\% level for all $z=0$ subhalos within $R_{\rm vir}.$ The error rate becomes increasingly relevant for subhalos at $r/R_{\rm vir} \lesssim 1/20.$ The error rate for disrupted subhalos is much higher: $\approx 25\%{-}30\%$ of all subhalos experience an error in their final snapshots. {\em Most} subhalos that disrupt at small radii ($r/R_{\rm vir} \lesssim 1/10$) experience such an error, even at high resolutions.

Counter-intuitively, as subhalo resolution increases, the RCT error rate also increases. There are two potential explanations for this. First, as seen in Fig.~\ref{fig:survival} and Fig.~\ref{fig:survival_vmax}, RCT disruption thresholds slightly increase with increasing resolution, meaning that it is possible that RCT performs slightly better at lower resolutions than it does at higher resolutions. Second, with our methodology, errors in RCT can only be caught if \textsc{Symfind} subhalos outsurvive their RCT counterparts. As shown in Fig.~\ref{fig:survival}, \textsc{Symfind} can track high-resolution subhalos to much lower masses than low-resolution subhalos, meaning that errors will be more easily caught in this regime.

\refadd{Throughout this paper, we compute $m_{\rm peak}$ using only pre-infall, in contrast to the standard method of computing this quantity, denoted here as $m_{\rm peak,0}$, which allows peak mass to be reached after infall. As seen in Fig.~\ref{fig:example_halo} and Fig.~\ref{fig:additional_halos}, $m_{\rm peak,0}$ can be achieved during this class of error, meaning that a subhalo population selected at a fixed $m_{\rm peak}$ will actually be a mixture of the target population and smaller subhalos that have erroneously scattered into the bin. This is not too large of an effect when considering the entire subhalo population, though: using $m_{\rm peak,0}$ to select samples instead of $m_{\rm peak}$ would only have increased the size of our samples by $\approx$ 5\%. There are some indications (Wan et al., in prep) that this effect becomes much more significant for subhalo populations selected at small radii, but we defer an in-depth discussion of this to future work.}

More importantly, existence of these errors is highly problematic for orphan models. The vast majority of these models create orphans at the last moment that the input halo catalog can track the subhalo (e.g., \citealp{moster_2013_galactic,behroozi_2013_consistent}; see \citealp{pujol_2017_nifty} for a wider review). This means that 20\% to 30\% of all orphan galaxies are initialized with incorrect positions, velocities, and masses, leading to similar errors in the inferred properties of galaxy populations. Nearly the entire low-radius orphan population will be generated with incorrect properties.

\subsection{Is Symfind ``good enough?''}
\label{sec:are_we_good_enough}

For the purposes of finding subhalos that are likely to still host satellite galaxies, yes. Fig.~\ref{fig:survival_vmax} shows that according to current galaxy disruption models, improving subhalo finding techniques would be unlikely to improve estimates of the abundance of resolved subhalos that are likely to still host satellite galaxies.

At low particle counts, there are some galaxy-hosting subhalos with low $m/m_{\rm peak}$ that our method cannot find (12\% and 18\% of $n_{\rm peak} = 300$ subhalos would be lost from our catalogs before meeting the more conservative of the \cite{smith_2016_preferential} and \cite{behroozi_2019_universemachine} lower limits, respectively). However, all such subhalos are so low-resolution that their mass loss rates and $v_{\rm max}$ values are not resolved, meaning that they are not reliable analysis targets for abundance studies, regardless of whether they can be found.

At high particle counts, some subhalos with very low $m/m_{\rm peak}$ are resolved but cannot be found with our method. However, the $m/m_{\rm peak}$ for these subhalos is so low that current models of galaxy disruption predict that the galaxies inside these subhalos should have either disrupted or have lost a substantial amount of mass, meaning that they would either not contribute to observed satellite galaxy abundances or contribute very weakly (see point 2 below).

In Fig.~\ref{fig:shmf} and Fig.~\ref{fig:radius_cdf}, we showed that our method finds increasingly large subhalo abundances at a fixed $m_{\rm peak}/M_{\rm vir}$ and increasingly concentrated subhalo radial number density distributions as the resolution is increased, respectively. These increases occurred across all subhalo masses. There is no conflict between these Figures and our conclusion in this Section. These Figures include all subhalos regardless of $m/m_{\rm peak}$, including subhalos that are unlikely to still host galaxies. Our testing shows that instituting cuts on $m/m_{\rm peak}$ that mimic the galaxy disruption models shown in Fig.~\ref{fig:survival_vmax} results in population statistics that are converged above $n_{\rm peak} \gtrsim 10^{3.5}$, but we defer quantitative analysis of this point until future work (see point {\em (ii)} below).

There are several caveats to this conclusion that we outline below.
\begin{enumerate}
\item This analysis has been performed on $\Lambda$CDM subhalos and not on subhalos run in hydrodynamic simulations or in alternative cosmologies. As discussed in Section \ref{sec:intro}, both conditions can lead to decreased central densities that can in turn lead to true disruption for low-mass subhalos for certain models. In this case, the condition for being ``good enough'' for satellite galaxy observations transforms from being able to follow all resolved, galaxy-hosting subhalos to being able to follow all resolved, galaxy-hosting, and undisrupted subhalos. This would likely relax the requirements for subhalo finders unless the changes in internal density also relaxed resolution requirements by, e.g., increasing the typical relaxation timescale. It is also possible that \textsc{Symfind}, specifically, could encounter trouble in cosmologies or hydrodynamics implementations that cause physical diffusion of highly bound particles to large radii. As such, domain-specific testing should be done before applying our method to non-$\Lambda$CDM simulations.
\item Although it is common to discuss galaxy ``disruption,'' as a discrete event, galaxy mass loss is continuous \citep{smith_2016_preferential} and it may be possible that some asymptotic stellar remnant exists even after substantial halo mass loss. This is why we have not made any predictions for the properties of satellite galaxy populations in this paper and defer such analysis to future work. That said, the disruption approximation is not a poor one and is sufficient for the purposes of this paper: galaxies generally maintain their mass until some critical point after which mass loss is rapid \citep{smith_2016_preferential}. Once this rapid mass loss begins, the disrupting galaxy is pushed into successively smaller stellar mass regimes. The steepness of the stellar mass function means that the fraction of galaxies at a fixed stellar mass that are heavily disrupted high-mass subhalos should rapidly decrease with decreasing $m_*/m_{*,{\rm infall}}$. 
\item Some observational probes of subhalos are sensitive to present-day subhalo masses and not the stellar masses of their satellite galaxies. {\em A priori} arguments \refadd{(see Section \ref{sec:good_enough})} suggest that these probes should have more lenient requirements for subhalo finders: at a fixed subhalo mass, the steepness of the subhalo $m_{\rm peak}$ function means that most subhalos will have large $m/m_{\rm peak}$ ratios and will thus be more easily tracked. Resolution requirements should also be more lenient and more easily applied, requiring only that the target population of subhalos instantaneously exceeds Eq.~\ref{eq:n_lim_vmax} or Eq.~\ref{eq:n_lim_mass}. That said, gravitational probes of mass can be very sensitive to the mass definitions adopted by halo finders \citep[e.g.,][]{mao_2018_resolution_discrepancy}, a problem that we do not address in this paper. Dedicated analysis is required to resolve this issue quantitatively.
\item As discussed in Section \ref{sec:galaxy_disruption}, there are a number of limitations in the galaxy disruption models used here. In particular, these limits were derived for high-mass galaxies, which likely survive for longer than low-mass galaxies due to significant self-gravity and adiabatic contraction. Addressing these limitations could potentially raise or lower the typical masses at which rapid galaxy mass loss begins.
\end{enumerate}

\subsection{Recommendations for subhalo finders and orphan models}
\label{sec:recommendations}

We make no claim that our method is the only existing subhalo finder that is ``good enough'' to model satellite galaxy abundances. It is certainly possible that some existing methods are able to track subhalos for similarly long periods of time, and we identify several candidate methods that seem likely to be able to do this in Section \ref{sec:method_comp}. However, we believe it is important for this to be demonstrated explicitly and for subhalo finders to quantify their reliability limits. This will allow users to make strict, quantitative statements about how large of an impact halo finder biases have on their analysis.

We encourage interested authors to do all of the following:
\begin{itemize}
\item Using the methodology described in Section \ref{sec:survial}, construct disruption threshold curves and compare them against numerical limits and galaxy disruption limits, as in Fig.~\ref{fig:survival_vmax}.
\item Recreate Fig.~\ref{fig:vacc_macc} and confirm that the $v_{\rm max}/v_{\rm max,infall}$ versus $m/m_{\rm infall}$ relation converges correctly at high particle counts. A subhalo-finding method that incorrectly loses mass in the outskirts of subhalos could lead to \refadd{subhalos which appear to disrupt at incorrectly low $m/m_{\rm peak}$ values and could thus lead to seemingly-low disruption thresholds which would overestimate how long subhalos stay intact. This would appear as curves that are higher than the expectations of idealized models in the  $v_{\rm max}/v_{\rm max,infall}$ -- $m/m_{\rm infall}$ plot.} While unfortunate, such an error would not be fatal: one could alternatively characterize survival curves in terms of $v_{\rm max}/v_{\rm peak}.$
\item Recreate Fig.~\ref{fig:mass_loss_conv} and confirm that $m/m_{\rm peak}$ is converged out to similar or larger mass ratios. A subhalo finder that experiences sudden drops in mass during its final snapshot could lead to overly optimistic disruption thresholds. Such a scenario would manifest as earlier non-convergence in $m/m_{\rm peak}$. This would not be a fatal problem either: it would merely raise the amplitude of the dashed curve in Fig.~\ref{fig:survival_vmax} for such a finder.
\item Include a technique that can identify and remove subhalos that were erroneously connected to a branch. We find that using the locations of particles that were highly bound at infall works well, although we also experimented with identifying sharp jumps in masses and positions, and this seems to also work fairly well as long as it is carefully calibrated against manually inspected property evolution tracks.
\end{itemize}
One major benefit of these tests compared to, say, comparing the subhalo mass functions measured by two separate subhalo finders is that the correct behavior of low mass subhalos in $\Lambda$CDM is known for all four tests: small subhalos that have not yet sunk to the centers of their hosts should survive indefinitely \citep[e.g.,][]{vandenbosch_2018_disruption,errani_navarro_2021_asymptotic}, $v_{\rm max}$ should evolve in a predictable manner with decreasing $m$ \citep{green_vdb_2019_tidal}, mass loss rates should agree with higher resolution mass-loss rates that are known to be converged, and highly bound particles should stay localized within the subhalo.

Some of this suggested analysis is non-trivial, so code implementing the first three tests will be made available to readers upon request. The fourth test must be implemented directly within one's subhalo finder\refadd{/merger tree codes}. Two simpler, but less conclusive tests would be to construct resolution-dependent subhalo $m_{\rm peak}$ functions (Section \ref{sec:shmf}) and radial distributions\footnote{As discussed in Section \ref{sec:radius_cdf}, the mass-dependent radial distribution of subhalos can be a good but not perfect proxy for the resolution-dependent radial distribution of subhalos at low subhalo masses.} (Section \ref{sec:radius_cdf}) selected by $m_{\rm peak}$ and not by present-day subhalo mass. Subhalo finders that do not falsely converge should find that the amplitude of the subhalo $m_{\rm peak}$ function increases with increasing resolution and that radial distributions selected by $m_{\rm peak}$ become more concentrated with increasing resolution.

Our analysis also has some implications for the creators and users of ``orphan'' models.

\begin{itemize}
\item Some subhalo finder+merger tree combinations can return subhalo branches that out-survive the regime where subhalos are numerically reliable. In these cases, it is possible that a well-calibrated orphan model may result in more accurate predictions than a subhalo finder (although this would need to be explicitly demonstrated). One should consider generating orphan galaxies at this numerical limit rather than at disruption.
\item A subhalo's final snapshot --- the time when many orphan models generate their mock subhalos tracers --- is the time when its properties are characterized the least accurately. Orphan methods would benefit from a methodology that can identify errors in the underlying subhalo catalogs and from generating orphan galaxies significantly before the final snapshot.
\end{itemize}

Finally, because of the existence of post-infall errors that can cause subhalo masses to spike, we recommend that all authors, not just authors of subhalo finders, who are interested in the peak mass of subhalos define that peak mass using only the mass accretion history of subhalos prior to their first infall onto a host, as we do in this paper. Doing so requires special post-processing of merger trees  to identify times when central halos are erroneously and temporarily classified as subhalos (see Section \ref{sec:tree_post_processing}).

\section{Comparison with other subhalo finding methods}
\label{sec:method_comp}

\subsection{Single-epoch subhalo finders}
\label{sec:single_epoch}

The most common subhalo-finding strategy is to identify objects in each snapshot of a simulation individually and then run a separate merger tree code that connects subhalos across snapshots, usually relying heavily on the IDs of particles contained by halos in each snapshot. This is the general structure used by RCT and dozens of other widely used tools \citep[e.g., most methods listed in][]{knebe_2011_MAD,srisawat_2013_sussing}. A detailed comparison of all these methods is beyond the scope of this paper, so we focus on a broader overview.

A number of papers have run a range of subhalo finders/merger trees on the same simulations and compared summary statistics to make inferences about the performance of subhalo finders \citep{knebe_2011_MAD,onions_2012_notts,onions_2013_notts_spin,srisawat_2013_sussing,avila_2014_sussing,behroozi_2015_major_mergers,elahi_2019_velociraptor}. All the papers listed above included comparisons with RCT (typically one of the two components of the pipeline, \textsc{Rockstar} or \textsc{consistent-trees} independently), and one might initially hope that this would allow for a sort of transitive comparison with our RCT tests in this paper.

But we do not believe such an extension of our results is warranted.

\citet{knebe_2011_MAD} found that most halo finders predict qualitatively similar mass/velocity functions, large-scale correlation functions, and bulk velocity PDFs for central halos, but did not compare subhalo statistics. \citet{onions_2012_notts} found that most of their tested subhalo finders recovered modestly fewer subhalos at a fixed instantaneous $m_{\rm 200c}$ than \textsc{Rockstar}, and that \textsc{Rockstar} recovered slightly more subhalos at small radii than most other subhalo finders. \citet{onions_2013_notts_spin} focused on subhalo spin, a quantity that we do not consider in this paper. \cite{srisawat_2013_sussing} found that many statistical summaries of the properties of merger trees are similar between \textsc{consistent-trees} and many other popular merger tree codes (with the notable exception of \textsc{HBT}, see Section \ref{sec:other_particle_tracking}). They also identify several merger tree features that reduce the chances of errors when present. All these features are present in \textsc{consistent-trees}. \citet{avila_2014_sussing} found that \textsc{consistent-trees} is able to correct for some errors that other merger trees miss, but generally concluded that subhalo finders had more influence on the evolution of subhalo properties than tree codes did and that the different tree codes broadly agreed with one another. \citet{behroozi_2015_major_mergers} focused on major mergers, finding that many subhalo finders struggle to correctly follow major mergers, and ultimately recommend methods based on particle tracking that are careful to remove objects that have truly sunk to the centers of their hosts (e.g., condition 2 in Appendix \ref{sec:subhalo_disruption}). \citet{elahi_2019_velociraptor} showed that at a fixed instantaneous subhalo mass, \textsc{Rockstar} finds roughly the same number of subhalos as several other subhalo finders and finds roughly the same radial distribution of subhalos at a fixed instantaneous mass. 

This literature is often summarized as meaning that RCT performs at least as well as most other subhalo finders. If this reading is correct, it would in turn imply that the majority of subhalo finders suffer from problems similar to RCT. In particular, RCT's radial distribution of subhalos has converged to a false solution (Fig.~\ref{fig:radius_cdf}) as has its SHMF at different radii (Fig.~\ref{fig:shmf_converge}), so one might expect that agreement between RCT and other halo finders on quantities related to the radial profile would imply the same for these finders. While such a pessimistic scenario is certainly possible, the tests described above are actually not very sensitive to subhalo disruption issues. Almost all of these tests compare subhalo populations {\em at a fixed instantaneous mass} (at least in part to try to isolate issues in merger trees from issues in subhalo finders) and such selections can be misleading. As discussed in Section \ref{sec:radius_cdf}, it is well known that subhalo profiles at fixed instantaneous masses converge quickly to a profile that is much less concentrated than the dark matter halo itself, while subhalos selected by $m_{\rm peak}$ have more complicated convergence behavior and result in radial profiles more in line with satellite galaxy observations \citep[e.g.,][]{nagai_kravtsov_2005_radial}. Many authors have also focused on the radial distributions of all the subhalos output by their subhalo finders, meaning that the signal is dominated by the most poorly resolved and error-prone objects. To further underscore the insensitivity of these tests, many of the aforementioned papers compared \textsc{Subfind} and \textsc{Rockstar}, but as far as we can tell, the fact that these two subhalo finders have wildly different convergence behavior (Fig.~\ref{fig:radius_cdf} and Fig.~\ref{fig:subfind_radius_cdf}) was not reported. In general, the fact that subhalos can lose such a large amount of mass prior to the disruption of their satellite galaxy means that the properties of instantaneous-mass samples are often not predictive of galaxy behavior.

To further complicate matters, while some of the aforementioned studies performed idealized tests where the difference in mass definitions between subhalo finders could be compared, tests performed on full cosmological simulations generally did not match subhalos to one another across subhalo finders. This meant that a single shared mass definition could not be used across all the subhalo finders, leading to some additional ambiguity in interpreting many of these tests: when a sample of subhalos is created across multiple subhalo finders, is the same set of subhalos being selected? When considering two different SHMFs, does the difference arise because one finder is identifying more subhalos than the other, or does it arise because they are finding the same subhalos but assigning different masses to them? This issue is avoided in our study: particle-tracking and RCT subhalos are explicitly matched against one another, and $m_{\rm peak}$ is defined using only the pre-infall mass accretion histories computed by RCT. The exact same samples of subhalos are compared in all of our tests and $m_{\rm peak}$ is the same value regardless of the subhalo finder.

We recommend that authors interested in investigating disruption-related issues in other subhalo finders follow the testing procedure that we lay out in Section \ref{sec:recommendations}, which focuses on survival thresholds and multi-resolution tests and is very sensitive to subhalo disruption issues. This procedure also has the benefit that one knows what the theoretical predictions of $\Lambda$CDM are prior to running the tests. One does not know {\em a priori} how $\Lambda$CDM predicts the subhalo mass function or radial distribution of subhalos should behave, meaning that even if one detects a difference between subhalo finders using these statistics, it is hard to interpret.

\subsubsection{The Caterpillar variant of Rockstar}
\label{sec:caterpillar}

As part of the Caterpillar zoom-in simulation suite, \citet{griffen_2016_caterpillar} created a variant of \textsc{Rockstar} after noticing that a large number of visually apparent, high-resolution, low-radius subhalos were missing from \textsc{Rockstar} catalogs. This variant changes the criteria used internally by \textsc{Rockstar} to be less aggressive about removing partially unbound phase-space overdensities. This change leads to the creation of a large number of spurious subhalos, so \citet{griffen_2016_caterpillar} also changed \textsc{Rockstar}'s single unbinding pass to a full iterative unbinding and introduced a criterion that removed subhalos very close to the centers of their hosts. This change modestly increases the number of subhalos at a fixed instantaneous mass \citep{griffen_2016_caterpillar} and leads to profile resolution dependence that is qualitatively similar to Fig.~\ref{fig:radius_cdf} \citep{manwadkar_kravtsov_2022_grumpy}. So it is very likely that this variant has better convergence behavior than standard \textsc{Rockstar}, and that \textsc{Rockstar}'s false convergence could be related to its procedure for deciding which phase-space overdensities are truly subhalos. It would be interesting to quantify the behavior of this variant with the tests described in Section \ref{sec:recommendations}.

\subsection{Other particle-tracking methods}
\label{sec:other_particle_tracking}

Tracking particles to find bound substructure at later times is an old concept and was used in a number of early subhalo studies \citep[e.g.,][]{tormen_1998_survival}. Two recent, cutting-edge software packages within this tradition are HBT+ \citep{han_2018_hbt_plus}, a successor to HBT (\citealt{han_2012_hbt}; see also \citealt{springel_2021_gadget_4}), and the particle-tracking model recently added to the post-processing framework for \textsc{Sparta} \citep{diemer_2017_sparta_i,diemer_2023_haunted}.

All methods of this class have the same basic structure: some set of membership rules are used to associate particles with subhalos prior to infall, and then a second method is used to find the centers and masses of those subhalos later in time using all the tracked particles. HBT+ uses iterative unbinding to find subhalo masses and calculates subhalo properties from the bound remnant. Excessively unbound particles are removed with each snapshot, but a buffer of modestly unbound particles is kept within the tracking set to prevent runaway stochastic mass loss and to allow for some mass flow into and out of the subhalo. A variant of HBT+'s algorithm has also been implemented in \textsc{Gadget-4} as \textsc{Subfind-HBT} \citep{springel_2021_gadget_4}, although it is safest to treat this as a separate algorithm, as some design decisions are different from HBT and HBT+.

\textsc{Sparta} uses a similar technique but does not explicitly calculate boundedness. Instead, it measures masses from raw overdensity radii and keeps a buffer of higher-radii particles. It is unlikely that an unbound subhalo would be able to maintain a well-defined overdensity radius, and the tests in \citet{diemer_2023_haunted} suggest that for the majority of subhalos, bound masses are fairly close to these overdensity masses. More work needs to be done to understand how these masses compare for different subhalo populations (e.g., highly stripped, low-radius subhalos with large tidal tails). If a comparison with the methodology outlined in Section \ref{sec:recommendations} is to be done, it is also important to better understand the connection between these overdensity masses and bound masses close to the moment of disruption. 

Our method deviates a bit from this pattern. We continue tracking all particles regardless of previous status, identify substructure within the tracked particles using a traditional subhalo finder (in our case, \textsc{Subfind}), and use particles that were highly bound at infall to select the correct density peak to use as the subhalo's center. Once that center is located, halo properties are calculated independently of the intermediate subhalo finder. There are some small trade-offs associated with this choice (our method can recover from temporary errors easily, but also needs to be more careful about accidentally identifying matter that has been stripped and tidally mixed as the true subhalo center), but we don't believe there is any {\em a priori} reason to expect that any one of these particle-tracking algorithms significantly out-performs the others based purely on the high-level description of these algorithm choices.

We consider it plausible that these other particle-tracking-based subhalo finders would also pass the criteria outlined in Section \ref{sec:good_enough}, but this has not yet been explicitly demonstrated (see Section \ref{sec:recommendations}). We list some assorted notes on the performance of these methods below:
\begin{itemize}
\item The original HBT code has been compared against other subhalo finders in a number of previous subhalo finder comparison projects (see overview in Section \ref{sec:single_epoch}). As with RCT, a tempting way to read these tests is that HBT performs comparably or better than most other halo finders on these tests, but, as discussed in Section \ref{sec:single_epoch}, the tests performed in many of these papers are not very sensitive to the types of disruption issues discussed in this paper and missed substantial differences between \textsc{Subfind} and \textsc{Rockstar}, meaning that it is difficult to draw strong conclusions about the disruption behavior of HBT from these tests alone. Interestingly, \citet{srisawat_2013_sussing} found that positions of HBT subhalos are substantially more consistent with their previous trajectories than essentially all other tested merger tree codes. It is hard to interpret this result as anything other than a success for HBT.
The improved HBT+ was written after many of these tests were performed, but \citet{han_2018_hbt_plus} compared HBT+ against the \textsc{Subfind} halo finder and the \textsc{DTree} merger tree code \citep{jiang_2014_dtree}. They found that HBT+ recovered somewhat more ($\approx 10\%$) low-mass subhalos and 200\% to 300\% more high-mass ($m/M_{\rm vir}\gtrsim 0.2$) subhalos at a fixed {\em instantaneous} mass than \textsc{Subfind}. HBT+ also leads to more concentrated subhalo profiles at fixed instantaneous mass. However, this work also finds that \textsc{Subfind}+\textsc{DTree} has a large {\em unevolved} subhalo mass function (i.e. the $m_{\rm peak}$ mass function combining both surviving and disrupting subhalos). \citet{han_2018_hbt_plus} convincingly argue that these differences are due to a combination of \textsc{Subfind} missing mass in the outskirts of large subhalos \citep{vdb_2016_staistics_ii} and the fragmentation of \textsc{Subfind}+\textsc{DTree} branches.
\item Tests in \citet{springel_2021_gadget_4} seem to indicate that \textsc{Subfind-HBT} and \textsc{Subfind} subhalos tend to disrupt at qualitatively similar times (Fig.~36 in \citealp{springel_2021_gadget_4}), that the two methods lead to similar subhalo mass functions (Fig.~38 in \citealp{springel_2021_gadget_4}), and that the typical low-to-moderate resolution subhalo found by RCT has a longer main branch than a similar subhalo found by \textsc{Subfind-HBT} (Fig.~40 in \citealp{springel_2021_gadget_4}) while the opposite is true at higher resolutions. This may be evidence that \textsc{Subfind-HBT} performs similarly to traditional single-epoch subhalo finders, or it may indicate that this set of tests is not very sensitive to subhalo finder performance issues.
\item \citet{diemer_2023_haunted} demonstrated that the subhalos found by \textsc{Sparta} substantially out-survive RCT subhalos, and showed that the difference between RCT subhalo mass functions and \textsc{Sparta} subhalo mass functions is qualitatively similar to the difference between RCT and our method (Fig.~\ref{fig:shmf}).
\end{itemize}

One problem that all particle-tracking methods suffer from is detecting when a subhalo has sunken into the center of its host. Even after completing such a merger, the subhalo can remain fully self-bound until the end of the simulation. In our method, any subhalo that is within its own half-mass radius of the host's center and that never leaves the center of the host after this point is considered disrupted. \textsc{Sparta} counts objects that have been within 0.05 $R_{\rm 200m,host}$ of the host's center for more than half a dynamical time as merged, with a correction factor to account for Poisson noise \citep{diemer_2023_haunted}. HBT+ removes subhalos that are too close to the center of the host in phase space, with the distance metric set by the position and velocity dispersion of the 20 most-bound host particles \citep{han_2018_hbt_plus}. This metric was motivated by the fact that merged subhalos appear as a distinct population in this space. All these methods are effective at removing the vast majority of merged subhalos, although it is possible that studies focused on very low-radius subhalos or on tracking the end states of major mergers could benefit from revisiting them carefully and performing detailed comparisons.

\subsection{``Orphan'' subhalo models}
\label{sec:orphan}

Even older than particle-tracking methods or single-epoch halo finders are ``orphan'' subhalo models \citep[e.g.,][]{white_1987_galaxy_distribution}. In these models, a subhalo is followed until the point of disruption (or accretion), and then one of several methods is used to estimate the location of the subhalo's center thereafter \cite[see review in][]{pujol_2017_nifty}. Most commonly, the most-bound particle is used \citep[e.g.,][]{white_1987_galaxy_distribution,wang_2006_clustering}, but some authors use many highly bound particles instead \citep[e.g.,][]{Carlberg_Dubinski_1991_cluster_infall,korytov_2023_cores}. Some models eschew particles altogether and estimate the location of a test particle orbiting through an idealized potential \citep[e.g.,][]{somerville_2008_sam,behroozi_2019_universemachine}.

There are two defining differences between a catalog of orphan subhalos and a catalog of subhalos found by a true subhalo finder. The first is that the properties of orphan subhalos are inferred from some post-processing model rather than being directly taken from a simulation. The positions and velocities of the orphan subhalo are the most obvious example of this, but the modeling can extend to other properties as well. For example, many authors have measured subhalo mass loss rates for some set of subhalos that a traditional halo finder is able to identify and then use those mass loss rates to compute $m(t)$ \citep[e.g.,][]{jiang_2016_statistics,behroozi_2019_universemachine,sultan_2021_cores}. Finally, orphan models must have a method for disrupting the subhalo, either by removing it once it gets too close to the center of its host \citep[e.g.,][]{white_1987_galaxy_distribution}, through monitoring the internal state of the tracked particles \citep{korytov_2023_cores}, or by removing the orphan once the modeled $m(t)$ crosses some limiting value  \citep{somerville_2008_sam,reddick_2013_connection, behroozi_2019_universemachine}.

The introduction of an additional modeling layer to infer subhalo properties allows for additional free parameters and researcher degrees of freedom, introduces systematic uncertainty into predictions, and opens up the question of whether any incorrect predictions of the orphan model are a failure on the part of $\Lambda$CDM, the galaxy formation model used, or the decisions that went into tracking orphan subhalos. This is concerning because most orphan models predict that $\approx 20\%$ to 30\% of all low-mass satellites need to be represented by orphans \citep{pujol_2017_nifty}. There is some controversy over whether orphan modeling is needed to reproduce the radial distributions of satellite galaxies, as we discuss in Section \ref{sec:radius_cdf}. Given how sensitive radial distributions are to the disruption physics of satellites, the fact that it is unclear whether orphan modeling is even needed for this class of observation further underscores how much systematic uncertainty there is in this regime.

One great strength of orphan models is that they are able to make predictions for subhalo populations that the simulation is not able to resolve. In contrast, a subhalo finder can only report the results of the simulation it is run on, resolution errors and all. Nonetheless, many common approaches used for orphan models are problematic. Many orphan models generate orphans based on the properties of the corresponding subhalo at the snapshot that subhalo was last found in the catalog, but as shown in Fig.~\ref{fig:rockstar_error_rate}, a large fraction of RCT subhalos are actually errors during their final snapshot, especially at small radii. This would lead to incorrect orphan subhalo properties. Many orphan models use the most-bound particle to model the location of the subhalo center, but as discussed in Appendix \ref{sec:parameter_selection}, in a non-trivial fraction of subhalos the most-bound particle can numerically diffuse out of the center before the subhalo actually disrupts. Population-averaged mass loss models do a good job at reproducing average $m$ functions, but do a poor job of matching the individual properties of subhalos due to the dependence of subhalo mass loss rates on orbital parameters, the scatter in individual mass loss rates, and possible evolution in subhalo mass loss rates over time (see discussion in \citealp{jiang_2016_statistics} as well as Fig.~7 of \citealp{sultan_2021_cores}). This becomes concerning when comparing sub-populations of halos with different orbital parameter distributions such as one does, e.g., when constructing a number density profile.

In light of these limitations, the orphan modeling techniques described in \citet{heitmann_2021_last_journey,sultan_2021_cores,korytov_2023_cores} are very promising. This method uses the $N$ most-bound ``core'' particles (20 in \citealp{korytov_2023_cores}) at the moment of infall and uses a density estimate to identify a central particle, which is then taken as the orphan's position and velocity. Starting the orphan tracking at the moment of infall avoids the issue of subhalo finder errors during the final snapshot, and using many particles lessens the impact of numerical diffusion. Additionally, this method scales well to $\gtrsim10^{12}$ particle simulations, a feat that few sophisticated subhalo identification/modeling techniques can boast. However, the use of multiple particles encounters its own difficulties, particularly if properties of those particles are used to estimate disruption times (e.g., the effective radius of the core particles, as explored in \citealp{korytov_2023_cores}). As illustrated in Fig.~\ref{fig:conv_limits}, the most-bound particles used for core-tracking are usually unconverged before infall even starts (the halo in this Figure has $8\times 10^4$ particles at infall, and the 800 particles closest to the center of the subhalo have passed their individual relaxation times; this is typical for a subhalo at this resolution level). Thus, tying the evolution of subhalo properties to the evolution of core particles --- beyond just using them to estimate a location in phase space --- may not avoid the convergence issues that orphan models are intended to avoid. Our method also makes use of the same unconverged core particles, but only to select between density peaks; in this method, it is fine if they numerically diffuse outwards, as long as at least some of them stay within the subhalo.

\section{Conclusions}
\label{sec:conclusions}

One of the major outstanding questions in the theory of $\Lambda$CDM \refadd{is what this model predicts as the true behaviour of subhalos}. These small orbiting objects are complex enough that effective modeling usually involves simulations, but are deceptively difficult to simulate and to identify once the simulation has been run. The uncertainty in how durable subhalos are places limitations on a number of cutting-edge cosmological probes, particularly those that depend on large-scale clustering of galaxies and the properties of satellite galaxies.

In this paper, we present a new method for identifying simulated subhalos (Section \ref{sec:methods}), with a particular focus on robustness tests. This method follows a subhalo's particles prior to infall and then uses an existing subhalo finder (currently \textsc{Subfind}; \citealp{springel_2001_populating}) along with the subhalo's most bound particles at infall to identify the subhalo within this tracked set of particles. 
\begin{itemize}
\item We perform extensive testing on the reliability limits of our method and on the popular \textsc{Rockstar} halo finder and find that our method substantially outperforms \textsc{Rockstar}, tracking subhalos to orders-of-magnitude lower masses (Section \ref{sec:survial}; Fig.~\ref{fig:survival}).
\item Our method recovers $15\%$ to $45\%$ more subhalos at a fixed $m_{\rm peak}$ than \textsc{Rockstar} (Section \ref{sec:shmf}; Fig.~\ref{fig:shmf}) in our fiducial simulations, particularly subhalos close to the centers of their hosts ($35\%$ to $120\%$ more subhalos within $R_{\rm vir}/4$), and the number of recovered subhalos increases substantially with increasing resolution.
\item Subhaloes found with our method are so long-lived that --- when combined with reasonable galaxy--halo models --- they do not require the use of ``orphan'' subhalos to follow subhalos until the point of likely galaxy disruption once a relatively modest resolution limit of $n_{\rm peak} > 4\times 10^3$ is met (Section \ref{sec:good_enough}; Fig.~\ref{fig:survival_vmax}). This limit is set by the resolution of the simulation, not by failures of the subhalo finder.
\item We also outline a concrete set of steps that can be used to determine whether other subhalo finders meet the same criteria (Section \ref{sec:recommendations}). We discuss caveats associated with this conclusion in Section \ref{sec:are_we_good_enough}.
\end{itemize}

The longer survival times of our subhalos allow us to quantitatively test some predictions of idealized simulations into the deep mass loss regime (Sections \ref{sec:numerical_limits}, \ref{sec:structural}, and \ref{sec:mass_loss_rates}).
\begin{itemize}
\item We find that these idealized simulations and the numerical limits derived from them do a good job of describing the $v_{\rm max}$ values of disrupting subhalos (Section \ref{sec:structural}; Fig.~\ref{fig:vacc_macc}).
\item These numerical criteria are quite restrictive and put significant limits on the ability of simulations to study the \refadd{internal} structure of subhalos ($n_{\rm peak} > 3\times 10^4$ for $v_{\rm max}$ to be resolved until the point of likely galaxy disruption; Fig.~\ref{fig:survival_vmax}).
\item Fortunately, subhalo mass loss rates --- and therefore abundances -- are resolved to much lower resolutions ($n_{\rm peak} > 4\times 10^3$ for $m(t)$ to be resolved until the point of likely galaxy disruption; Section \ref{sec:mass_loss_rates}; Figs.~\ref{fig:mass_loss_conv} and~\ref{fig:survival_vmax}).
\item As part of this testing, we demonstrate that simple techniques for estimating stacked mass loss rates suffer survivor bias and that this bias is strong enough to qualitatively change the shapes of mass loss curves (Fig.~\ref{fig:survivor_bias}; see also \citealp{han_2016_unified}). We lay out statistical techniques that avoid this problem (Section \ref{sec:mass_loss_rates}).
\end{itemize}

We also demonstrate that the \textsc{Rockstar} halo finder falsely converges with increasing resolution (Section \ref{sec:survial} and \ref{sec:radius_cdf}; Figs.~\ref{fig:survival} and~\ref{fig:radius_cdf}), giving the impression of numerical reliability to populations of subhalos that are not reliably tracked. We perform some limited analysis/discussion on the performance of other popular subhalo finding tools (Section \ref{sec:method_comp} and Appendix \ref{sec:subfind_radius_cdf}) but defer making strong statements to future work. We also identify several errors in \textsc{Rockstar}+\textsc{consistent-trees} merger trees and outline procedures for addressing them in Appendix \ref{sec:tree_post_processing}. The simplest of these corrections is that we advocate for measuring $m_{\rm peak}$ and $v_{\rm peak}$ for subhalos using only pre-infall masses because many subhalos experience large spikes in mass during disruption.

\refadd{All data used in this paper, along with analysis libraries and extensive documentation can be found at \url{http://web.stanford.edu/group/gfc/symphony/} \citep{symlib}.}

\section*{Acknowledgments}

We would like to thank Bryn\'e Hadnott for contributions to the website and tutorials associated with this paper and Tara Dacunha, Althea Hudson, and Azana Queen for helping to debug our analysis libraries. We also thank Benedikt Diemer, Keith Mansfield, Andrey Kravtsov, Peter Behroozi, Jiaxin Han, Frank van den Bosch, Jelle Aalbers, Tom Abel, Andrew Hearin, and Viraj Manwadkar for useful discussions and comments which helped improve the quality of this work. P.M. thanks Bella Shipp for support while working on this manuscript.

This work was supported by the Kavli Institute for Particle Astrophysics and Cosmology and by U.S. Department of Energy SLAC Contract DE-AC02-76SF00515. This work used data from the Symphony suite of simulations (\url{http://web.stanford.edu/group/gfc/symphony/}). 

This paper made heavy use of the \textsc{NumPy} \citep{NumPy}, \textsc{SciPy} \citep{SciPy}, and \textsc{matplotlib} \citep{Matplotlib} libraries.



\bibliographystyle{aasjournal}
\bibliography{main} 




\appendix

\section{Subhalo-Tracking Algorithm}
\label{sec:subhalo_algo_appendix}

Our subhalo-finding method tracks the particles that belonged to a subhalo prior to its first infall and uses only those particles to find the subhalo at later times.
\begin{enumerate}
\item We use an existing halo finder and merger tree code to track halos before they become subhalos. We perform post-processing to remove errors in the tree (Appendix \ref{sec:tree_post_processing}).
\item For each subhalo, we find all the particles that accreted onto the subhalo prior to infall, as well as the most-bound particles at infall (Appendix \ref{sec:particle_associations}).
\item We use an existing halo finder to identify density peaks within the set of tracked particles (Appendix \ref{sec:identifying_substructure}).
\item We use the most-bound particles to select which density peak is the true center of the subhalo. (Appendix \ref{sec:subhalo_center})
\item We calculate halo properties based on this center using all the tracked particles. (Appendix \ref{sec:subhalo_properties})
\end{enumerate}
At the highest level of abstraction, this strategy is similar to the codes \textsc{HBT}/\textsc{HBT+} \citep{han_2012_hbt,han_2018_hbt_plus} and \textsc{Sparta} \citep{diemer_2023_haunted}, although in detail the methods are quite different. We discuss this in more detail in Section \ref{sec:method_comp}.

\subsection{Merger Tree Post-Processing}
\label{sec:tree_post_processing}

Merger trees frequently contain errors: common errors include spurious subhalos that flit in and out of existence, or aphysical jumps in halo mass during major mergers and subhalo disruption \citep[e.g.,][]{srisawat_2013_sussing,behroozi_et_al_2014_mergers,rodriguez_gomez_2015_sublink,mansfield_2020_three}. We apply a post-processing step to our input halo catalogs that (a) removes subhalos that are almost certainly artifacts (b) accurately identifies the snapshot of first infall while accounting for mass jumps, and (c) accurately calculates $m_{\rm peak}$ while accounting for mass jumps.

First, we remove any merger tree branches that are likely numerical artifacts: branches that disrupt without merging with another halo and branches that are already subhalos in their first snapshot. In a simulation with coarse snapshots, this latter condition can occur innocently, but for a simulation with fine snapshots --- like the halos in the Symphony suite --- it essentially cannot occur naturally. In Appendix \ref{sec:false_mergers in Rockstar}, we discuss the frequency and properties of these branches and elaborate on the argument that these branches are not physical.

Merger tree errors make it difficult to determine when a halo first became a subhalo and have been a consistent thorn in the side of splashback subhalo analysis \citep{mansfield_2020_three,diemer_2021_flybys}. The most important issue is that during a major merger between two similar-mass halos, one halo can be identified as the central in one snapshot and then the other as the central in the next. This means that by the time the merger has been completed with a tree-merger, the branch containing the central will often have been temporarily labeled as a subhalo of the branch that it just destroyed. In general, this leads to estimating times of first infall that are too early and subhalo fractions that are too high. The current version of RCT attempts to handle this issue for infall times by identifying jumps in mass accretion histories larger or smaller than a certain fraction of a halo's current mass and using those jumps to inform the decision of whether to flag a branch as having been the subhalo of another branch. This general type of method works fairly well in most cases, but we find that it leads to a large number of complicated edge cases and can still misidentify some mergers even when parameters are carefully tuned.

We account for these events with an improved version of the algorithm used in \citet{mansfield_2020_three}. For each branch, $i$, we calculate, $m_{\rm peak,0,i}$, the maximum value of $m$ ever achieved by that branch. Note that this is different from the typical $m_{\rm peak}$ definition used throughout this paper, which is only measured prior to first infall. The snapshot of first infall cannot be found until subhalo associations are determined, so using it at this stage would not be possible. We find all other branches, $j$, that were ever hosts of $i$ and for which $m_{\rm peak,0,j}>m_{\rm peak,0,i}$. These are ``candidate host branches.'' A candidate host branch is removed from consideration if $j$ merges with $i$ in a later snapshot or if $j$ is a subhalo of $i$ {\em after} the last snapshot during which $i$ was a subhalo of $j$.  Candidate host branches that pass these checks become ``confirmed host branches,'' and the snapshot of first infall is the first snapshot where branch $i$ was a subhalo of a confirmed host branch. The use of $m_{\rm peak,0}$ to rank branches at this stage can lead to errors in cases where a major merger has not yet finished by the time the simulation completes. But in these cases, there is no clear way to determine which of the two merging branches would eventually become the true central anyway.

\subsection{Particle Associations}
\label{sec:particle_associations}

Next, we flag all the particles that have ever been accreted onto each branch in the merger tree. We assume that particles cannot be accreted from a larger branch to a smaller branch. This is similar to the philosophy behind the criteria used by \textsc{Sparta} \citep{diemer_2023_haunted} in its particle-tracking subhalo finder, which enforces a series of criteria that prevent the vast majority of host halo particles from being included in the tracked set of subhalo particles. \textsc{HBT} \citep{han_2012_hbt} and \textsc{HBT+} \citep{han_2018_hbt_plus} allow the user to supply whatever particles they wish. Given the complexity of particle tracking, in practice, this means that the simplest thing for most users would be to pass the particles identified by the input halo finder at the moment of infall, but this is not a fundamental feature of the \textsc{HBT+} algorithm.

The assumption that particles do not accrete from a host halo to a subhalo is a reasonable one. An unbound, dissipationless particle approaching an isolated halo can only accrete onto that halo if the halo's potential deepens due to mass growth during the particle's first passage through the halo
\citep[e.g.,][]{fillmore_goldreich_1984_self_similar,bertschinger_1985_self_similar}, and subhalos are actively losing mass throughout their orbits. Additionally, the difference in velocity between the subhalo and the typical host halo will be of the order of the host's $V_{\rm max}$ --- which will be much higher than the escape velocity of the subhalo --- and the host's tidal field will substantially limit the spatial region in which particles are capable of being bound to the subhalo. Even in cases where a host particle's preexisting orbit brings it close enough to be bound to a subhalo, it is quickly lost again and has little impact on the long-term mass evolution of the subhalo \citep{han_2012_hbt}.

We also flag all tracked particles according to whether they were ``smoothly accreted'' or accreted as part of an in-falling subhalo. For each particle, we find all central halos for which that particle was ever within $R_{\rm vir}$ of. To do this quickly, we use the grid-based linked-list method described in \citet{mansfield_2020_three} and originally implemented in \citet{mansfield_2017_shellfish}. We say that a particle is {\em smoothly accreted} onto a branch if a halo within that branch was the first halo that ever contained the particle. A particle is {\em accreted} onto the branch if it was contained by one of that branch's halos before being contained by any other branch with a larger $m_{\rm peak}$. Smaller halos cannot, by construction, accrete particles from bigger ones. Particles can smoothly accrete onto at most one branch but can accrete onto an arbitrary number of branches.

The choice of halo boundary ($R_{\rm vir}$ in our case) impacts particle-tracking in three ways beyond merely increasing/decreasing the number of subhalos in the catalog.
\begin{enumerate}
\item If too large a halo boundary is chosen, particles that never orbited the subhalo can be counted in subhalo property calculations.
\item If too large a halo boundary is chosen, some particles which never orbited a larger halo can be excluded from accretion onto subhalos prior to infall.
\item If too small a halo boundary is chosen, some particles which are truly orbiting the subhalo will not be tracked.
\end{enumerate}
Effect 1 is not very important as long as some reasonable definition of halo radius is chosen: this choice of radius certainly changes the amount of splashback mass in isolated halos \citep{diemer_2021_flybys}, but this mass is lightly bound and only 2\% of the mass outside the instantaneous virial radius is bound for the median subhalo in the SymphonyMilkyWay suite. Effect 2 is not relevant for our choice of $R_{\rm vir}$: essentially all isolated halos have splashback radii that are larger than $R_{\rm vir}$ (e.g., Fig. 6 in \citealp{mansfield_2017_shellfish}). Effect 3 is also not very important for our choice of $R_{\rm vir}$ for the same reason that Effect 1 is weak: the virial radius is large compared to the extent of bound matter in a typical subhalo.

As a final step before identifying subhalos, we use a tree code to calculate the potential energy of all tracked particles during the snapshot of first infall, and the velocity of the RCT subhalo at this snapshot to calculate kinetic energies. We then identify and store the IDs of the 32 most-bound particles so they can be found in later snapshots.

\subsection{Identifying Density Peak Candidates}
\label{sec:identifying_substructure}

Once the tracked set of particles has been identified, we attempt to find a subhalo among the tracked particles in every post-infall snapshot. To do this, we must locate the density peak that makes up the subhalo's center in each of these snapshots. This density peak will be used to calculate subhalo properties in Appendix \ref{sec:subhalo_properties}. First, we find candidate density peaks in the subhalo's smoothly accreted matter. Later, highly-bound particles will be used to select one candidate peak as the true center (Appendix \ref{sec:subhalo_center}).

We identify candidate locations for the subhalo by running a conventional halo finder on the smoothly accreted particles associated with that subhalo to identify all density peaks, both bound and unbound. For implementation simplicity, we identify peaks with the \textsc{Subfind} algorithm, as described in \citet{springel_2001_populating}. Briefly, \textsc{Subfind} estimates the density at every particle using an SPH (smoothed-particle hydrodynamics) kernel over $k$ neighboring particles. From here, a density threshold is lowered through the range spanned by the particles. As each particle passes through the threshold, it is either labeled as a density peak, a particle associated with a previously found peak, or a saddle point between two adjacent peaks. At each saddle point, a heuristic is used to determine which of the two peaks is the dominant one, and for the purposes of future particle additions, the two peaks are joined under the banner of the winning peak. We take the dominant peak to be the one with more particles.\footnote{\citet{springel_2021_gadget_4} report that using information from previous snapshots about the total size of all the substructures the particle has ever been associated with performs substantially better than this heuristic and solves some long-running issues with \textsc{Subfind}-based merger trees. This would be an interesting  direction for future experimentation.}

We find density peaks {\em using only smoothly accreted particles} to avoid accidentally centering the subhalo on one of its ``sub-subhalos,'' objects that were orbiting the subhalo before its moment of infall. Because much of a central halo's mass is accreted as subhalos, unbinding cannot be performed at this stage because the smoothly accreted matter is not necessarily self-bound. Even if this were not true, restricting the analysis to only bound density peaks would not be desirable because sub-subhalos can become unbound after their subhalo does, meaning that sometimes the correct behavior is to associate the subhalo with an unbound density peak even if a bound alternative exists.

Because unbinding has not yet been performed, the positions of each candidate subhalo location are the positions of the highest-density particle within the peak, and the velocities are estimated with an SPH kernel over the nearest $k$ particles to that position. One alternative method for finding the velocity would be to take the subhalo's velocity to be the average velocity of the particles that were bound in the previous unbinding step, as HBT+ does \citep{han_2018_hbt_plus}. This likely works well for HBT+, which discards tracked particles that have become too unbound, but since we are performing analysis on all tracked particles, this procedure tends to pull the mean velocity towards the velocity of the host halo.

In our procedure, $k$ sets a floor for the mass of substructure we can track in later steps: once the number of bound particles in the subhalo is smaller than $k/2$, subhalo positions and velocities become inaccurate, and subsequent unbinding passes may incorrectly remove the subhalo because kinetic energies are overestimated relative to an incorrect bulk velocity. In practice, our subhalos become unconverged at higher particle counts anyway: see Section \ref{sec:mass_loss_rates}.

It is likely that a more sophisticated peak-finding algorithm would meaningfully improve the performance of this algorithm (for example, in Appendix \ref{sec:subfind_radius_cdf} we show that \textsc{Subfind} is less capable than \textsc{Rockstar} at finding low-mass subhalos in dense environments), but we defer this to future work due to the complexity of re-implementing/interfacing with \textsc{Rockstar}. All subhalo properties are calculated independently of the peak-finding algorithm, meaning that the primary impact of the choice of peak-finder is whether the peak can be found at all and what bulk velocity is associated with the peak. As we demonstrate in Section \ref{sec:good_enough}, using \textsc{Subfind} to identify peaks leads to subhalo survival times which are long enough to follow all resolved subhalos up to the point when their satellite galaxies would likely disrupt, so improving subhalo survival times beyond the current algorithm is not a high priority.

Following the testing in Appendix \ref{sec:parameter_selection}, we set $k=16$.

\subsection{Selecting the Subhalo Center}
\label{sec:subhalo_center}

Since candidate density peaks were found using only smoothly accreted particles, there would ideally be exactly one density peak: the center of the halo. However, sub-subhalos below the limit where particle tracking is performed can still form density peaks, as can phase-mixed tidal tails orbiting the host halo's center, fragmented low-density portions of unmixed tidal tails, and the gravitational wakes of large sub-subhalos.

To make matters worse, there is no way to unambiguously tell these false centers apart from the true subhalo center using a single snapshot of information. For example, consider subhalo A --- a large subhalo that has experienced dynamical friction and merged with its host but still has a surviving, untracked sub-subhalo that continues to orbit the host --- and subhalo B --- a subhalo whose center is still bound but has large, phase-mixed tidal tails that have reached an equilibrium distribution that forms a density peak at the center of the host halo. Our testing revealed many examples of both types of subhalos. Subhalo A has truly disrupted, and its true center is a low-density, high-mass, unbound peak at the center of the halo and its false center is a low-mass, high-density, bound peak far from the center. For subhalo B, the opposite is true.

To resolve this ambiguity, for each subhalo branch, we identify the $N_{\rm core}$ most-bound ``core'' particles at the snapshot before first infall. For each of these particles, we identify the density peak that it belongs to. As part of the process of identifying \textsc{Subfind} peaks, each particle has already been associated with a peak. The density peak that has the largest number of these core particles associated with it is chosen as the subhalo center.

Following the testing in Appendix \ref{sec:parameter_selection}, we set $N_{\rm core}=32$.

\subsection{Calculating Subhalo Properties}
\label{sec:subhalo_properties}

Once the center of the subhalo has been identified, we perform an ``iterative unbinding'' pass on all the particles associated with the branch, including particles that were not smoothly accreted. An iterative unbinding pass means that the gravitational potential energy and kinetic energy of every particle are calculated in a loop. With each iteration of the loop, unbound particles are removed until the loop converges and either all particles have been removed or no additional particles are removed. The velocity identified in Appendix \ref{sec:identifying_substructure} is used as the center-of-mass velocity when computing kinetic energy. This is different from the approach used in \textsc{HBT}/\textsc{HBT+} \citep{han_2012_hbt,han_2018_hbt_plus} --- which iteratively updates the peak velocity to be the bound center of mass velocity --- and \textsc{Sparta} \citep{diemer_2023_haunted} --- which does not perform unbinding.

We find that for almost all subhalos, two unbinding passes will get masses within $\approx 1\%$ of the full iterative unbound masses (see Appendix \ref{sec:iterative}), but because we are working with rather small simulations in this paper, we iterate until convergence. We calculate potentials using a fast $k$d-tree code.\footnote{\url{https://github.com/phil-mansfield/gravitree}} This code is similar in philosophy to \textsc{fast3tree} from \citet{behroozi_2013_rockstar} but uses an alternative memory layout and implicit node connectivity scheme that substantially improves memory footprint and performance. Counter-intuitively, our testing indicates that just assuming spherical symmetry and calculating iterative potentials from an enclosed mass profile leads to very similar masses (Appendix \ref{sec:iterative}), but we opt not to take advantage of this fact in this work.

We note that this method tends to estimate masses slightly smaller than \textsc{Rockstar} (see Appendix \ref{sec:iterative}; this is also visible in Fig.~\ref{fig:example_halo}). This is mostly because \textsc{Rockstar} only performs a single unbinding pass, but part of the difference also comes from a small amount of host matter being incidentally associated with the subhalo and included within \textsc{Rockstar}'s 6D friends-of-friends groups despite not having been accreted to the subhalo prior to infall. This latter problem is exacerbated by \textsc{Rockstar}'s lack of iteration.

For each halo, we also compute the ``tidal mass'' of the halo. Tidal masses/radii are not used in this paper but are included in the published dataset. To do this, We perform iterative unbinding with an additional step. After flagging all bound particles within a given iteration we also compute the tidal radius, $r_{\rm tidal},$ and additionally remove particles outside $r_{\rm tidal}$ before iterating. We define the tidal radius as
\begin{equation}
    \label{eq:r_tidal}
    r_{\rm tidal} = r\,\left( \frac{m_{\rm sub}(<r_{\rm tidal})/M(<r)}{2 + (\omega/\omega_{\rm circ})^2 - \frac{d\ln M(<r)}{d\ln r}|_r} \right)^{1/3}.
\end{equation}

Here, $r$ is the radius of the subhalo, $M(<r)$ is the mass of the host halo enclosed within $r,$ $\omega_{\rm circ} = \sqrt{GM(<r)/r^3}$ is the angular speed the subhalo would have if it were on a circular orbit, and $\omega=|\vec{v}\times\vec{r}|/|\vec{r}|$ is the instantaneous angular speed of the subhalo. In the rare cases where the denominator of $r_{\rm tidal}$ is non-positive, we take $r_{\rm tidal}$ to be infinite. This definition of $r_{\rm tidal}$ is one of many that has been used in the literature \citep[see review in][]{vandenbosch_2018_disruption}, and makes several non-trivial assumptions: that the host and subhalo mass profiles are spherically symmetric, that the only important pseudo-force acting on the subhalo is the centrifugal force, and that the distance between the host center and the subhalo is large compared to the size of the subhalo. More generally, although tidal shears are well-defined instantaneous quantities, the concept of tidal radii and the stability of particles orbiting a subhalo is poorly defined for non-circular or non-equilibrium orbits. 

We also compute several secondary halo properties:
\begin{itemize}
\item The half-mass radius, $r_{\rm half}$. 
\item The maximum rotation curve, $v_{\rm max}.$
\item $r_{\rm max}$, the radius at which $v_{\rm max}$ occurs.
\end{itemize}
All three quantities are computed using only bound particles.

\subsection{Subhalo Disruption}
\label{sec:subhalo_disruption}

We count a subhalo as disrupted (1) if it is completely unbound (2) if the distance to the center of the host halo is smaller than the subhalo's $r_{\rm half}$, or (3) if none of the $N_{\rm core}$ particles are within $r_{\rm half}$. There are several other conditions that can cause a subhalo to disrupt in some rare edge cases (see below).

In practice, condition (1) is almost never relevant and disruption is entirely determined by conditions (2) and (3). The one exception to this is rare instances where subhalo bulk velocities are incorrectly estimated (see below). Condition (2) is necessary for cases where a subhalo quickly sinks to the center of its host through dynamical friction while its particles remain formally bound. It is almost certainly too conservative and would need to be investigated further for studies of major mergers or subhalos at pericenter. In practice, subhalos that are incorrectly marked as disrupted through this condition prior to the final snapshot of the simulation are re-inserted into the simulation through interpolation once they leave their hosts' centers. Condition (3) is the primary way that low-mass subhalos disrupt. This condition protects against accidentally selecting the wrong density peak.

Occasionally, a subhalo will be identified as intact for a single snapshot long after the initial disruption due to a statistical fluctuation in the density field combined with a random conjunction of phase-mixed core particles. To protect against this, any subhalo that has disrupted at snapshots $i-1$ and $i+1$ according to conditions (1) - (3) is also considered disrupted at snapshot $i$.

Lastly, if RCT has identified a subhalo during the same snapshot for this branch, we compare the locations of the subhalos found by the two methods and count the number of core particles within them. If the two subhalos are outside one another's half-mass radii and there are more core particles within the RCT subhalo, we count the particle-tracking subhalo as disrupted (see Appendix \ref{sec:parameter_selection}). This condition is rarely triggered.

\subsection{Interpolation}

Subhalo disruption does not stop a subhalo from being searched for in subsequent snapshots. Occasionally criterion (2) triggers unnecessarily for a subhalo close to pericenter or an incorrect velocity center will be identified for a subhalo, causing too much mass to become unbound, leading to condition (3) triggering incorrectly for a snapshot. In these cases, continuing to search at later snapshots prevents subhalo branches from terminating too quickly.

When a subhalo temporarily disrupts and is re-found in later snapshots, we interpolate its properties during its period of temporary disruption. Positions are interpolated with cubic splines to allow for more accurate pericenter estimates, masses are found through piecewise-power-law interpolation, and all other quantities are found through piecewise-linear interpolation. Splines are not used in the latter two cases because the noise inherent in our halo quantity measurements leads to excessive ``ringing'' in some cases.

In rare cases ($\lesssim 0.7\%$ of subhalos), a subhalo cannot be found during its infall snapshot but can be found in later snapshots. In these cases, we use the properties of the corresponding RCT halo during the infall snapshot as an interpolation anchor. In cases where information about the halo's profile is needed to compute a property that is not in the default RCT catalogs, we assume an NFW profile with the  same $v_{\rm max}$ value as the subhalo.

\section{False mergers}
\label{sec:false_mergers in Rockstar}

\begin{figure}
\hspace{-3.5mm}
\includegraphics[width=0.475\textwidth]{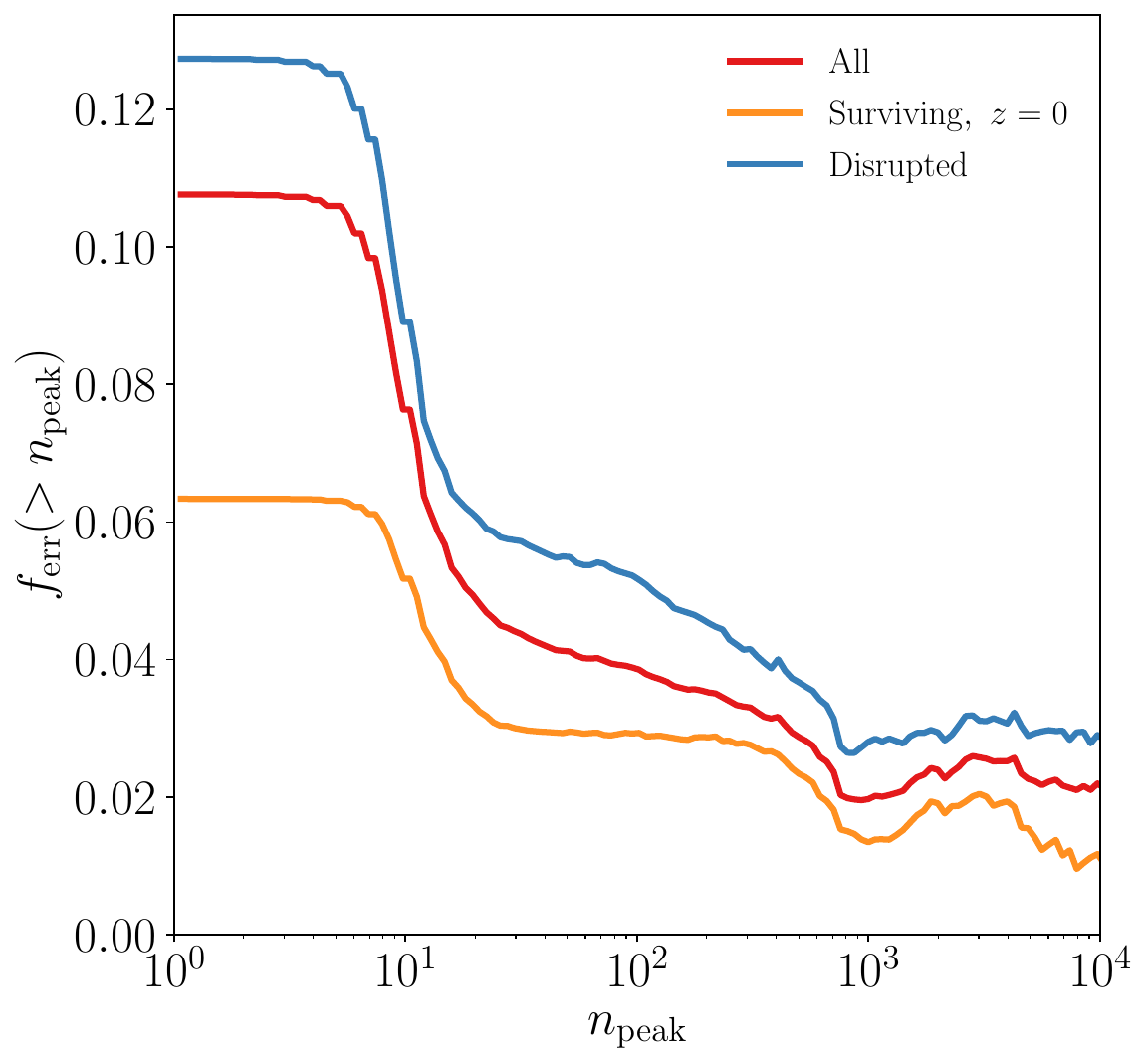}
\caption{The fraction of \textsc{Rockstar} branches which started out as subhalos. Symphony's snapshot cadence is fine enough that these branches are likely numerical artifacts: either subhalos that are flitting in and out of existence, or subhalos that \textsc{Rockstar} had previously lost track of and was unable to match with its true progenitors. The red curve shows all the branches that have ever been subhalos of one of Symphony's main hosts, the orange curve shows the subset of those halos that survive to $z=0,$ and the blue curve shows all the branches that disrupt before $z=0$. The frequency of this class of error decreases with increasing resolution, reaching the few-percent level for subhalos with hundreds of particles. It is more common in disrupting branches than surviving ones.}
\label{fig:false_mergers}
\end{figure}

In our analysis, we remove subhalos that appear to form within the virial radius of their host. 

In $\Lambda$CDM, halos form due to the collapse of low-density perturbations \citep[e.g.,][]{gunn_gott_1972_infall} over timescales that are much longer than the orbital timescale of the resulting halo. Density perturbations within halos cannot undergo collapse like this because the tidal forces of the host and the shearing force within the rotating frame of the perturbation are stronger than the perturbation's self-gravity. This means that the only physical way in which a subhalo could experience its first snapshot within the virial radius of another halo is if it formed and was accreted within the time between two adjacent snapshots. The largest gap between two snapshots in our simulation suite is 0.020$\times T_{\rm orbit},$ i.e. one-fiftieth of the time it would take a particle to orbit the virial radius of a halo at that snapshot. This means that any such subhalo would have to form immediately outside the virial radius of their host. Tidal forces cause in-falling subhalos to stop growing well outside the radius of the host halo that they will eventually orbit \citep[e.g.,][]{behroozi_et_al_2014_mergers,mansfield_2020_three}, meaning that this should only occur for very poorly resolved objects that are flitting above-and-below the RCT's detection limit and which are possibly Poisson fluctuations rather than bona fide halos. The numerical reliability of such objects is highly suspect.

Therefore, it is reasonable to assume that any subhalo which appears to form within its host in our simulations is the result of an RCT error. There are two general ways this could happen: first, the subhalo could be a real subhalo that was lost by RCT and then found some number of snapshots later. One of the great strengths of \textsc{consistent-trees} is its ability to connect subhalos that have been temporarily lost \citep{behroozi_2013_consistent}, but even \textsc{consistent-trees} must do this heuristically and is not perfect. The second possible way this could happen is if the halo finder misidentifies a transient overdensity (such as an aftershock or shell generated by a major merger) in the halo as a subhalo. The former error can lead to double-counting the subhalo in some forms of analysis and an incorrect $m_{\rm peak}$ value, while the latter leads to an entirely erroneous catalog entry. We do not attempt to differentiate these two types of errors.

In Fig.~\ref{fig:false_mergers}, we show the fraction of merger tree branches that start off their first snapshot as a subhalo. We show these fractions for all branches (red), branches that survive until the end of the simulation (orange), and branches that disrupt before the end of the simulation (blue). Generally, the error rate decreases with increasing subhalo resolution and reaches the $\approx3{-}5\%$ level at our sample limit of 300 particles. There are sharp features at $n_{\rm sub,peak}\approx15$ and $n_{\rm sub,peak}\approx 800$. We opt not to investigate the causes of these features other than to note that RCT mass functions of isolated halos are unconverged below 100 particles \citep{behroozi_2013_rockstar}, meaning that a 15 particle halo cannot be reliably detected or tracked across time by RCT.

To summarize, these halos do not make up a large fraction of the overall population but are certainly incorrect and easily removed.

\section{Parameter Selection}
\label{sec:parameter_selection}

\begin{figure*}
\hspace{-3.5mm}
\includegraphics[width=0.48\textwidth]{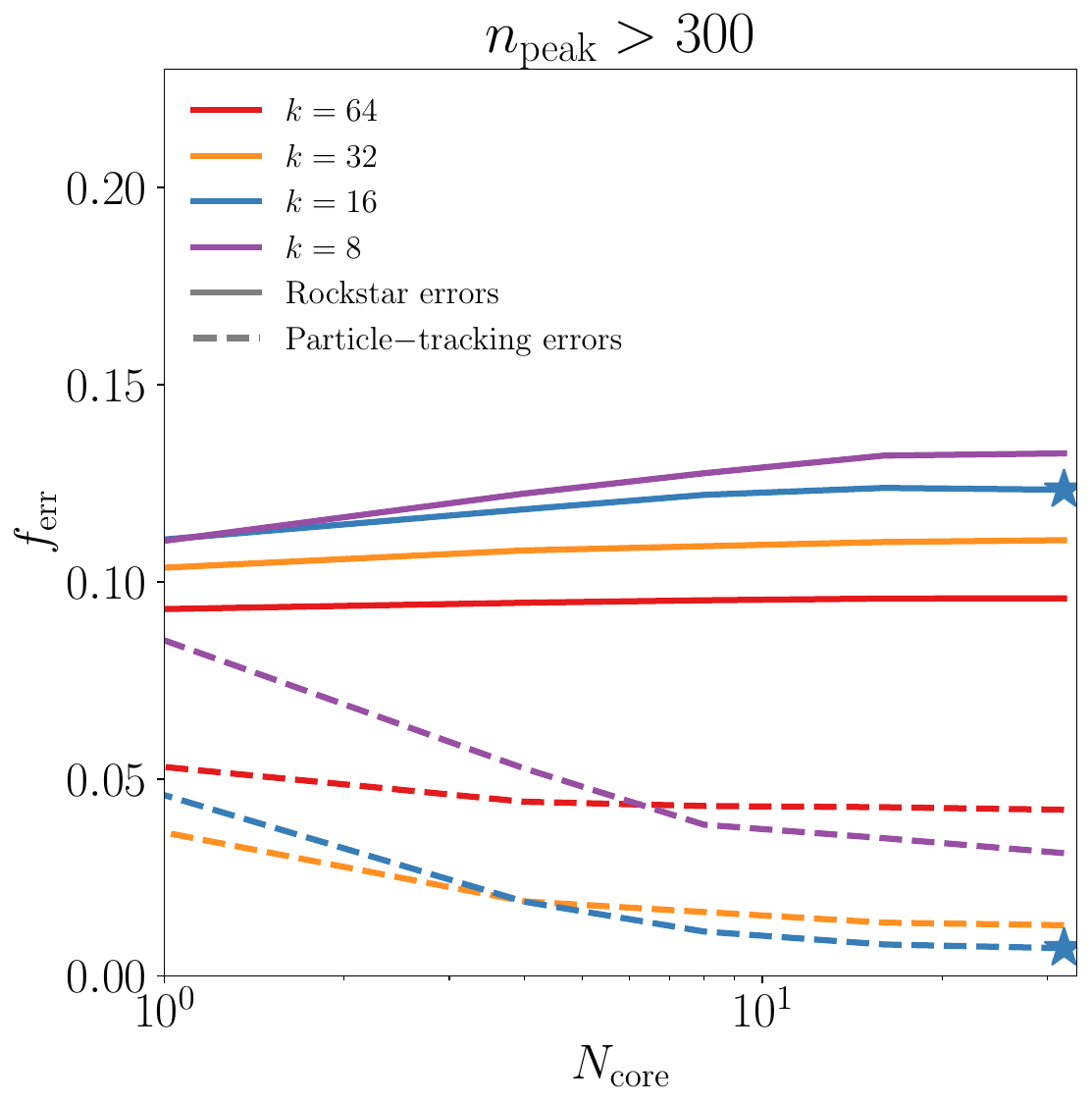}
\includegraphics[width=0.48\textwidth]{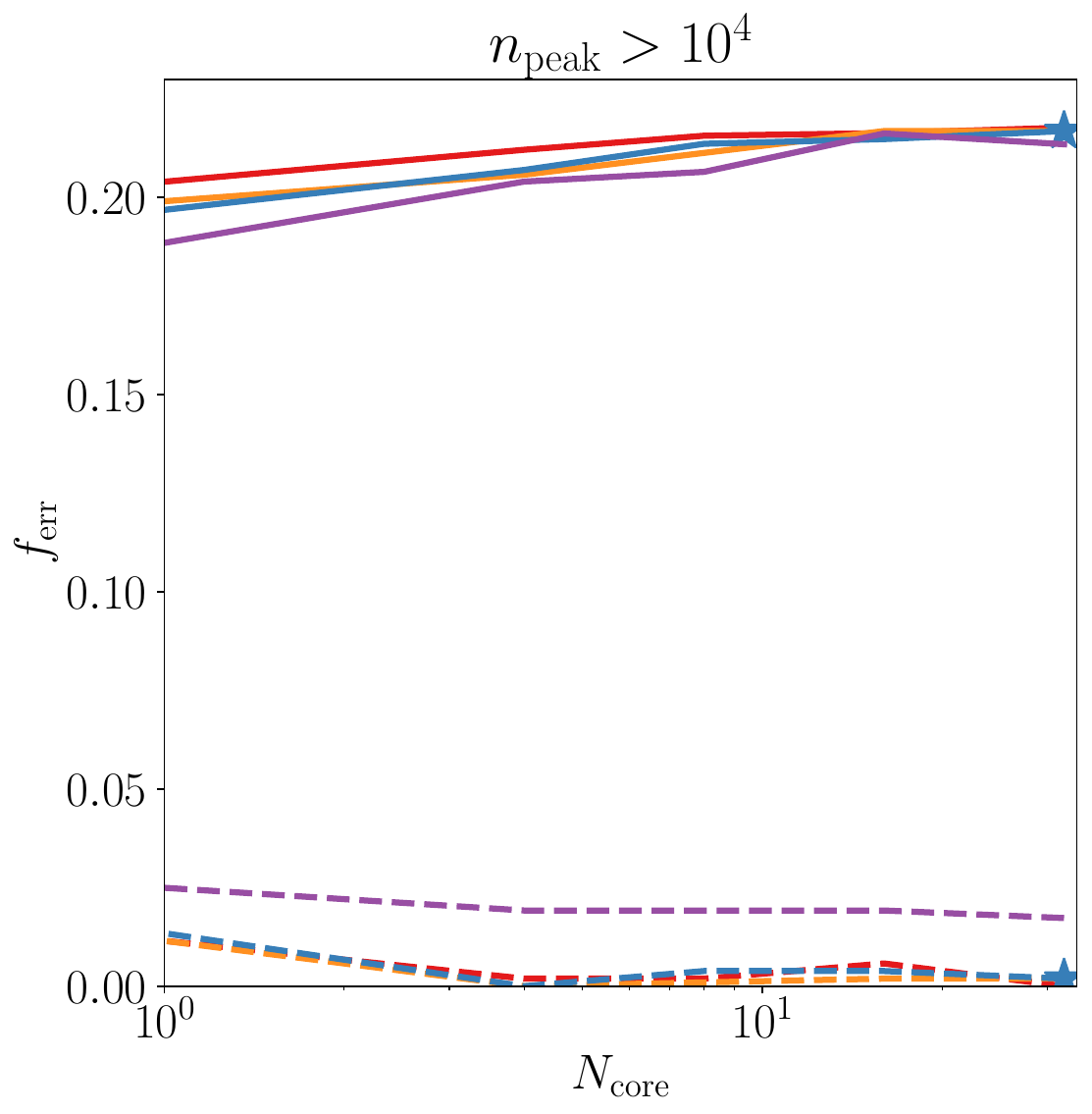}
\caption{A comparison of \textsc{Rockstar} and \textsc{Symfind} error rates as a function of our algorithm's two free parameters, $k$ and $N_{\rm core}$. The left panel shows low-resolution subhalos and the right panel shows moderate-resolution subhalos. The $y$-axis shows $f_{\rm err}$, the fraction of halos misidentified or missed entirely by either catalog. The best performance is achieved when the error rate in the core catalogs (dashed lines) is smallest and the error rate in the \textsc{Rockstar} catalogs (solid lines) is highest. The blue stars show the parameter choices used in this paper, $k=16$ and $N_{\rm core}=32$.}
\label{fig:parameter_tests}
\end{figure*}

\textsc{Symfind} has two free parameters. First, a set of highly bound  particles at infall (``core'' particles) are tracked, and second, some smoothing scale needs to be chosen for the \textsc{Subfind} sub-routine that identifies density peaks. The number of core \refadd{particles} is $N_{\rm core}$ and the smoothing scale is set by the $k$-nearest particles to a given point. 

To test the impact of different parameter choices, we estimate $f_{\rm err}$, the fraction of subhalos that have been incorrectly identified by a particular method. A subhalo is ``incorrectly identified'' if the subhalo truly exists, but the finder could not find it, or if it identifies an incorrect density peak as the subhalo's center.  It is impossible to directly measure $f_{\rm err}$, but it is possible to estimate its value by comparing two subhalo catalogs generated with different methods on the same set of subhalos. In this Appendix, we compare RCT subhalo catalogs and \textsc{Symfind}. We will \textsc{Symfind} select algorithm parameters that minimize $f_{\rm err}$.

We consider a particular finder to have made an error if (a) it does not find a subhalo that is found by the other method or (b) if it and the other method both find the subhalo, but the two subhalos are in inconsistent locations and the other finder has a more reliable identification. We consider a pair of subhalos to have inconsistent locations if their centers are separated by a distance larger than the smaller of their two $r_{1/2}$ values. In the case where there are inconsistent locations, we consider the subhalo with more core particles within its $r_{1/2}$ to be more reliable.

In Fig.~\ref{fig:parameter_tests}, we show how $f_{\rm err}$ depends on $N_{\rm core}$ and $k$ for both \textsc{Symfind} and RCT. The default parameters for our algorithm are set to be the pair that minimizes the particle-tracking $f_{\rm err}$ and are shown as stars. These values are $k=16$ and $N_{\rm core}=32$ and result in $f_{\rm err}= 0.6\%$ for $n_{\rm peak}>300$ and $f_{\rm err} = 0.2\%$ for $n_{\rm peak}>10^4.$ This is a negligible fraction. RCT has a higher error rate of 12\% and 22\% for these two resolution bins, respectively. These errors are almost entirely caused by RCT subhalos disrupting before particle-tracking subhalos. We analyze this behavior extensively and with more intuitive statistics in Sections \ref{sec:survial} and \ref{sec:shmf}. 

We performed the same analysis on the first snapshot after infall. 0.7\% of RCT subhalos with $n_{\rm peak}>300$ do not have a corresponding \textsc{Symfind} subhalo during this snapshot. We allow \textsc{Symfind} to find subhalos in subsequent snapshots, even if none was found in a previous snapshot, and 97\% of these initially missing subhalos are found later and survive for at least as many snapshots as their RCT counterparts. Because the first snapshot after infall is the {\em best} resolved snapshot, $\approx 0.7\%$ is a good estimate for the instantaneous error rate in finding a subhalo with \textsc{Symfind}. A consequence of this is that the $0.6\%$ error rate during the last snapshot of the simulation is entirely consistent with being caused by the sorts of short-lived errors, which would be corrected if our simulations were run past $z=0$.

Finally, we note that these tests may have implications for some ``orphan'' models, i.e., those that attempt to estimate the trajectory and properties of a subhalo and its hypothetical galaxy after a subhalo finder loses track of it (see Section \ref{sec:orphan}). One popular method is to take the subhalo's most bound particle during the last surviving snapshot or at infall. As we discuss in Section \ref{sec:stitching_errors}, the former method runs afoul of the aforementioned errors that RCT subhalos encounter during their last snapshots. The latter method would likely perform similarly to the dashed curves at $N_{\rm core}=1$, which corresponds to only using the most-bound particle. The only difference is that particle-tracking uses this particle to select which overdensity the particle is associated with, rather than assigning the subhalo's location to the particle itself. Thus, the behavior of such an orphan model is better approximated by smaller $k$ values that would cause this overdensity to be closer to the particle. At the smallest tested $k=8$, the $N_{\rm core}$ error rate is 8\% at $n_{\rm peak}>300$ and 2\% at $n_{\rm peak}>10^4.$ This is quite impressive, given the simplicity of the method. However, we caution readers that the similarly modest-seeming error rates of RCT lead to false convergence (Section \ref{sec:survial}) and order-unity biases in some subhalo statistics (Sections \ref{sec:shmf} and \ref{sec:radius_cdf}), so we would advise more explicit testing.

Some readers may be surprised that the most-bound particle can leave the center of its subhalo before less-bound particles. Briefly, the reason is that numerical effects can cause this particle to diffuse outwards to larger orbits over time. More specifically, this is due to the non-convergence of the innermost portions of these dark matter halos. Any self-gravitating system made up of discrete particles has a characteristic ``relaxation time,'' after which small-angle scattering between particles begins to influence the evolution of the system \citep[see][and references therein for discussion of this in the context of dark matter halo simulations]{ludlow_2019_numerical}. Once a dark matter system has been simulated for longer than its relaxation time, the approximation that its particles are collisionless fails, and numerical effects alter its properties. The relaxation times of the inner regions of dark matter halos are shorter than the relaxation times of their outskirts, and in practice, a substantial portion of the inner region in any long-lived subhalo is unconverged (see Fig.~\ref{fig:conv_limits}). For a system with equal-mass particles, the relaxation time also serves as the ``segregation time'' \citep[see][and references therein]{ludlow_2019_equipartition}, the timescale over which particle populations that had previously been segregated to different radial distribution will begin to exchange particles. For low-energy particles, this always manifests as an increase in orbital energy and distance from the center of the halo. 

This numerical scattering places a fundamental limit on how long any method based on tracking highly bound particles over time can truly track the center of a subhalo. Using a larger pool of particles, as we do here, can help mitigate the problem but cannot solve it.

\section{Iterative unbinding and future algorithm performance}
\label{sec:iterative}

\begin{figure}
\hspace{-3.5mm}
\includegraphics[width=0.475\textwidth]{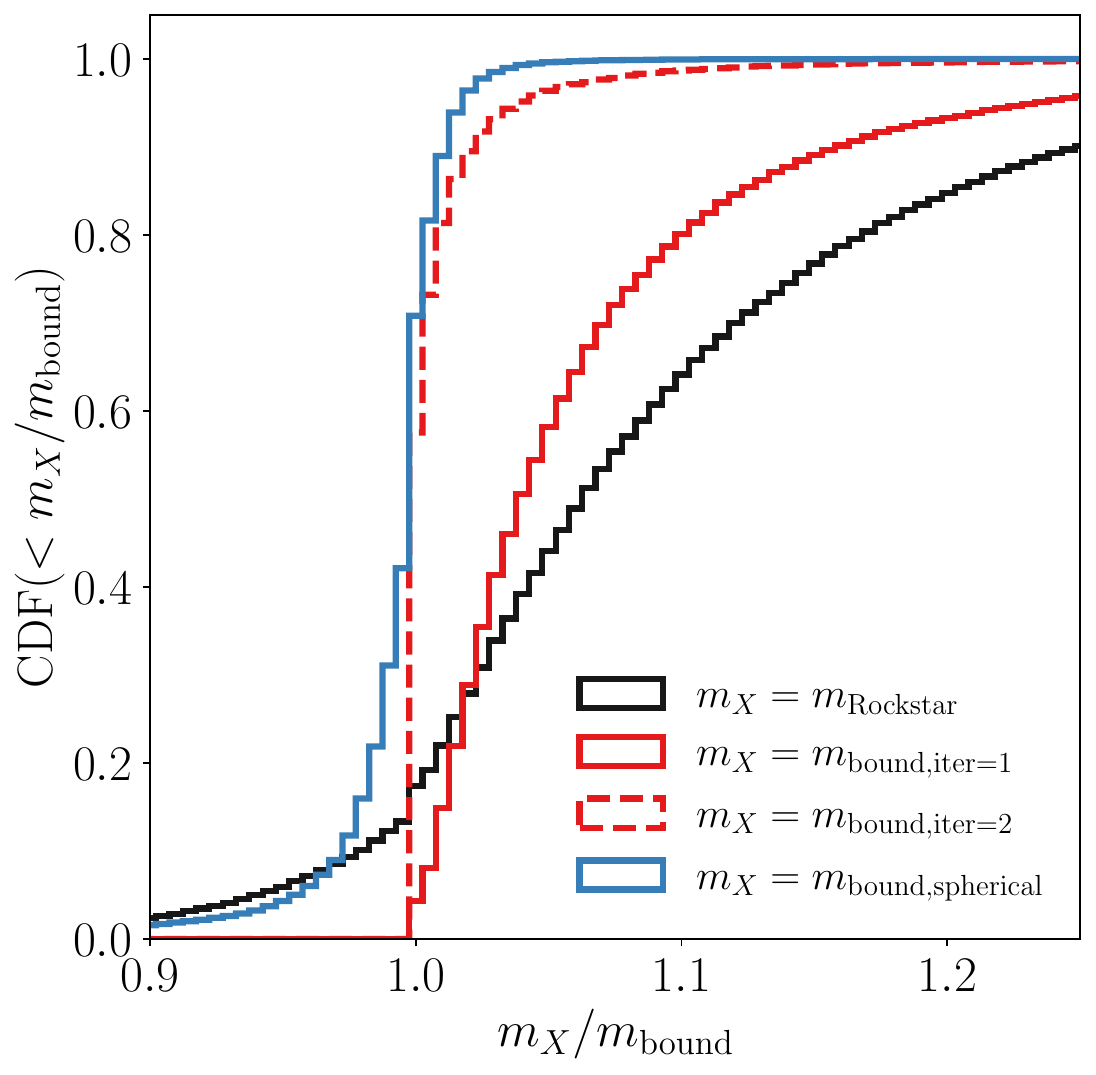}
\caption{The relationship between subhalo masses and $m_{\rm bound}$, the bound mass calculated from iteratively unbinding the tracked particles. The black curve compares \textsc{Rockstar} mass to $m_{\rm bound}$, the two red curves compare a limited number of iterative unbindings on tracked particles, and the blue curve compares against full iterative unbinding where the potential is calculated assuming spherical symmetry.}
\label{fig:mass_comp}
\end{figure}

The masses analyzed in this paper are calculated by performing iterative unbinding with a $k$-d tree-code\footnote{\url{https://github.com/phil-mansfield/gravitree}} on all the particles that were initially associated with the subhalo. This tree code is relatively efficient: all data is stored in a single flat array with node topology stored implicitly and has been aggressively line-optimized in critical regions. Our code can compute the potential of an Einasto point distribution in $\approx60\%$ the runtime taken by \textsc{Fast3tree}, the tree code used by \textsc{Rockstar}, when accuracy parameters are held fixed.

Nonetheless, iterative unbinding is a time-consuming calculation. The creation and walking of the underlying $k$-d tree is inherently expensive, requiring frequent conditional branching and non-local memory access. The cost of evaluating the potential of $n$ particles through the tree requires $\mathcal{O}(n\,\log\,n)$ operations and $\mathcal{O}(n\,\log\,n)$ memory, meaning that while the scaling at large-$n$ is not poor, it is also slower than a linear algorithm. Many subhalos take more than ten iterations to fully complete the unbinding process. To make matters worse, while it is relatively straightforward to thread-parallelize this operation on a single shared-memory machine, efficient parallelization across machines is quite involved \citep[e.g.,][]{potter_2017_pkdgrav4}. The Symphony suite only has $\gtrsim 10^9$ tracked particles across all its hosts, meaning that none of these performance issues are important for this work. But they would be serious barriers to scaling our approach up to modest-to-large-sized simulations.

One potential optimization would be to cap the number of iterations used during unbinding rather than requiring strict convergence. This could reduce the total runtime by an appreciable constant factor. Another potential optimization would be to assume that the subhalo is spherically symmetric, calculate the subhalo's enclosed bound mass profile, estimate escape velocities from this profile, and use these escape velocities to calculate boundedness. The mass profile and escape velocities can be calculated in $\mathcal{O}(n)$ runtime with small constant-time costs.

In Fig.~\ref{fig:mass_comp}, we compare the masses estimated by full iterative unbinding and these optimizations. This mass is referred to as $m$ in the main text of this paper, but we differentiate it from other mass estimates by referring to it as $m_{\rm bound}$ in this Appendix. To aid in comparison with RCT masses, we only analyze subhalos that have a surviving RCT subhalo, in which neither the RCT nor \textsc{Symfind} halos would be considered errors via the criteria described in Appendix \ref{sec:parameter_selection}, and use the RCT velocity to calculate the kinetic energies of particles. The results in this Section are not strongly sensitive to this choice. We compare $m_{\rm bound}$ against the bound mass calculated with a single iteration, $m_{\rm bound,iter=1}$, the bound mass after two iterations, $m_{\rm bound,iter=2}$, the iterative bound mass under the assumption of spherical symmetry, $m_{\rm bound,spherical},$ and the bound mass reported by RCT, $m_{\rm Rockstar}$.

If only a single unbinding pass is used, The median subhalo mass is about 4\% too high. This is a small amount, but there is a long tail to the bias, and one runs the risk that this bias is correlated with subhalo properties. However, if two or more unbinding passes are used, masses agree very closely with full unbinding. This is consistent with the convergence behavior of the two subhalos tested in Appendix A of \citet{han_2012_hbt}. Assuming spherical symmetry also results in masses close to $m_{\rm bound}$, with small tails. As with small-iteration unbinding, it is possible that errors in mass are correlated with the orbital state of the halo, so we would strongly recommend further testing before assuming spherical symmetry during unbinding. One could also imagine making higher-order approximations of the subhalo's density field (such as its shape tensor) and including this information in the potential calculation to reduce the size of the tails in the mass error distribution.

The median RCT mass is biased about 7\% high with a small tail to lower masses and a very large tail to larger masses. This is qualitatively similar to the distribution of $m_{\rm bound,iter=1}/m_{\rm bound}$, meaning that most of this difference probably comes from the fact that RCT only performs a single unbinding pass. However, the fact that the even the single-iteration bound mass distribution is less biased than RCT masses, means that means that RCT is likely also assigning a meaningful number of host particles to most subhalos.

Lastly, we consider applying a future version of \textsc{Symfind} to very large simulations. One of the reasons why subhalo finders find it difficult to run on massive simulations is that they usually either require that all a halo's particles be stored on the same shared-memory node or that large amounts of per-particle data be transferred between nodes, leading to complex and expensive communication patterns. But in exchange for a hopefully modest decrease in accuracy, \textsc{Symfind} would not need large communication volumes or for all particles to be stored on the same node. Halo positions and velocities could be estimated using only the core particles themselves, a technique which has been shown to scale well to very large simulations \citep[][but see also caveats in Section \ref{sec:orphan}]{heitmann_2021_last_journey,sultan_2021_cores,korytov_2023_cores}. If potentials are calculated from approximations of the subhalo's underlying density field, very little data would need to be transferred between the nodes that a subhalo was split across (e.g., the difference between transferring the coefficients in a low-order representation of a density profile compared to the transfer of potentially millions of particles). The key question plaguing such a setup would be whether the velocity estimated from the core particles would be reliable enough to estimate particle kinetic energies without excessively biasing bound mass calculations. We experimented with several methods for estimating subhalo velocities using only the properties of core particles during the development of \textsc{Symfind} but did not come up with a satisfactory solution. This remains an interesting topic for future work.

\section{Additional Subhalo Trajectories}
\label{sec:additional_halos}

\begin{figure*}
\hspace{-3.5mm}
\includegraphics[width=0.24\textwidth]{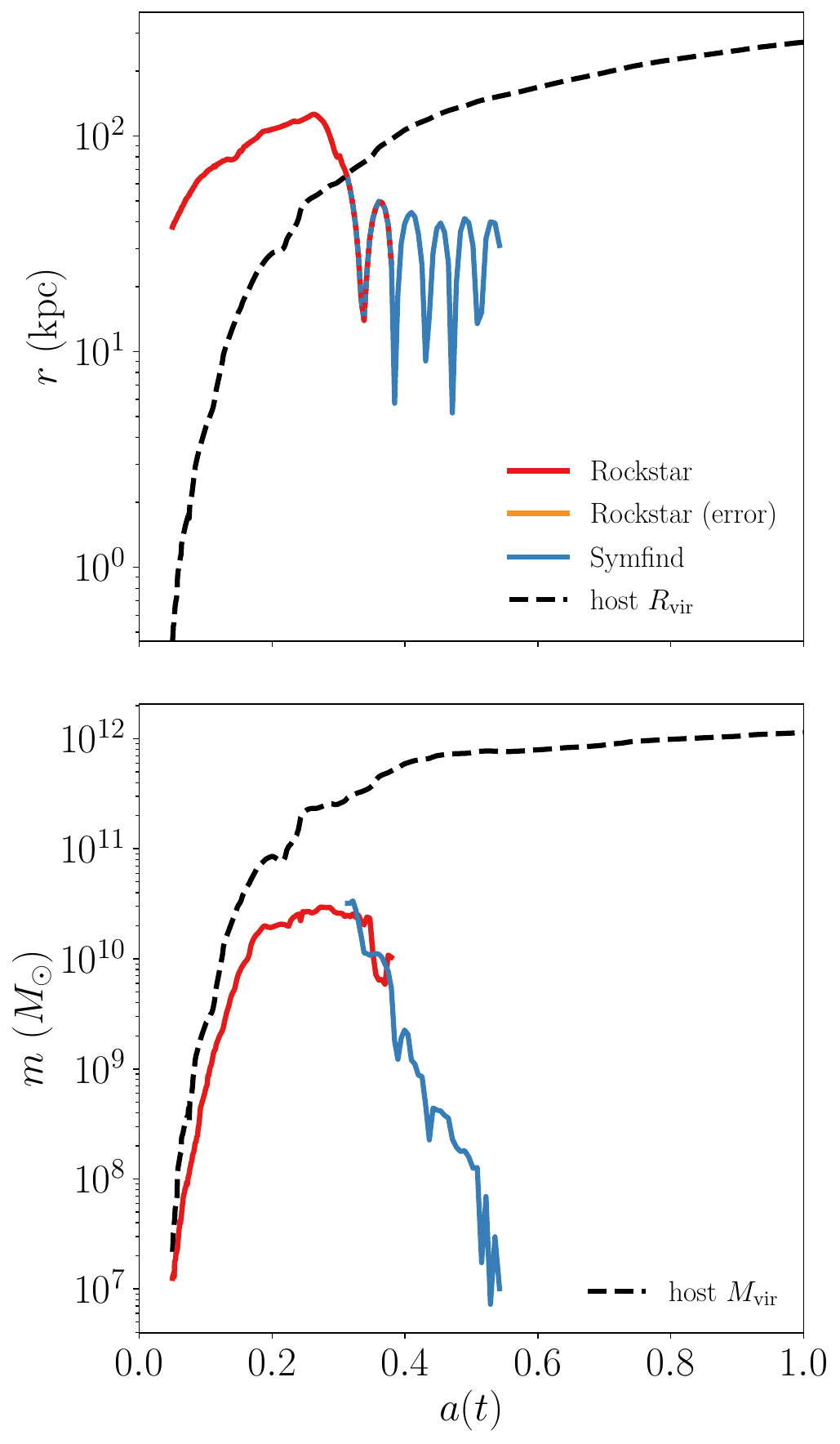}
\includegraphics[width=0.24\textwidth]{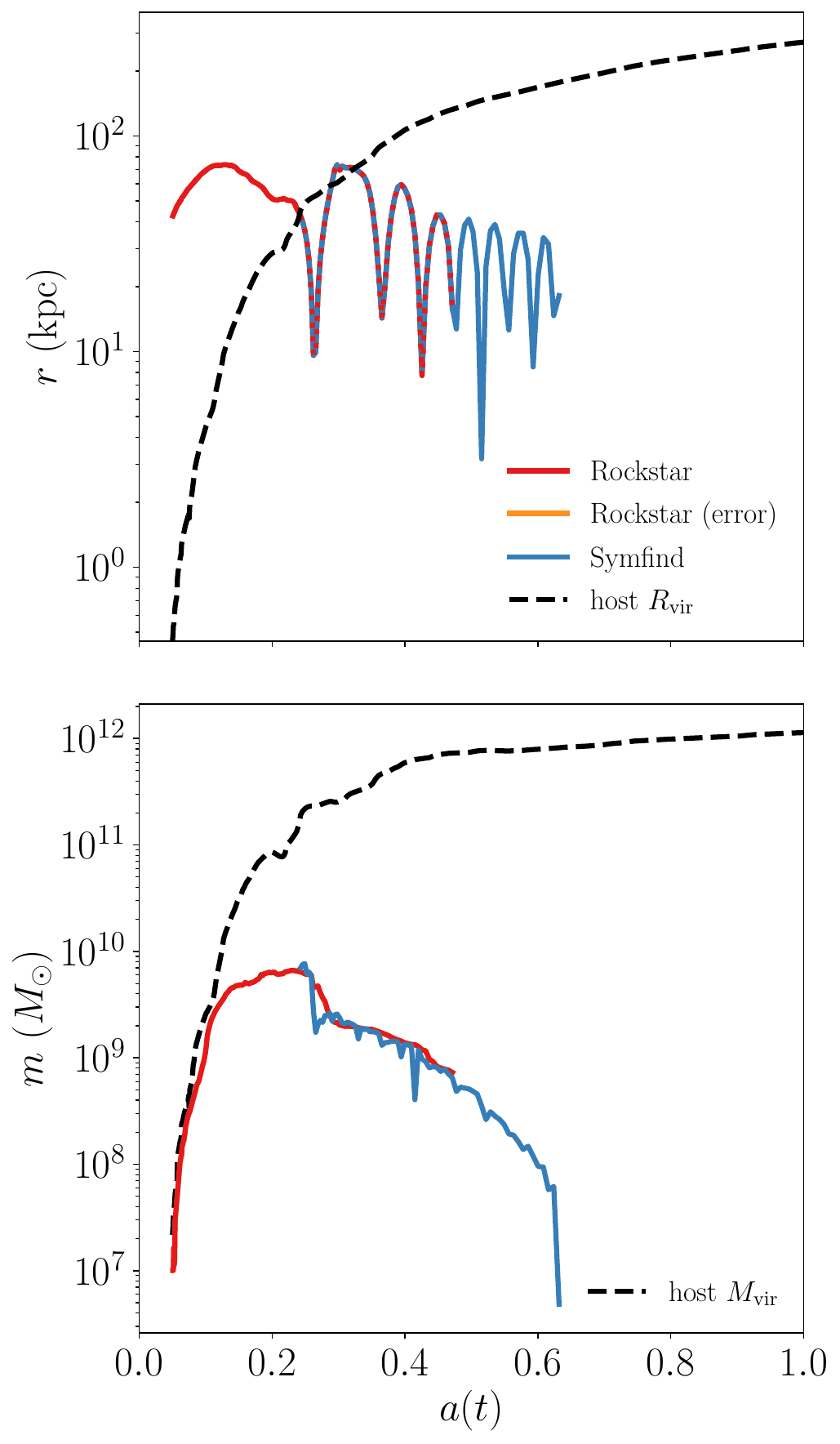}
\includegraphics[width=0.24\textwidth]{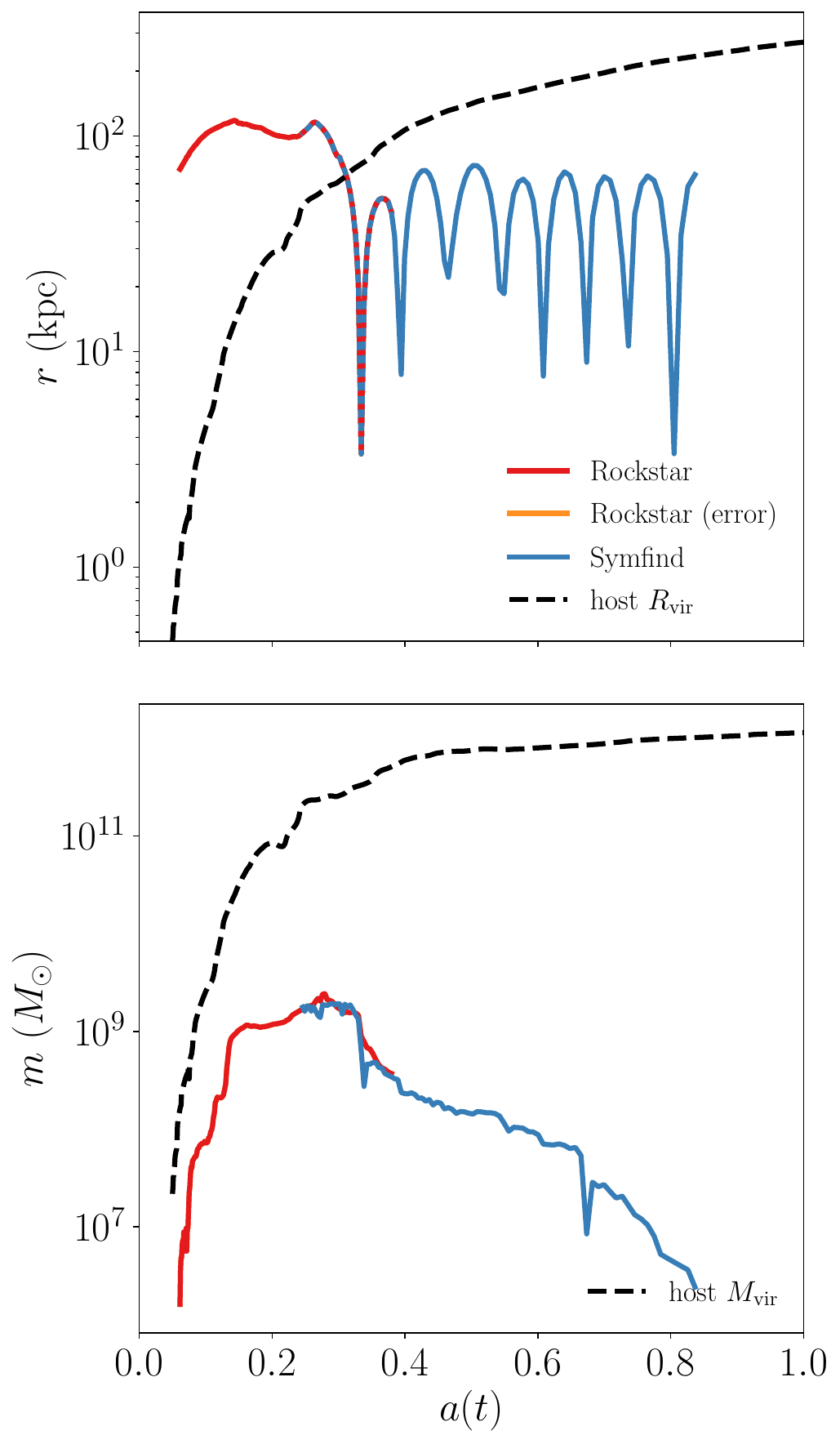}
\includegraphics[width=0.24\textwidth]{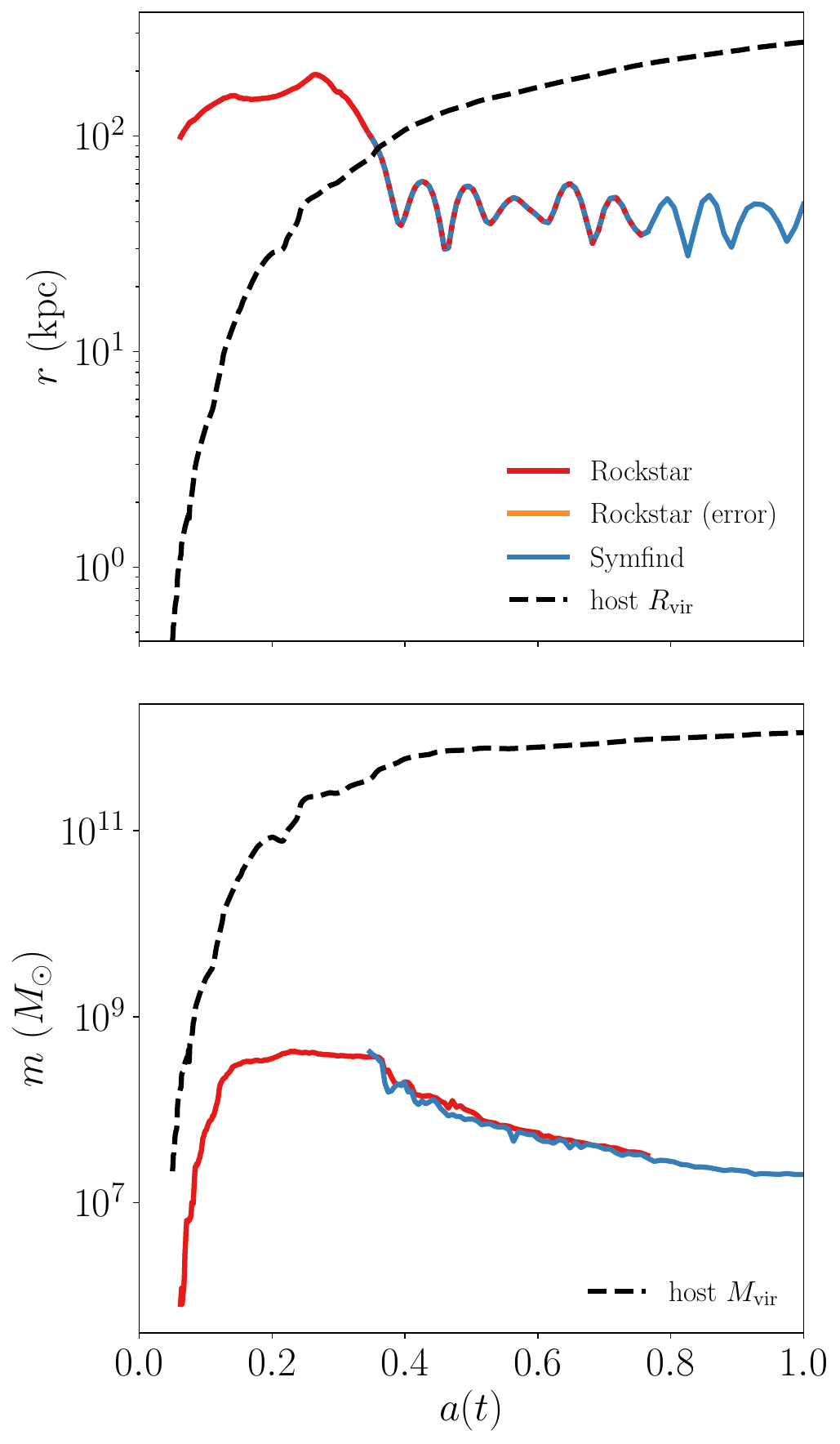} \\\vspace{15pt}
\includegraphics[width=0.24\textwidth]{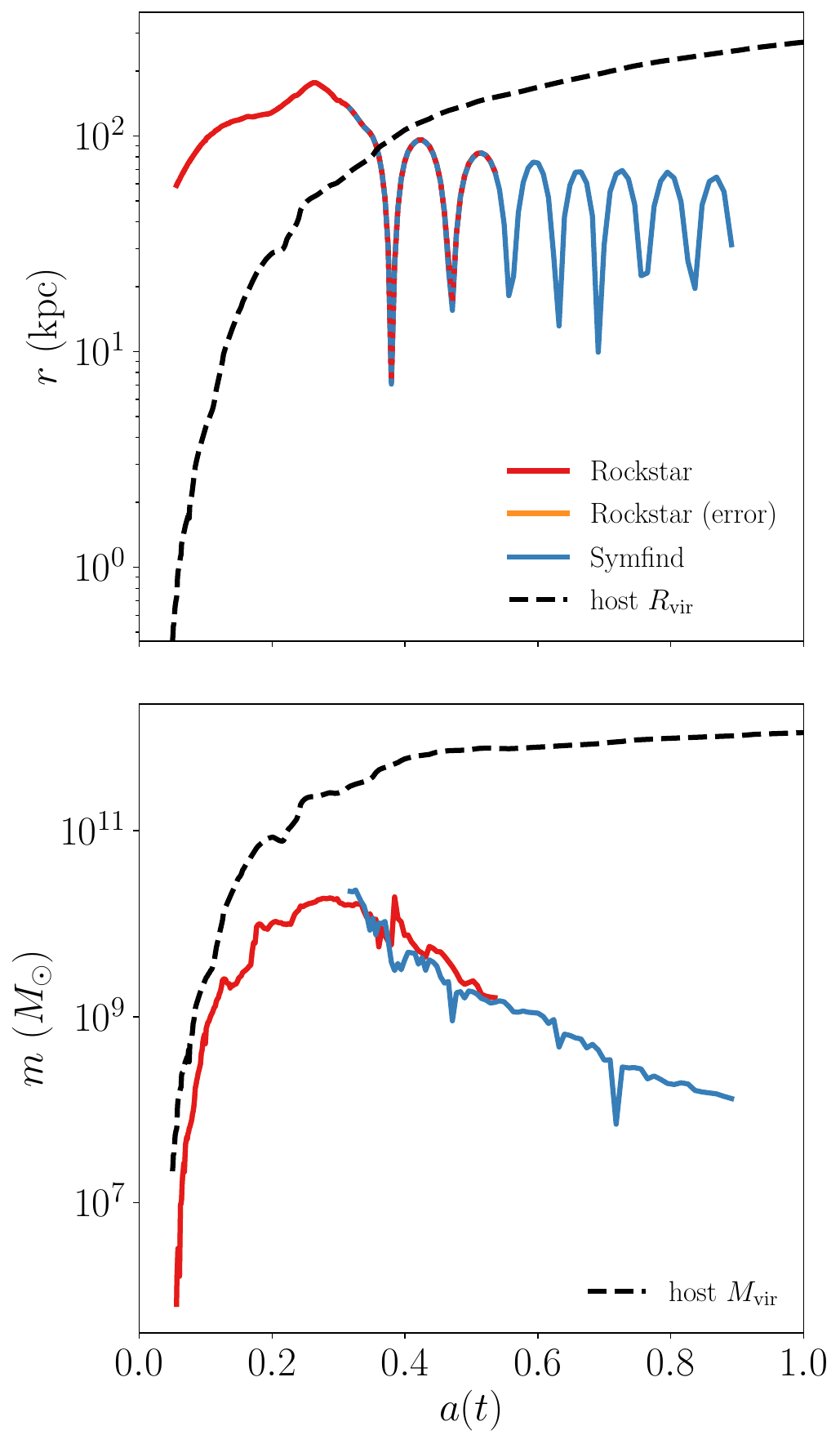}
\includegraphics[width=0.24\textwidth]{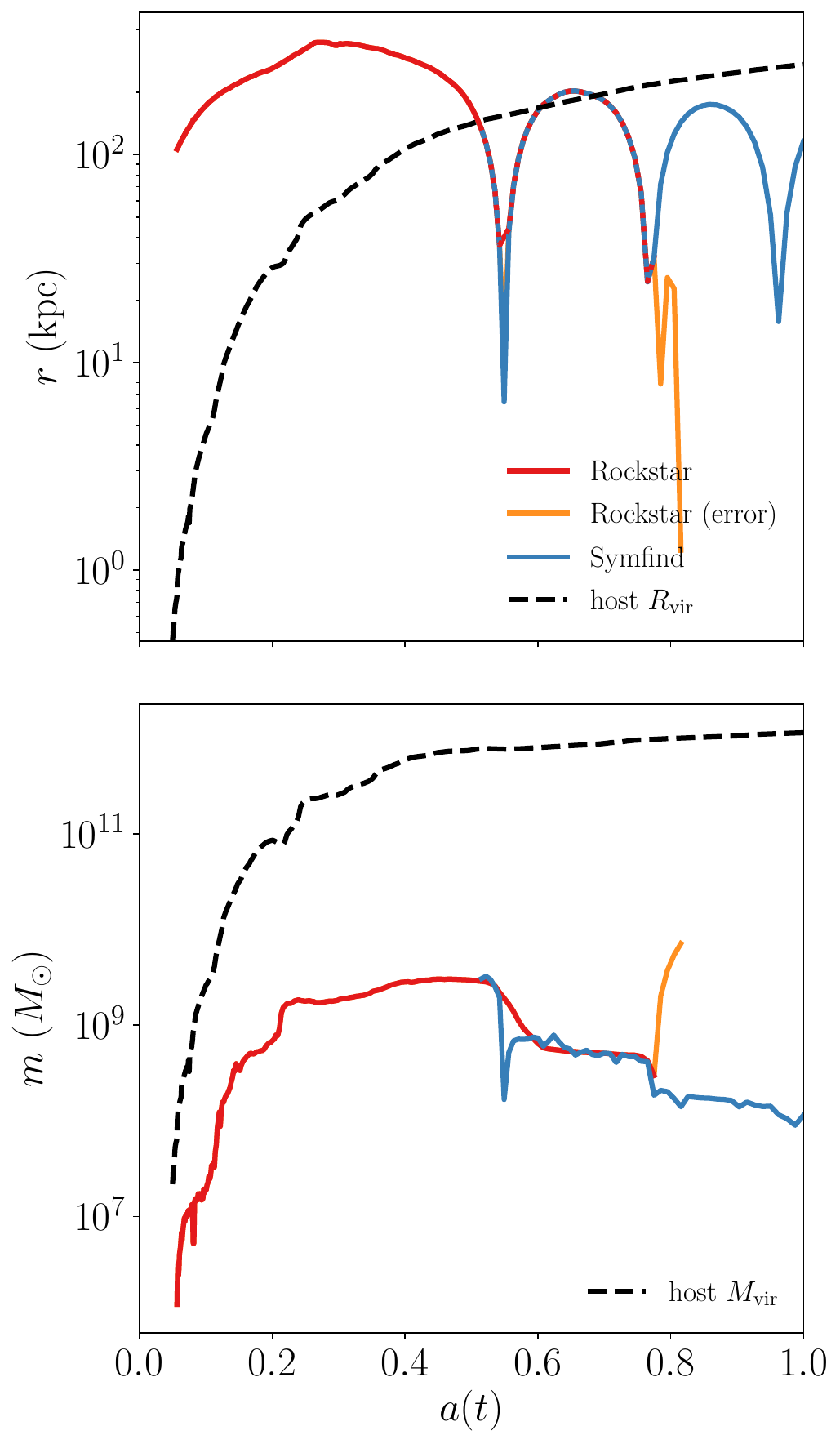}
\includegraphics[width=0.24\textwidth]{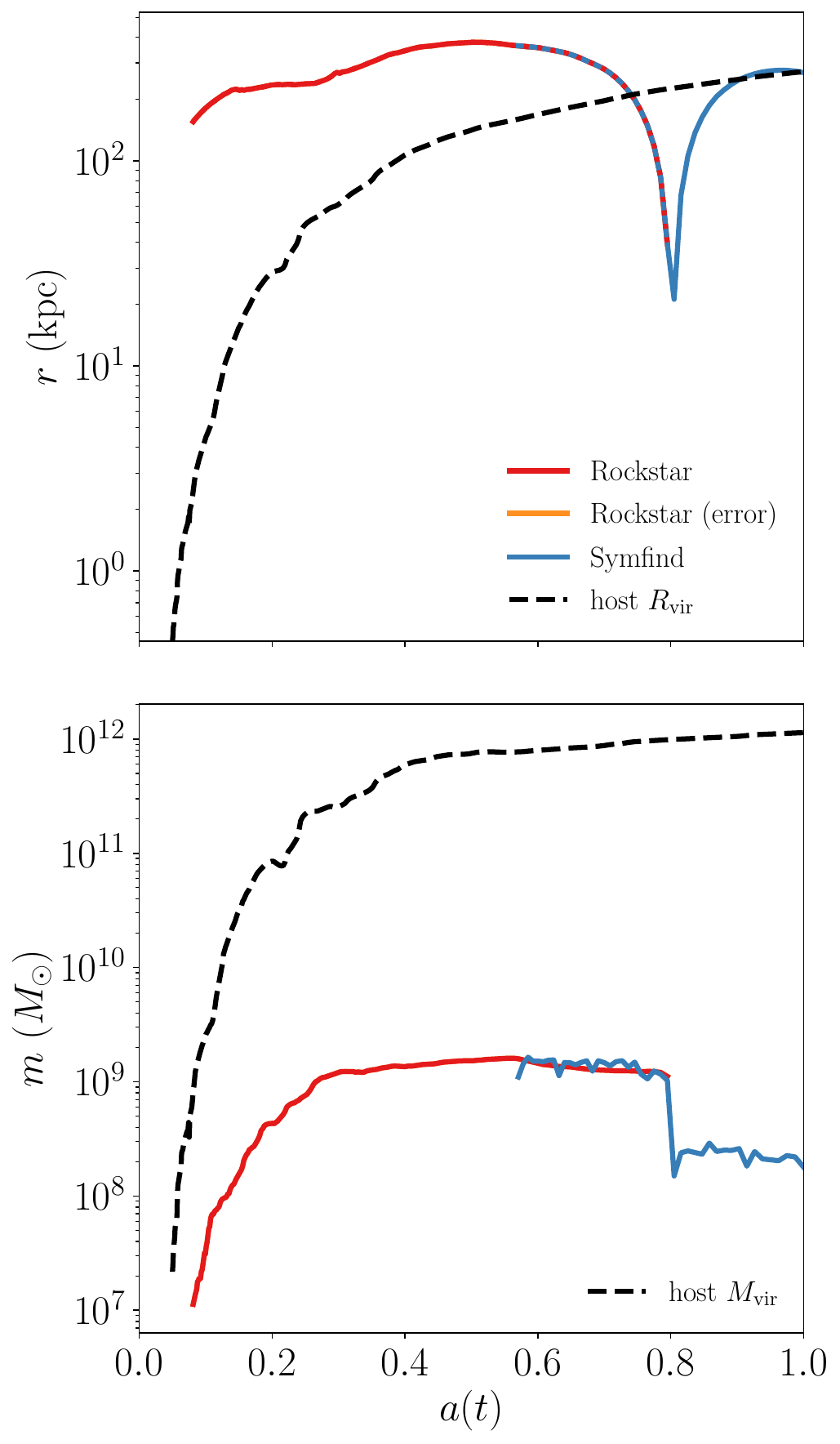}
\includegraphics[width=0.24\textwidth]{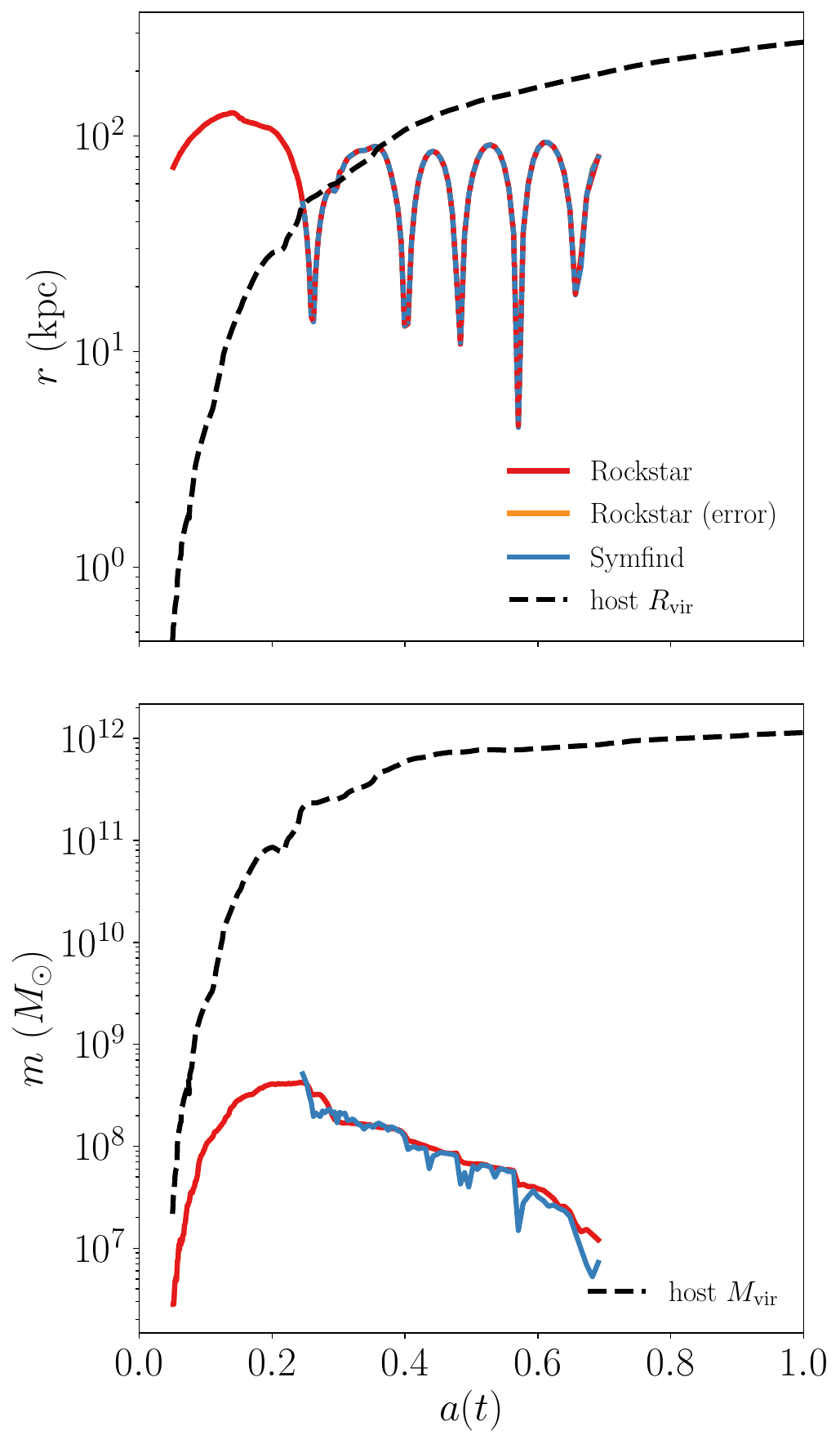}
\caption{The evolution of eight randomly selected subhalos from the first host halo, Halo023, in SymphonyMilkyWay, using the same representation as Fig.~\ref{fig:example_halo}. Each column corresponds to a different resolution range: $n_{\rm peak}\approx 5\times 10^4$ for the first column and $n_{\rm peak}\approx10^4$, $5\times 10^3$, and $10^3$ for the next three columns, respectively. Within each column, the top two rows correspond to one halo, and the bottom two rows are another halo. The first and third rows show the distance between the subhalo and its host, and the second and fourth show the mass evolution of the subhalo. \textsc{Rockstar} branches are shown in red, and particle-tracking branches are shown in blue. \textsc{Rockstar} is shown as orange when it encounters an error (see Appendix \ref{sec:parameter_selection}). The virial radius of the host halo is shown as a dashed black line in the first/third rows and the virial mass is shown as a dashed black line in the second and fourth rows. See Appendix \ref{sec:additional_halos} for discussion.}
\label{fig:additional_halos}
\end{figure*}

In Fig.~\ref{fig:additional_halos} we show eight additional randomly selected subhalos using the same format as Fig.~\ref{fig:example_halo}. These subhalos were selected from across resolutions ranging from $n_{\rm peak} \approx 5\times 10^4$ to $n_{\rm peak}\approx 10^3$ and were only selected if the subhalo disrupted in either the RCT or \textsc{Symfind} catalog. We also did not include major mergers with $m_{\rm peak}/M_{\rm vir}(t_{\rm infall}) > 0.1$. The same qualitative behavior that was seen in Fig.~\ref{fig:example_halo} is seen here: subhalos followed with \textsc{Symfind} typically outlast RCT and survive to much smaller masses before disrupting. This is consistent with the survival analysis performed in Section \ref{sec:survial}, which showed that across a large statistical sample, particle-tracked subhalos tend to out-survive RCT subhalos by about a factor of 10{-}30 in mass. Before disruption, the two methods are in good agreement with one another. Several other interesting features can be seen in these subhalos.

First, RCT has a particularly difficult time following subhalos that have lost large amounts of mass during a pericentric passage. This can be seen in the bottom halos of columns 2 and 3; the top halo in column 3 also disrupts shortly after a catastrophic pericenter. However, this is not the only way these objects disrupt. Most of the RCT disruptions occur at more-or-less random points along their orbits and can occur even if the mass is being lost very slowly.

Second, some RCT subhalos (e.g., the bottom subhalo in column 4) survive as long as their particle-tracked equivalents (and although we don't show an example here, can sometimes out-survive them). This is consistent with the analysis performed in Section \ref{sec:survial}. In Fig.~\ref{fig:survival}, we show that there is some overlap in disruption masses even at very high values of $n_{\rm peak}$ and that at low resolutions (like the aforementioned subhalo, which only has $n_{\rm peak}\approx 10^3$), those two distributions become closer and the amount of overlap increases.

Third, some subhalos rapidly lose mass during their final few snapshots (e.g., the top subhalo in column 2). This is likely due to runaway mass loss from numerical effects. It is possible that the very low $m/m_{\rm infall}$ tail of the left panel of Fig.~\ref{fig:survival} corresponds to other objects like this, meaning that it is unclear how reliable this region of the survival curve is. Even if true, this would not affect our arguments in Section \ref{sec:good_enough} about the overall impact of our method on galaxy populations, as these limits were derived from the high $m/m_{\rm infall}$ tail.

Lastly, we caution readers against over-interpreting this qualitative sample. The small size means that certain features are over/underrepresented. For example, only one RCT subhalo ends in an error, but as we show in Section \ref{sec:stitching_errors}, this is fairly common. Additionally, the fact that we have selected only subhalos that have disrupted in at least one catalog will bias this sample towards more quickly disrupting objects, meaning that one would visually overestimate the mass ratios where a typical subhalo disrupts due to right-censoring. We analyze disruption masses with a full statistical treatment that accounts for these biases in Section \ref{sec:survial}.

\section{The dependence of survival curves on subhalo mass}
\label{sec:survival_mass_dependence}

\begin{figure}
\hspace{-3.5mm}
\includegraphics[width=0.475\textwidth]{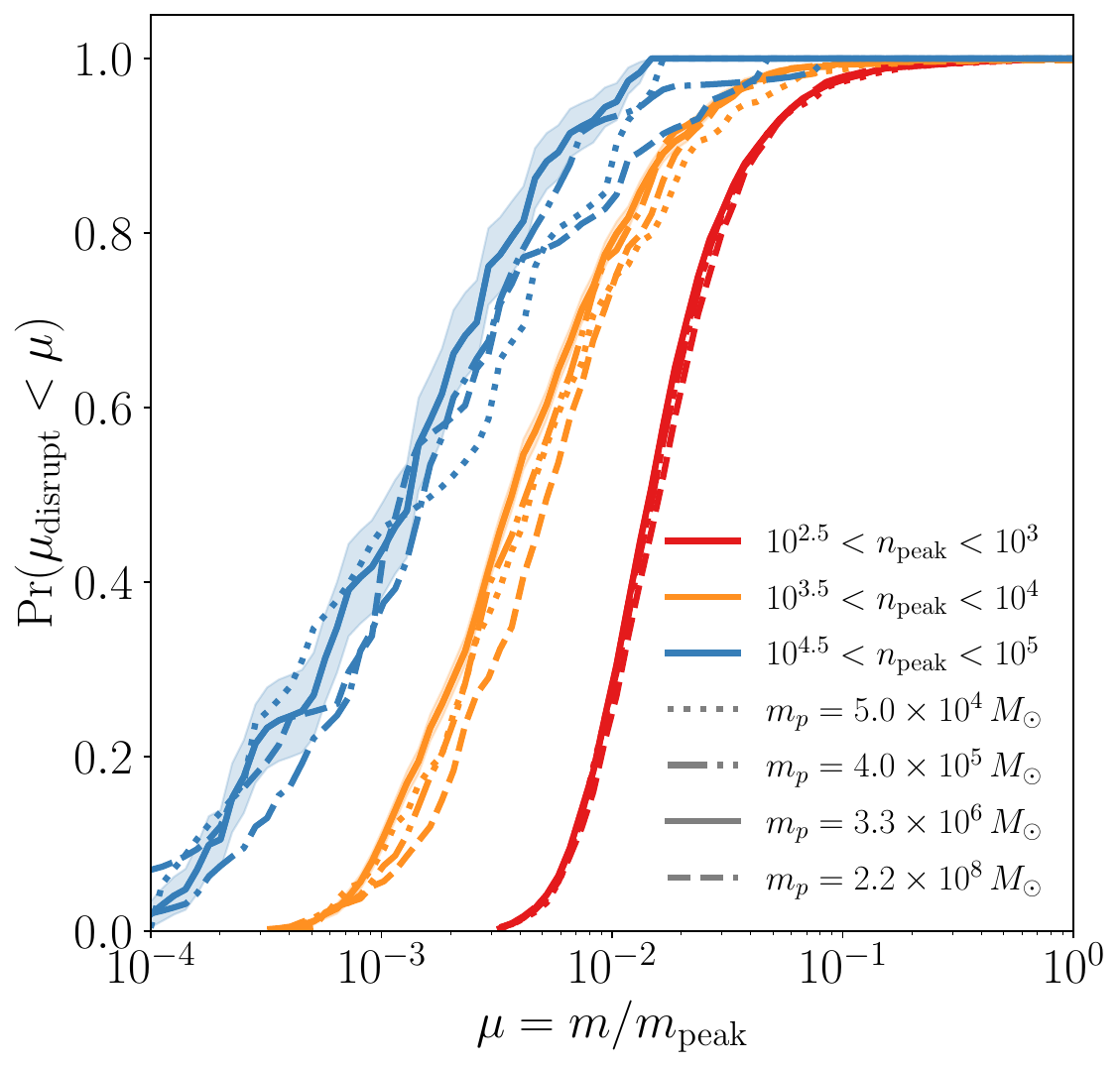}
\caption{Survival curves for several $n_{\rm peak}$ bins using a variety of particle masses (see Fig.~\ref{fig:survival}). Different $n_{\rm peak}$ bins are shown in different colors, and different particle masses are shown as different line styles. Confidence intervals estimated through Greenwood's formula are shown for $m_p=3.3\times 10^6$ and omitted for other curves. Despite the particle masses tested here spanning nearly four orders of magnitude, there is no appreciable systematic mass trend to disruption masses.}
\label{fig:survival_mass_dependence}
\end{figure}

In this Appendix, we test whether the survival curves of subhalos found with \textsc{Symfind} depend on subhalo mass/particle mass or whether they depend primarily on subhalo resolution. To do this, we extract subhalos in three resolution bins ($10^{2.5}<n_{\rm peak}<10^3$, $10^{3.5}<n_{\rm peak}<10^4$, and $10^{4.5}<n_{\rm peak}<10^5$) from the SymphonyLMC, SymphonyMilkyWay, SymphonyGroup, and SymphonyL-Cluster simulation suites ($m_p=5.0\times10^4\,M_\odot$, $m_p=3.3\times 10^6\,M_\odot$, and $m_p=2.2\times10^8\,M_\odot$, respectively.) We follow the procedure described in Section \ref{sec:survial} to construct survival curves for the nine resultant subhalo populations and show them in Fig.~\ref{fig:survival_mass_dependence}.

Fig.~\ref{fig:survival_mass_dependence} shows no appreciable dependence on $m_p,$ and consequently no dependence on $m_{\rm peak}$. Since subhalos are binned by $n_{\rm peak},$ this means that there is no dependence on $m_{\rm peak}$ either. Thus, our subhalo survival criteria are well-characterized by $n_{\rm peak}$ alone, and we are justified in combining simulation suites when computing the $n_{\rm peak}$-dependent disruption threshold for our finder.

This does not imply that the subhalo mass loss rate is independent of subhalo mass, only that subhalo finder disruption criteria are.

\section{Resolution dependence of the subhalo mass function}
\label{sec:shmf_converge}

\begin{figure*}
\hspace{-3.5mm}
\includegraphics[width=0.475\textwidth]{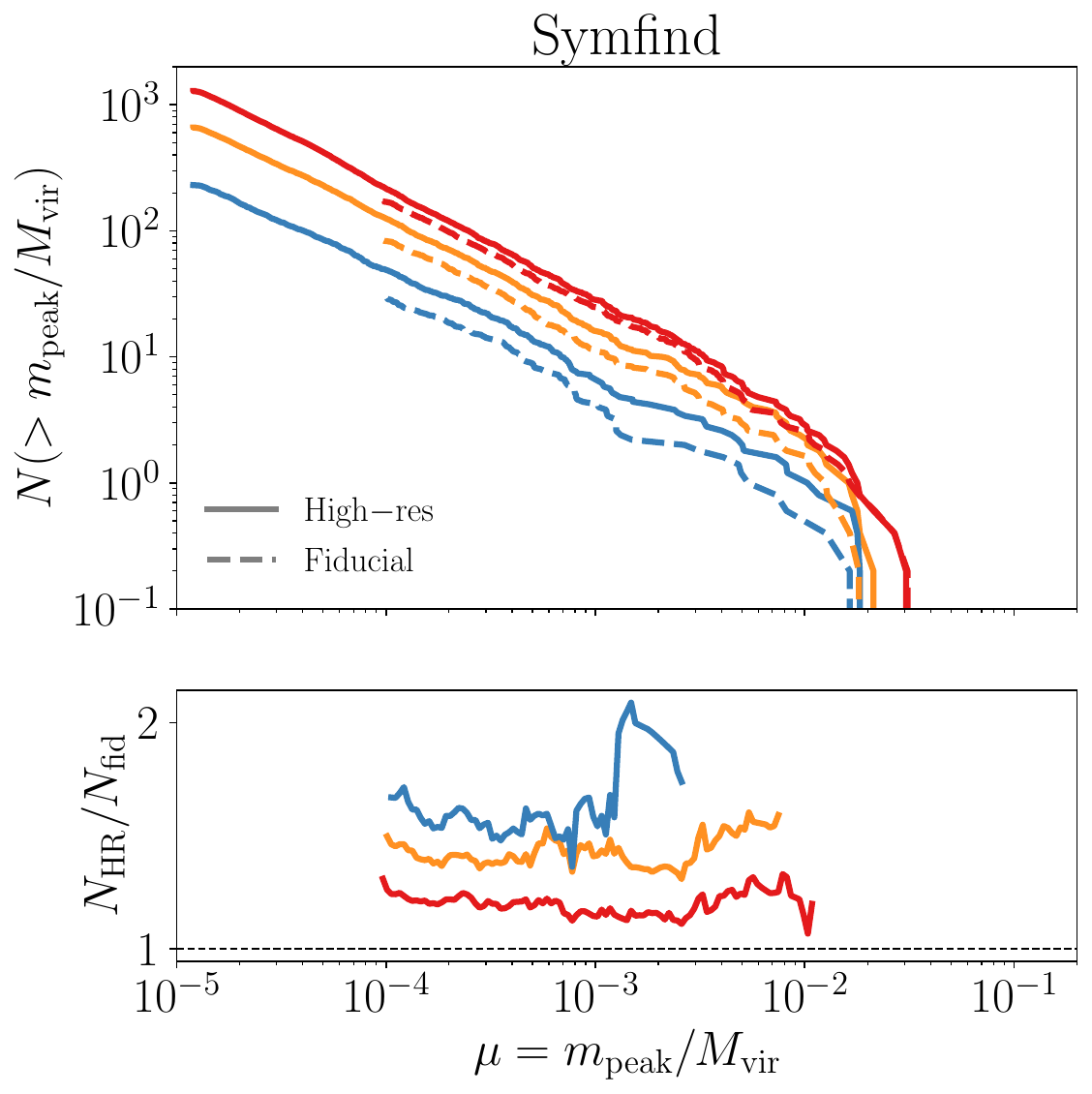}
\includegraphics[width=0.475\textwidth]{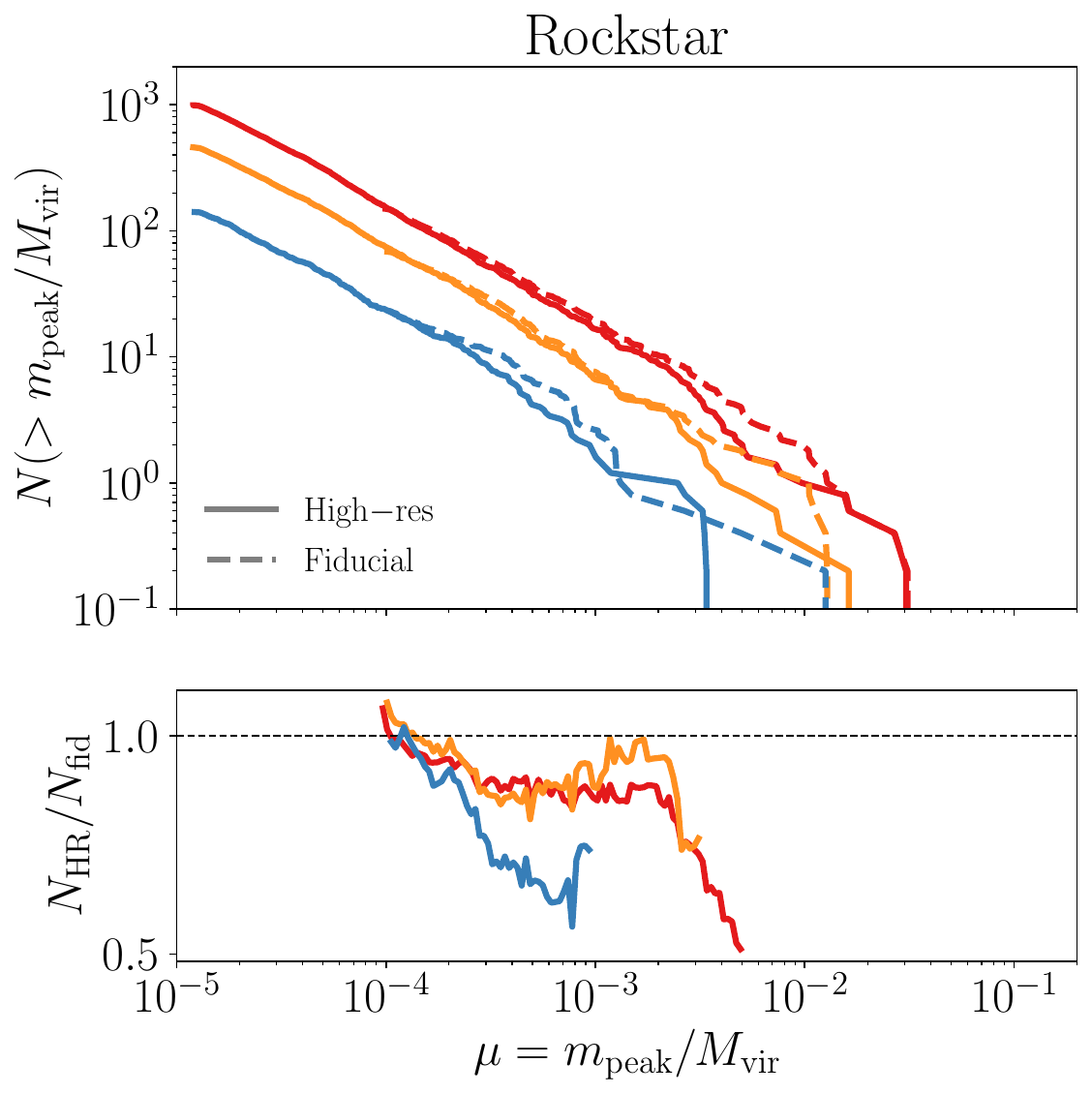}
\caption{
The $z=0$ SHMF for both fiducial-resolution and high-resolution simulations. The left panel shows \textsc{Symfind} and the right panel shows \textsc{Rockstar}. The format of these figures is the same as Fig.~\ref{fig:shmf}. \textsc{Symfind} SHMFs finds more subhalos as resolution is increased, while \textsc{Rockstar} does not. In fact, as one would expect from Fig.~\ref{fig:survival}, increasing resolution modestly {\em decreases} the number of high-mass subhalos found by \textsc{Rockstar}.
}
\label{fig:shmf_converge}
\end{figure*}

In Fig.~\ref{fig:shmf_converge}, we show the resolution dependence of the $m_{\rm peak}$ SHMF in \textsc{Symfind} and \textsc{Rockstar}. We compare the high-resolution SymphonyMilkyWayHR suite against its fiducial resolution resimulations. All other analysis choices are the same as in Section \ref{sec:shmf}.

With an ideal subhalo finder and ideal simulation, one would expect \refadd{the abundance of subhalos at a fixed $m_{\rm peak}$ to increase with increasing resolution} because subhalo remnants will be able to survive to smaller $m/m_{\rm peak}$ ratios. This effect should be stronger at small radii because the distribution of $m/m_{\rm peak}$ skews to lower values at smaller radii \citep[e.g.][]{han_2016_unified}.

When \textsc{Symfind} is used to find subhalos, increasing resolution increases the number of recovered subhalos, ranging from $\approx20\%$ increase in subhalo abundance within $R_{\rm vir}$ to a $\approx 50\%$ increase within $R_{\rm vir}/4$. This agrees with {\em a priori} expectations for the resolution dependence of the $m_{\rm peak}$ SHMF.

In contrast, RCT does not find more subhalos with increasing resolution. For most of the subhalo mass range considered here, increasing resolution actually {\em decreases} subhalo abundances by about $\approx 5\%{-}10\%$. The $r<R_{\rm vir}/4$ SHMF is consistent with there being a larger decrease in abundances at small radii, but number statistics are poor in this bin, particularly at $m_{\rm peak}/M_{\rm vir} \gtrsim 10^{-3}$, so it is difficult to make strong statements. \refadd{The convergence behavior of low particle-count and high-particle count subhalos is different. Instead of trending downwards with increasing resolution like high particle count subhalos, very low particle-count subhalos ($n_{\rm peak} \lesssim 300$) increase in abundance. Our sample cut at 300 particles removes much of the low-resolution trend, but the behavior can be seen clearly in the convergence tests in \citet{nadler_2022_symphony}. Note that \citet{nadler_2022_symphony} performs tests on the differential mass function instead of the cumulative mass function, pushing the mass scale where the transition between upwards non-convergence at low resolution and downwards non-convergence at high resolution occurs at somewhat higher masses.}

As discussed in Section \ref{sec:sims}, we have removed one of the high-resolution/fiducial host pairs which was present in the \citet{nadler_2022_symphony} tests because numerous high-mass subhalos in the high-resolution run were never accreted in the fiducial-resolution run. Given the small number of hosts in the sample, this may also impact the exact mass where the transition occurs.

The behavior seen in our RCT tests and in the \citet{nadler_2022_symphony} tests can be explained by a combination of several effects. In the absence of numerical effects, convergence of the $m_{\rm peak}$ SHMF is driven by the $n_{\rm peak}$ dependence of the survival curves shown in Fig.~\ref{fig:survival}. Flat survival curves would lead to the same number of subhalos being recovered. Decreasing survival curves (as is seen for \textsc{Symfind}) lead to more recovered subhalos as resolution increases, and increasing survival curves (as is seen for RCT) lead to fewer recovered subhalos as resolution is increased. The amount of increase/decrease also depends on the uncensored $m_{\rm peak}$-dependent $m/m_{\rm peak}$ distribution, which cannot be predicted {\em a priori}.

\refadd{Another important effect is the impact of numerical biases, such as those discussed in Sections \ref{sec:numerical_limits} and \ref{sec:mass_loss_rates}. Given that increasing resolution decreases the strength of the numerical effects that lead to early subhalo disruption, and that the mass dependence of numerical effects is stronger than the mass dependence of increased subhalo finder completeness (Fig.~\ref{fig:survival_vmax}), one might be surprised that subhalo finder incompleteness apparently dominates. However, as Fig.~\ref{fig:survival_vmax} shows, at the $n_{\rm peak}$ values considered in this study, numerical effects only become relevant at relatively low $m/m_{\rm peak}$, meaning that they are only a concern for subhalo finders that are sensitive to large degrees of mass loss. (In a sense, numerical issues are lurking in the hard-to-reach parts of mass loss space, analogous to how a deep-sea kraken might be a serious problem for a submarine, but not a battleship.)

Quantitatively, the exact convergence behavior of a subhalo finder would need to be calculated through a combination of the true, fully converged $m/m_{\rm peak}$ distribution, the $m/m_{\rm peak}$-dependent completeness of the halo finder, and systematic biases impacted by numerical effects. We do not attempt to perform this modeling here.}


Can the behavior of RCT be referred to as ``false convergence?'' In the sense that it would be easy for a researcher to look at the $r<R_{\rm vir}$ curves and conclude that the low-mass SHMF isn't changing much, yes. Strictly, however, RCT $m_{\rm peak}$ functions are unconverged {\em in the wrong direction}. 

\section{The radial distribution of SUBFIND subhalos}
\label{sec:subfind_radius_cdf}

\begin{figure}
\hspace{-3.5mm}
\includegraphics[width=0.475\textwidth]{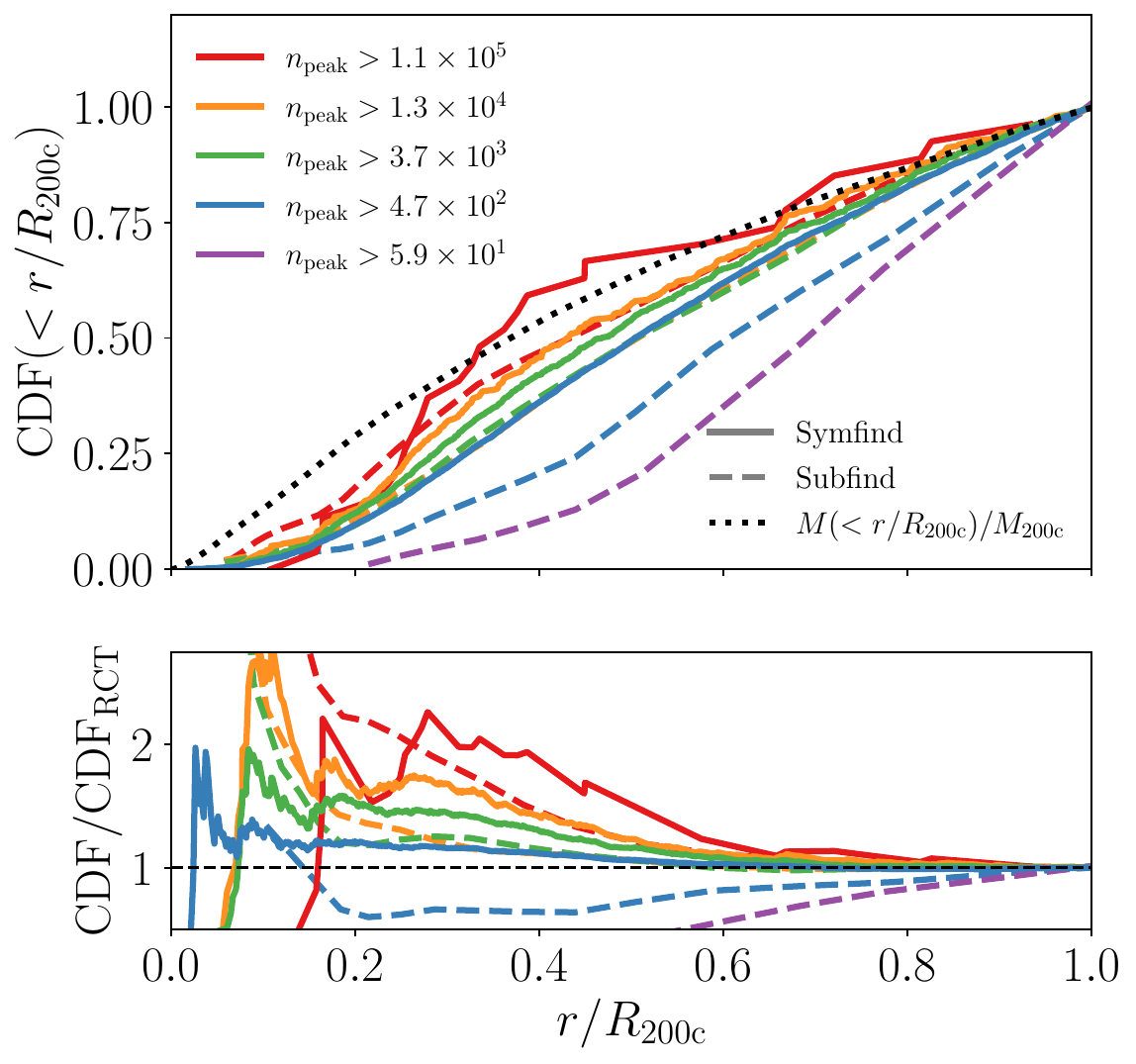}
\caption{Comparison between the radial distribution of subhalos as measured with \textsc{Subfind} and particle-tracking. This Figure is the same as Fig.~\ref{fig:radius_cdf}, except that the dashed curves show \textsc{Subfind} subhalos from Aq-A-1 (adapted from \citealp{han_2016_unified}), and mass definitions and analysis bins have been altered to match the choices made in \citet{han_2016_unified} (see Appendix \ref{sec:subfind_radius_cdf} for discussion). Unlike RCT, \textsc{Subfind} strongly depends on subhalo resolution, although it converges more slowly than particle-tracking.}
\label{fig:subfind_radius_cdf}
\end{figure}

In Section \ref{sec:radius_cdf}, we studied the radial distribution of subhalos with \textsc{Symfind} and RCT as a function of $n_{\rm peak}.$ We found that increasing resolution led to no change in the radial distribution of RCT subhalos. Increasing resolution caused particle-tracking profiles to become more concentrated and approach the radial distribution of dark matter around the host center. We argued that this was due to false convergence in RCT. This leads to a natural follow-up question: is false convergence specific to RCT or is it also seen in other halo finders? In this Appendix, we perform the same test on the \textsc{Subfind} halo finder \citep{springel_2001_populating} and the \textsc{DTree} merger tree code \citep{jiang_2014_dtree}. We consider this a preliminary (as opposed to definitive) attempt at answering this question and defer a more exhaustive approach to future work. For lexical simplicity, in this Section, we will use the term ``\textsc{Subfind}'' to refer to the combination of both codes.

We use data from the Aquarius project \citep{springel_2008_aquarius}, a series of high-resolution, dark-matter-only simulations of Milky Way-mass halos. The highest resolution of these simulations, Aq-A-1, contains particles with masses of $1.7\times 10^3\,{\rm M}_\odot$ and was resimulated at four coarser resolutions, spaced by factors of 8 in particle mass. We adapt Fig.~5 in \citet{han_2016_unified}, which measured subhalo radial profiles for Aq-A-1 and its resimulations as a function of resolution, to the same format as our Fig.~\ref{fig:radius_cdf}. \citet{han_2016_unified} make several different analysis choices compared to Fig.~\ref{fig:radius_cdf}:
\begin{enumerate}
\item The radius of the host halo is taken to be the overdensity radius corresponding to 200 times the critical density of the universe, replacing $R_{\rm vir}$ with $R_{\rm 200c}$.
\item A different set of resolution bins are used.
\item No cut is made to remove subhalos with high subhalo-to-host mass ratios at infall.
\item The set of resolution tests from \citet{han_2016_unified} that we compare against are performed at a fixed mass across resimulations rather than across different mass bins within the same simulation.
\item Profiles are measured for a single host halo, rather than averaged across a large population of host halos.
\item Aq-A-1 is higher mass than any of the hosts in SymphonyMilkyWay.
\item The locations of the most-bound particles at infall were not used to confirm that late-time subhalos were truly the descendants of the initial halos they are connected to.
\end{enumerate}

With these differences in mind, we show the dependence of the radial distribution of satellites on resolution in Fig.~\ref{fig:subfind_radius_cdf}. \textsc{Subfind} is shown as dashed lines, while the results for particle-tracking are shown as solid lines. Profiles are normalized relative to Eq.~\ref{eq:cdf_rct}, an accurate fit to the radial distribution of >300-particle satellites in RCT. Analysis in the particle-tracking curves has been altered to use the same conventions as \citet{han_2016_unified} for points 1 through (3 above. We cannot match choices 4 through 6 due to the difference in simulation suites. We opt not to remove restriction 7 from the particle-tracking data. We discuss the limitations that these differences impart on our analysis later in this Appendix. We only apply particle-tracking to subhalos with $n_{\rm peak} > 300.$ This means we cannot include a particle-tracking curve for the lowest-resolution, 59-particle bin.

Unlike RCT, \textsc{Subfind} appears not to converge to a false profile. As resolution improves, subhalo profiles become increasingly centrally concentrated. However, below $\lesssim 2.7\times 10^3{-}4.6\times10^2$ particles, \textsc{Subfind} substantially underestimates the abundance of subhalos at small radii, and improvements above RCT only begin once subhalos reach high resolutions $n_{\rm infall} \gtrsim 1.3\times 10^{4}{-}1.1\times 10^5$.

This is very roughly consistent with Fig.~40 in \citet{springel_2021_gadget_4}, which found that \textsc{Subfind-HBT} subhalos have longer main branch lengths than RCT subhalos at high resolutions and shorter main branches at low resolutions and that the performance of \textsc{Subfind} and \textsc{Subfind-HBT} are able to track subhalos for comparable amounts of time.

Unfortunately, this apparent (and desirable) non-convergence is highly dependent on the last, highest-resolution bin. The $n_{\rm infall} > 3.7 \times 10^3$ and $n_{\rm infall} > 1.3 \times 10^4$ bins agree with one another, meaning that without the highest-resolution bin, the impression given by this analysis would be that at low resolutions, \textsc{Subfind} suppresses inner structure and converges to a profile very similar to \textsc{Rockstar} as resolution is increased. Therefore, interpreting this plot depends sensitively on how the three highest-resolution bins are accounted for. We lay out several possible explanations for this behavior.

The simplest possibility is that the limited size of the sample leads to either an upwards fluctuation in the highest-resolution bin or a downward fluctuation in the second-highest bin. The \citet{han_2016_unified} cuts correspond to $m_{\rm infall} > 1.9\times 10^8$ across all resolution bins, which would lead to approximately $1{-}2\times10^3$ subhalos, given Aq-A-1's host mass. The Kolmogorov-Smirnov statistic between the two highest-resolution bins is 0.12, which has a less than one-in-$10^6$ chance of occurring if those two bins were truly drawn from the same underlying distribution. The chances of the $n_{\rm infall} > 1.3\times 10^4$ bin fluctuating downwards from some hypothetical intermediate position between the $n_{\rm infall} > 1.1\times 10^5$ and $n_{\rm infall} > 3.7\times 10^3$ bins is higher but still unlikely: the probability of observing a KS statistic larger than 0.06 between two 1000-point samples drawn from the same distribution is less than 3\%. So this type of fluctuation, while possible, is unlikely.

That said, it is still possible that a {\em correlated} fluctuation could cause the red curve to skew high. This would not be accounted for by the Kolmogorov-Smirnov statistic, which assumes that all samples from both distributions are independent. Subhalos are often accreted in subgroups that contain many objects, which would increase the probability of large fluctuations in the CDF.

Another possibility is that the lack of additional correctness checks (as suggested in Section \ref{sec:recommendations} and Appendix \ref{sec:tree_post_processing}) on \textsc{Subfind} outputs leads to high-resolution subhalos being incorrectly linked with unrelated subhalos, merger remnants, and other low-radius detritus. As we show in Fig.~\ref{fig:rockstar_error_rate} and Fig.~\ref{fig:parameter_tests}, such behavior is common in RCT, and there is no {\em a priori} reason to expect it to expect it to be less common in \textsc{Subfind}. Such errors have already been encountered by \citet{han_2018_hbt_plus} when comparing the $m_{\rm peak}$ mass functions of \textsc{HBT+} and \textsc{Subfind}.  Fig.~\ref{fig:rockstar_error_rate} shows that the behavior of such merger tree errors can have a complex relationship with radius and resolution: the effect of an RCT-like error in \textsc{Subfind} would be to increase the apparent relative abundance of high $n_{\rm infall}$ subhalos at small radii. However, these errors tend to be short-lived in RCT (e.g.,~Fig.~\ref{fig:example_halo}), so to make an appreciable impact on the radial distribution, they would need to be longer-lived in \textsc{Subfind}. The lack of these additional correctness checks is the reason we have not included survival curve analysis for \textsc{Subfind} subhalos based on public merger tree datasets in this paper.

The final possibility is that Fig.~\ref{fig:subfind_radius_cdf} is an accurate representation of the change in the radial distribution of real subhalos as a function of resolution. This would require that the survival function of \textsc{Subfind} improves continuously with resolution, ``stalls'' across a decade in resolution, and then begins to improve again. While odd, this is possible: halo finders and merger trees are complex algorithms, not physical systems, and their behavior can pick up non-obvious characteristic resolution scales (e.g., Fig.~\ref{fig:false_mergers}). This topic would benefit from further study, particularly with better number statistics and error-detection methods.

We have also overplotted the radial distribution of subhalos measured through our particle-tracking method. If taken at face value, the comparison implies that particle-tracking leads to radial distributions that are roughly as concentrated as would be measured by \textsc{Subfind} in a simulation with $\approx\times8$ higher resolution. However, our current comparison does not allow for a high degree of confidence in this statement. Our particle-tracking profiles are not measured at a fixed mass, meaning that higher-resolution bins experience more dynamical friction, which could lead to more concentrated radial distributions. While ideally this comparison should be done with apples-to-apples subhalo samples, neither RCT (Fig.~\ref{fig:radius_cdf}; see also \citealp{vdb_2016_segregation}) nor the distribution of most-bound particles in subhalos (Fig.~3 in \citealp{han_2016_unified}) shows evidence for dynamical friction having a strong effect on low-mass subhalo profiles. We have also compared radial CDFs between SymphonyMilkyWay and SymphonyMilkyWayHR and found them in good agreement with one another at fixed resolution levels, meaning that we do not see evidence for a mass dependence.

Additionally, rather than a statistical sample, Aq-A-1 is a single halo, allowing for correlated fluctuations in its subhalo population. Aq-A-1 has a slightly more centrally concentrated subhalo number density profile than the average host halo in its mass range \citep{hellwing_2016_coco}, meaning that with a fixed subhalo-finding method, Aq-A-1 will appear to approach large concentrations faster than an average SymphonyMilkyWay halo.

In light of this discussion, we summarize the findings of this Appendix as follows:
\begin{itemize}
\item There is a high likelihood that radial number density profiles measured with \textsc{Subfind}, RCT, and our particle-tracking at a given infall mass have very different convergence behavior. This is in conflict with a common reading of the subhalo finder comparison literature, as we discuss at length in Section \ref{sec:method_comp}.
\item There is a high likelihood that at moderate-to-low $n_{\rm peak}$ ($\approx 100'{\rm s}$ of particles), \textsc{Subfind} is missing a meaningful number of low-radius subhalos relative to RCT and our method.
\item Several factors prevent us from unambiguously determining whether \textsc{Subfind} converges to a false radial number density profile at a fixed $n_{\rm peak}$, as RCT does, but the most straightforward reading of Fig.~\ref{fig:subfind_radius_cdf} is that it does not.
\end{itemize}

\label{lastpage}
\end{document}